\documentclass[
aps,
preprintnumbers,
eqsecnum,
superscriptaddress,
showpacs,
nofootinbib,
notitlepage
]{revtex4-1}

\usepackage[T1]{fontenc}
\usepackage[utf8]{inputenc}
\usepackage{amsmath}
\usepackage{amsfonts}
\usepackage{amssymb} 
\usepackage{amsxtra}
\usepackage{array}
\usepackage{graphicx}
\usepackage{hepunits}
\usepackage{units}
\usepackage{xspace}
\usepackage[dvipsnames]{xcolor}
\usepackage{footmisc}
\usepackage{hyperref}
\usepackage[normalem]{ulem}

\DeclareMathOperator{\sgn}{sgn}

\DeclareMathOperator\arctanh{arctanh}

\newcommand{\refeq}[1]{Eq.~\eqref{#1}}
\newcommand{\refeqs}[2]{\textrm{Eqs}.~(\ref{#1})-(\ref{#2})}
\newcommand{\reftab}[1]{Tab.~\ref{#1}}
\newcommand{\reffig}[1]{Fig.~\ref{#1}}
\newcommand{\refcite}[1]{Ref.~\cite{#1}}
\newcommand{\refscite}[1]{Refs.~\cite{#1}}
\newcommand{\refsec}[1]{Sec.~\ref{#1}}
\newcommand{\refapp}[1]{App.~\ref{#1}}

\newcommand{\ie}{\textit{i.e. }}
\newcommand{\eg}{\textit{e.g. }}
\newcommand{\etal}{\textit{et al.\xspace}}
\newcommand{\lhs}{l.h.s.\xspace}
\newcommand{\rhs}{r.h.s.\xspace}
\newcommand{\apriori}{\textit{a priori}\xspace}

\newcommand{\ds}{Dyson-Schwinger\xspace}

\newcommand{\bs}{Bethe-Salpeter\xspace}

\newcommand{\ket}[1]{\ensuremath{\left|#1\right\rangle}\xspace}
\newcommand{\bra}[1]{\ensuremath{\left\langle #1\right|}\xspace}

\newcommand{\Radon}[1]{\mathcal{R}#1}
\newcommand{\fRadon}[3]{\mathcal{R}#1\left(#2, #3\right)}

\newcommand{\wl}[2]{\ensuremath{\left[#1;#2\right]}\xspace}

\newcommand{\vperp}[1]{\ensuremath{\textbf{#1}_\perp}}

\newcommand{\hbmks}[0]{h_{\textrm{BMKS}}}
\newcommand{\chibmks}[0]{\chi_{\textrm{BMKS}}}
\newcommand{\Fcbmks}[0]{F_{\textrm{BMKS}}}

\newcommand{\hp}[0]{h_{\textrm{P}}}
\newcommand{\chip}[0]{\chi_{\textrm{P}}}
\newcommand{\Fcp}[0]{F_{\textrm{P}}}

\newcommand{\hpw}[0]{h_{\textrm{PW}}}
\newcommand{\Fcpw}[0]{F_{\textrm{PW}}}
\newcommand{\Gcpw}[0]{G_{\textrm{PW}}}
\newcommand{\Dcpw}[0]{D_{\textrm{PW}}}
\newcommand{\HDGLAP}[0]{H_{|\textrm{DGLAP}}}

\newcommand{\htoymodelbmks}[0]{h_{\textrm{BMKS}}^{\textrm{Toy}}}
\newcommand{\htoymodelp}[0]{h_{\textrm{P}}^{\textrm{Toy}}}
\newcommand{\htoymodelpsing}[0]{h_{\textrm{P, Sing}}^{\textrm{Toy}}}
\newcommand{\htoymodelpreg}[0]{h_{\textrm{P, Reg}}^{\textrm{Toy}}}
\newcommand{\Htoymodel}[0]{H^{\textrm{Toy}}}
\newcommand{\HtoymodelDGLAP}[0]{H_{|\textrm{DGLAP}}^{\textrm{Toy}}}
\newcommand{\HtoymodelERBL}[0]{H_{|\textrm{ERBL}}^{\textrm{Toy}}}
\newcommand{\Htoymodelreg}[0]{H^{\textrm{Toy, Reg}}}
\newcommand{\HtoymodelregDGLAP}[0]{H_{|\textrm{DGLAP}}^{\textrm{Toy, Reg}}}
\newcommand{\HtoymodelregERBL}[0]{H_{|\textrm{ERBL}}^{\textrm{Toy, Reg}}}

\newcommand{\Fctoymodelp}[0]{F_{\textrm{P}}^{\textrm{Toy}}}
\newcommand{\Gctoymodelp}[0]{G_{\textrm{P}}^{\textrm{Toy}}}
\newcommand{\Fctoymodelpw}[0]{F_{\textrm{PW}}^{\textrm{Toy}}}
\newcommand{\Gctoymodelpw}[0]{G_{\textrm{PW}}^{\textrm{Toy}}}

\begin{document}

\title{Covariant Extension of the GPD overlap representation at low Fock states}

\author{N. Chouika}
\affiliation{IRFU, CEA, Université Paris-Saclay, F-91191 Gif-sur-Yvette, France}
\author{C. Mezrag}
\affiliation{Physics Division, Argonne National Laboratory, Argonne IL 60439,USA}
\affiliation{Istituto Nazionale di Fisica Nucleare, Sezione di Roma, P.le A. Moro 2, I-00185 Roma, Italy}
\author{H. Moutarde}
\affiliation{IRFU, CEA, Université Paris-Saclay, F-91191 Gif-sur-Yvette, France}
\author{J.~Rodr\'iguez-Quintero}
\affiliation{Dpto. Ciencias Integradas, Centro de Estudios Avanzados en Fis., Mat. y Comp., Fac. Ciencias Experimentales, Universidad de Huelva, Huelva 21071, Spain}

\begin{abstract}
We present a novel approach to compute Generalized Parton Distributions 
within the Lightfront Wave Function overlap framework.
We show how to systematically extend Generalized Parton Distributions computed
within the DGLAP region to the ERBL one, 
fulfilling at the same time both the polynomiality and positivity conditions. 
We exemplify our method using pion Lightfront Wave Functions inspired by recent results 
  of non-perturbative continuum techniques and algebraic nucleon Lightfront Wave
  Functions.
  We also test the robustness of our algorithm on reggeized phenomenological
  parameterizations.
  This approach paves the way to a better understanding 
  of the nucleon structure
  from non-perturbative techniques and to a unification of Generalized Parton
  Distributions and Transverse Momentum Dependent Parton Distribution Functions
  phenomenology through Lightfront Wave Functions.
\end{abstract}

\maketitle

\tableofcontents{}

\section{Introduction}

Generalized Parton Distributions (GPDs) were introduced two decades 
ago \cite{Mueller:1998fv, Ji:1996nm, Radyushkin:1997ki} and have been since then
a topic of strong interest in the hadron physics community, both on the 
theoretical and experimental side \cite{Ji:1998pc,
Goeke:2001tz, Diehl:2003ny, Belitsky:2005qn, Boffi:2007yc, Guidal:2013rya,
Mueller:2014hsa}.
 A significant amount of beam time of the upgraded Jefferson Laboratory 
 facility is dedicated to their experimental study and they are one of the core
 scientific cases for building the U.S. Electron-Ion Collider (EIC).
Not only are they an elegant way to unify both Parton Distribution Functions
(PDFs) and Form Factors into a single object, 
leading to a three-dimensional picture of hadrons \cite{Burkardt:2000za}, 
but they also provide genuine information and constraints through their 
skewness dependence.
If these constraints have been used to produce phenomenological 
models \cite{Guidal:2004nd,
Goloskokov:2005sd, Polyakov:2008aa, Kumericki:2009uq, Goldstein:2010gu}, none of
these models is built within a formalism guaranteeing \apriori all the
constraints to be fulfilled at the same time.
The same problems also appears from non-perturbative computations of GPDs 
using various techniques (see \eg \refcite{Mezrag:2016hnp, Tiburzi:2017brq}
and refs therein),
and only in very few cases particular models have taken care of both constraints (see \eg \refscite{Broniowski:2007si,Dorokhov:2011ew}).

In this paper, we focus on two of them, the so-called polynomiality and positivity 
properties.
The former states that the Mellin Moments of the GPDs are polynomials (of
definite degree) of the skewness. It comes from the
Lorentz structure of the matrix elements of local operators defining these Mellin moments.
On the other hand, positivity gives an upper bound on the absolute value of GPDs 
taken at a given kinematics in terms of PDFs.
This property can be derived from Hilbert space norms. Consequently, covariant
approaches usually successfully recover polynomiality but have hard time dealing 
with positivity.
On the other hand, quantum mechanical based techniques, like Fock states
expansions, have the positivity property ``built-in'' but usually fail at
getting polynomial Mellin moments. Facing this issue, model builders have
generally to favor one of these two fundamental properties, at the risk
of violating the other.

We present here a general solution to this problem.
Section \ref{sec:gpd-theory-modeling} will remind the reader the details and subtleties of modeling GPDs through the ways mentioned above.
Section \ref{sec:radon-inverse} introduces the theory behind our technique,
while in section \ref{sec:numerical-implementation} we present our  
method, based on the numerical inversion of the Radon transform with incomplete
data.
Section \ref{sec:applications} is then dedicated to the applications of our
algorithm to different examples of Lightfront Wave Functions (LFWFs). 
In section \ref{sec:discussion} we discuss 
our results and the phenomenological relevance of ambiguities related to the
inverse of the Radon transform.
Section \ref{sec:conclusion} concludes this paper.

\section{GPD Theory and Modeling \label{sec:gpd-theory-modeling}}

GPDs are defined as a Lightfront projection of a non-diagonal hadronic matrix element of a bi-local operator.
For the sake of simplicity, we will consider only the case of a twist-2
chiral-even quark GPD of a spin-0 hadron (\eg a pion):
\begin{equation}
  \label{eq:GPDDefinition}
  H^{q}\left(x,\xi,t\right)=\frac{1}{2}\int\frac{\mathrm{d}z^{-}}{2\,\pi}\,e^{i\,x\,P^{+}z^{-}}\left.\left\langle
  P+\frac{\Delta}{2}\right|\bar{\psi}^q\left(-z\right)\gamma^{+}\psi^q\left(z\right)\left|P-\frac{\Delta}{2}\right\rangle
  \right|_{z^{+}=0,\,z_{\perp}=0}\,,
\end{equation}
where $P$ (resp. $\Delta$) is the momentum average (resp. transfer)
of the hadron states, $t=\Delta^{2}$ and $x$ (resp.
$\xi=-\frac{\Delta^{+}}{2\,P^{+}}$) is the longitudinal momentum fraction average (resp. transfer) of
the quarks. 
$q$ classically stands for the quark flavor. 
Any four-vector $v$ is expressed in light cone coordinates through $v^{\pm} =
(v^0 \pm v^3)/\sqrt{2}$ and $v=(v^+, v_\perp, v^-)$.

From definition (\ref{eq:GPDDefinition}), it
is straightforward to realize that one gets back the PDFs in the $\Delta = 0$ limit, and that integrating over $x$ yields
the quark contribution to the electromagnetic form factor.
On top of these properties, the physical GPD domain obeys $t < t_{\rm min} = -\frac{4\xi^2 M_H^2}{1-\xi^2}$, $M_H$ being the mass of the considered hadron, and $\xi \in [-1, +1]$.
In this domain, the support is such that $x \in [-1,1]$.
Due to time
reversal invariance, the GPDs defined through the operator of \refeq{eq:GPDDefinition} are even in $\xi$.
We will use this property to restrict to $\xi \geq 0$ in this paper (unless
explicitly stated otherwise).

In this section, we remind the reader the
important frameworks that allow to fulfill either positivity or polynomiality, and their consequences on modeling.

\subsection{Lightfront Wave Functions and positivity}
\label{subsec:overlap}

Lightfront quantization allows the expansion of a hadron state
$\ket{P,\lambda}$ of momentum $P$ and polarization $\lambda$ on a Fock basis:
\begin{equation}
  \label{eq:FockDecomposition}
  \ket{H;P,\lambda} = \sum_{N,\beta} \int [\textrm{d}x]_N
  [\textrm{d}^2\vperp{k}]_N
  \Psi^\lambda_{N,\beta}\left(x_{1},\mathbf{k}_{\perp1},...,x_{N},\mathbf{k}_{\perp N}\right) \ket{N,\beta;k_1,\dots,k_N},
\end{equation}
where the $\ket{N,\beta;k_1,\dots,k_N}$ denote the $N$-particles partonic
states with each particle carrying a momentum $k_i$.
$\beta$ stands for the relevant quantum numbers.
These states are weighted by the LFWFs
$\Psi^\lambda_{N,\beta}$, containing the non-perturbative physics, and
normalized as follows:
\begin{equation}
\sum_{N,\beta}\int\left[\mathrm{d}x\right]_{N}\left[\mathrm{d}^{2}\mathbf{k}_{\perp}\right]_{N}\left|\Psi_{N,\beta}^{\lambda}\left(x_{1},\mathbf{k}_{\perp1},...\right)\right|^{2}=1\,.\label{eq:lcwf_normalization}
\end{equation}
The measure element in \refeq{eq:FockDecomposition} fulfills momentum
conservation by construction:
\begin{align}
  \label{eq:DefMeasuresLCx}
  [\textrm{d}x]_N  =  &\prod_{i=1}^N \textrm{d}x_i \ \delta \left(1 - \sum_{i=1}^Nx_i\right), \\
  \label{eq:DefMeasuresLCk}
  [\textrm{d}^2\vperp{k}]_N  = 
  &\frac{1}{(16\pi^3)^{N-1}}\left(\prod_{i=1}^N\textrm{d}^2\textbf{k}_{\perp i}\right)
  \delta^2 \left(\sum_{i=1}^N \textbf{k}_{\perp i} -\vperp{P}\right),
\end{align}
where $i$ labels the partons.
Using this Fock state expansion, one can express GPDs in terms of LFWFs \cite{Diehl:2000xz}.
However, the partonic picture and therefore the way the GPDs are related to
LFWFs depends on the considered kinematics.
In the so-called DGLAP region ($|x|\ge \xi$),
the GPD is given by an overlap of LFWFs having the \emph{same} number of constituents.
Keeping the example of the pion, in the region $x \ge \xi$, we have \cite{Diehl:2003ny}:
\begin{eqnarray}
H^{q}\left(x,\xi,t\right) & = &
\sum_{N,\beta}\sqrt{1-\xi}^{2-N}\sqrt{1+\xi}^{2-N}\sum_{a}\delta_{a,q}\int\left[\mathrm{d}\bar{x}\right]_{N}\left[\mathrm{d}^{2}\bar{\mathbf{k}}_{\perp}\right]_{N}\delta\left(x-\bar{x}_{a}\right)\label{eq:OverlapDGLAP}\\
 & \times & \Psi_{N,\beta}^{*}\left(\hat{x}_{1}^{'},\hat{\mathbf{k}}_{\perp1}^{'},...,\hat{x}_{a}^{'},\hat{\mathbf{k}}_{\perp a}^{'},...\right)\Psi_{N,\beta}\left(\tilde{x}_{1},\tilde{\mathbf{k}}_{\perp1},...,\tilde{x}_{a},\tilde{\mathbf{k}}_{\perp a},...\right)\,,\nonumber 
\end{eqnarray}
where $\bar{x}$ and $\bar{\mathbf{k}}_{\perp}$ are the average momentum variables of the LFWF in the GPD symmetric frame,
and $x_a$ denotes the momentum fraction of the struck quark 
(labeled here with $a$ for \emph{active}).
The ``hat'' and ``tilde'' variables are the corresponding momenta boosted from the incoming and outgoing hadron frames respectively,
and can be related to the ``bar'' variables through:
\begin{equation}
\begin{array}{rclcrclc}
{\displaystyle \tilde{x}_{i}} & {\displaystyle =} & {\displaystyle \frac{\bar{x}_{i}}{1+\xi}} & {\displaystyle ,} & {\displaystyle \tilde{\mathbf{k}}_{\bot i}} & {\displaystyle =} & {\displaystyle \bar{\mathbf{k}}_{\bot i}+\frac{\bar{x}_{i}}{1+\xi}\frac{\mathbf{\Delta}_{\perp}}{2}\,,} & \textrm{for }i\neq a\,,\\
\\
{\displaystyle \tilde{x}_{a}} & {\displaystyle =} & {\displaystyle \frac{\bar{x}_{a}+\xi}{1+\xi}} & {\displaystyle ,} & {\displaystyle \tilde{\mathbf{k}}_{\bot a}} & {\displaystyle =} & {\displaystyle \bar{\mathbf{k}}_{\bot a}-\frac{1-\bar{x}_{a}}{1+\xi}\frac{\mathbf{\Delta}_{\perp}}{2}\,,}
\end{array}\label{eq:tilde_boost}
\end{equation}
and
\begin{equation}
\begin{array}{rclcrclc}
{\displaystyle \hat{x}_{i}^{'}} & {\displaystyle =} & {\displaystyle \frac{\bar{x}_{i}}{1-\xi}} & {\displaystyle ,} & {\displaystyle \hat{\mathbf{k}}_{\bot i}^{'}} & {\displaystyle =} & {\displaystyle \bar{\mathbf{k}}_{\bot i}-\frac{\bar{x}_{i}}{1-\xi}\frac{\mathbf{\Delta}_{\perp}}{2}\,,} & \textrm{for }i\neq a\,,\\
\\
{\displaystyle \hat{x}_{a}^{'}} & {\displaystyle =} & {\displaystyle \frac{\bar{x}_{a}-\xi}{1-\xi}} & {\displaystyle ,} & {\displaystyle \hat{\mathbf{k}}_{\bot a}^{'}} & {\displaystyle =} & {\displaystyle \bar{\mathbf{k}}_{\bot a}+\frac{1-\bar{x}_{a}}{1-\xi}\frac{\mathbf{\Delta}_{\perp}}{2}\,,}
\end{array}\label{eq:hat_boost}
\end{equation}
where $\mathbf{\Delta}_{\perp}^{2}=-\left(1-\xi^{2}\right)\left(t-t_{\rm min}\right) \geq 0$.
In the other part of the DGLAP region ($x \le -\xi$), there is a
similar result.

If we restrain ourselves to the valence contribution to the pion,
\ie the first Fock sector $N=2$, this relation can be further simplified.
Let us consider the $\pi^{+}$ case in which the first Fock sector
would be $u\bar{d}$. We get the following GPD:
\begin{equation}
H_{\pi^{+}}^{u}\left(x,\xi,t\right)=\int\frac{\mathrm{d}^{2}\mathbf{k}_{\bot}}{16\,\pi^{3}}\Psi_{u\bar{d}}^{*}\left(\hat{x}^{'},\hat{\mathbf{k}}_{\bot}^{'}\right)\Psi_{u\bar{d}}\left(\tilde{x},\tilde{\mathbf{k}}_{\bot}\right)\,,\label{eq:overlap_ud}
\end{equation}
where the LFWF depends only on one set of momenta by virtue of Eq.
(\ref{eq:DefMeasuresLCx}), and the hat and tilde relations used
are those of either the active quark in Eqs. (\ref{eq:tilde_boost})-(\ref{eq:hat_boost})
or the spectator (by symmetry of the LFWF).
This two-body truncated GPD will be used in later sections. 

Equation~\eqref{eq:OverlapDGLAP} highlights the underlying Hilbert space structure,
as the GPD appears now to be an inner product between two vectors.
Using the Cauchy-Schwartz inequality, it is possible to show the above-mentioned
positivity property \cite{Pire:1998nw,
Diehl:2000xz, Pobylitsa:2001nt, Pobylitsa:2002gw}, relating the GPDs to the PDFs
denoted by the quark flavour $q$.
For instance in the case of the pion quark GPD $H$:
\begin{equation}
  \label{eq:Positivity}
  |H^q(x,\xi,t)|_{x \ge \xi} \le
  \sqrt{q\left(\frac{x-\xi}{1-\xi}\right)q\left(\frac{x+\xi}{1+\xi}\right)} \, ,
\end{equation}
which can be generalized to any hadron, although in more complicated forms.

In the other kinematic area, called ERBL ($\xi \ge |x|$), almost the same overlap structure appears,
but with LFWFs involving \emph{different} numbers of constituents $N$ and $N+2$.
Indeed, there is no trace on $\beta$ nor $N$ in this case.

The fact that the $N$ and $N+2$ particles LFWFs overlap in the ERBL region has significant consequences.
First, a lowest order truncation in the Fock space would result in a vanishing GPD in the ERBL region.
Then whatever the dependencies in $\widetilde{x}$, $\widetilde{\mathbf{k}}_\perp$, $\widehat{x}$ and $\widehat{\mathbf{k}}_\perp$ of the wave function are, once expressed in the kinetic variables of the GPD symmetric frame, non-polynomial dependencies in $\xi$ appear (see \refeq{eq:tilde_boost} and \refeq{eq:hat_boost}).
If we consider also the fact that we do not expect the LFWFs to be polynomials of momenta, it becomes very unlikely that one can obtain polynomial Mellin moments independently in both kinematic regions.
Therefore, we expect compensations between the
DGLAP and ERBL regions to occur, canceling non-polynomial dependencies in the Mellin moments. 
However, the fact that the overlap representation mixes different particles numbers in the ERBL region
makes these cancellations unlikely at any \emph{finite} truncation order of the Fock space.
In fact, as already stressed in \refcite{Diehl:2003ny}, independent descriptions
of the DGLAP and ERBL regions will almost certainly break polynomiality and
hence indicate a violation of Lorentz covariance.
This suggests that polynomiality strongly ties the DGLAP and ERBL
regions in a subtle way since it relates $N$-body LFWFs to $N+2$-body LFWFs.
To the best of our knowledge, there has been so far no GPD models built from a
consistent computation of 2- and 4-body LFWFs. 
More generally, polynomiality cannot be readily observed in a GPD representation which is manifestly positive.

This raises an important issue about GPD phenomenology.
As emphasized in \refcite{Kumericki:2016ehc}, GPD extractions from experimental
data face the \emph{curse of dimensionality}: fitters have to extract
(potentially many) functions of several variables depending on unknown
parameters to be determined from measurements. 
A general first principles GPD parametrization,
obeying \apriori all theoretical requirements, is still unknown at the time of
writing.
Consequently there is a risk of an unphysical GPD parametrization being
numerically favored in data fitting by lack of \apriori constraints. 
Beyond its own intrinsic interest, a general model-building procedure
satisfying both polynomiality and positivity is therefore of direct
phenomenological relevance.
This problem has been solved previously in a particular case
\cite{Hwang:2007tb, Mueller:2014tqa} by building a covariant extension of an
overlap of LFWFs from the DGLAP to the ERBL region.  We provide in this paper a
general solution to this problem. But before describing it, we clarify the
meaning of polynomiality in the following section.

\subsection{Double Distributions and polynomiality} \label{subsec:double-distributions}

Since flavor plays no explicit role in the following discussion of the
properties of the GPD $H^q$, the subscript $q$ will be systematically dropped.

\subsubsection{Connection to the Radon transform}
\label{sec:double-distribution-and-radon-transform}

The polynomiality property states that the Mellin moments
$\int_{-1}^{1}\mathrm{d}x\,x^{m}\,H\left(x,\xi,t\right)$ of the GPD $H$ are
polynomials of $\xi$ of degree $m+1$:
\begin{equation}
\label{eq:def-polynomiality}
\int_{-1}^{1} \mathrm{d}x \, x^{m} H\left(x, \xi, t\right) =
\sum_{\substack{k=0\\k\textrm{ even}}}^{m+1} c^{(m)}_k(t) \xi^k \;,
\end{equation} 
This results from Lorentz symmetry and the decomposition of the matrix elements
of local twist-2 quark operators
$\bra{P+\frac{\Delta}{2}} \bar{q}(0)\gamma^+(i\overleftrightarrow{D}^+)^m q(0)
\ket{P-\frac{\Delta}{2}}$ 
(where $\overleftrightarrow{D}$ stands for
the left-right covariant derivative) in terms of the generalized form factors $c^{(m)}_k(t)$.
The further restriction to even powers of $\xi$ is brought by the time reversal invariance of QCD.
To simplify the
following discussion, we will drop the explicit $t$ dependence.

Let us assume the existence of an odd function $D(z)$ with support $z \in [-1, +1]$,
such that:
\begin{equation}
\label{eq:hausdorff-moment-problem-D-term}
\int_{-1}^{+1} \mathrm{d}z \, z^m D(z) = c^{(m)}_{m+1} \;.
\end{equation}
For $\xi \in [-1, +1]$, $D(x/\xi)$ has support $x \in \left[-\left|\xi\right|, +\left|\xi\right|\right]$ and satisfies:
\begin{equation}
\label{eq:polynomiality-D-term}
\int_{-1}^{+1} \mathrm{d}x \, x^m D\left(\frac{x}{\xi}\right) = \sgn\left(\xi\right) c^{(m)}_{m+1} \, \xi^{m+1} \;,
\end{equation}
and the Mellin moments of $H(x, \xi) - \sgn\left(\xi\right) D(x/\xi)$ become polynomials in $\xi$ of
order $m$.
Changing the variables $x, \xi \in \mathbb{R}^2$ to $s \in \mathbb{R}$ and $\phi
\in [0, 2\pi]$ through\footnote{This mapping is not one-to-one since changing $s$ to $-s$ is equivalent to changing $\phi$ to $\phi \pm \pi$ but this has no consequences on our argument.}:
\begin{eqnarray}
x & = & \frac{s}{\cos \phi} \;, \label{eq:dictionnary-s-cos-phi-x} \\
\xi & = & \tan \phi \;, \label{eq:dictionnary-tan-phi-xi} 
\end{eqnarray}
recasts the polynomiality condition on $H(x, \xi) - \sgn\left(\xi\right) D(x/\xi)$ to:
\begin{equation}
\label{eq:polynomiality-radon-canonical-variables}
\int_{-1}^{1} \frac{\mathrm{d}s}{\cos \phi} \, s^{m}
\left[H\left(\frac{s}{\cos \phi}, \tan \phi\right) - \sgn\left(\tan \phi\right) D\left(\frac{s}{\sin
\phi}\right)\right] = \sum_{k=0}^{m} c^{(m)}_k \cos^{m-k} \phi \sin^k \phi \,.
\end{equation}
An order-$m$ Mellin moment is an homogeneous polynomial of degree $m$,
which is exactly the Lugwig-Helgason consistency condition
\refeq{eq:ludwig-helgason-consistency-condition}. 
This condition asserts that $\left[H(s/\cos \phi, \tan \phi) - \sgn\left(\tan \phi\right) D(s/\sin \phi) \right] / \cos \phi$ is in the range of the Radon transform.
This ensures the existence of a
distribution $F_D(\beta, \alpha)$ such that:
\begin{equation}
\label{eq:radon-transform-in-dd-plus-d-gauge-canonical-variables}
\frac{1}{\cos \phi}
\left[H\left(\frac{s}{\cos \phi}, \tan \phi\right) - \sgn\left(\tan \phi\right) D\left(\frac{s}{\sin \phi}\right)\right]
= \int_\Omega \mathrm{d}\beta \mathrm{d}\alpha \,
F_D(\beta, \alpha) \, \delta( s - \beta \cos \phi - \alpha \sin \phi ) \equiv \fRadon{F_D}{s}{\phi} \,,
\end{equation}
where $\Radon{}$ is the Radon transform operator defined in \refapp{app:Radon}.
Switching back to ordinary GPD variables $\left(x, \xi\right)$, we have\footnote{The factor $\frac{1}{\cos \phi} = \sqrt{1+\xi^2}$ gets reabsorbed in the \emph{Dirac} distribution.}:
\begin{equation}
\label{eq:radon-transform-in-dd-plus-d-gauge}
H(x, \xi) = \sgn\left(\xi\right) D(x/\xi) + \int_\Omega \mathrm{d}\beta \mathrm{d}\alpha \,
F_D(\beta, \alpha) \, \delta(x - \beta - \alpha \xi) \;.
\end{equation}
The support $\Omega = \left\{\left(\beta,\alpha\right) \in \mathbb{R}^2 / \,
|\beta|+|\alpha| \le 1\right\}$ reflects the physical domain of GPDs $(x, \xi
\in [-1, +1]$).
Now noticing that:
\begin{equation}
\label{eq:D-term-as-radon-transform}
\frac 1 {\left|\xi\right|} \ D\left(\frac{x}{\xi}\right) = \int_\Omega \mathrm{d}\beta \mathrm{d}\alpha \,
\delta(\beta) D(\alpha) \delta(x - \beta - \alpha \xi) \;,
\end{equation}
$H$ can be expressed by means of an integral transform:
\begin{equation}
\label{eq:gpd-in-general-F-and-G-gauge}
H(x, \xi) = \int_\Omega \mathrm{d}\beta \mathrm{d}\alpha \,
\big[ F(\beta, \alpha) + \xi \, G(\beta,\alpha) \big] \delta(x - \beta -
\alpha \xi) \;,
\end{equation}
with $F(\beta,\alpha) = F_D(\beta, \alpha)$ and $G(\beta, \alpha)= \delta(\beta)
D(\alpha)$.
Since $H$ is $\xi$-even, $F_D$ is $\alpha$-even while $D$ and $G$ are
$\alpha$-odd.

\subsubsection{Different representations}
\label{subsec:different-representations}

More generally, according to the discovery of Teryaev \cite{Teryaev:2001qm} and 
Tiburzi \cite{Tiburzi:2004qr}, $F_D$ and $D$ do not constitute a unique 
parameterization for the integral representation
(\ref{eq:gpd-in-general-F-and-G-gauge}) of a given GPD $H(x,\xi)$.
A function $\chi(\beta, \alpha)$, vanishing\footnote{See
 \refcite{Tiburzi:2004qr} for a thorough discussion of boundary conditions. The most general case does not change the main line of our argument but brings some technical complexity.} on the boundary of $\Omega$, can be used for the
following redefinition:
\begin{eqnarray}
F(\beta, \alpha) & \rightarrow & F(\beta, \alpha) + \frac{\partial
\chi}{\partial \alpha}(\beta, \alpha) \;, \label{eq:gauge-transform-F} \\
G(\beta, \alpha) & \rightarrow & G(\beta, \alpha) - \frac{\partial
\chi}{\partial \beta}(\beta, \alpha) \;,
\label{eq:gauge-transform-G} 
\end{eqnarray}
which, as can be immediately seen by partial integration, does not modify the result of \eqref{eq:gpd-in-general-F-and-G-gauge} 
and leaves the GPD $H$ unchanged. 
For this to happen, the $\alpha$-parity of $F$ and $G$ also requires $\chi$ to be $\alpha$-odd. 
Thus, owing to the polynomiality property, infinitely many pairs $(F, G)$ 
--as many as $\alpha$-odd functions $\chi(\beta,\alpha)$ vanishing on $\Omega$-- 
exist and yield the same GPD.
Conversely, any function $H$ such as in \refeq{eq:gpd-in-general-F-and-G-gauge}
satisfies the polynomiality condition (\ref{eq:def-polynomiality}). 

$F$ and $G$ are called Double Distributions (DDs).
$F$ has been independently introduced by M\"uller \etal \, \cite{Mueller:1998fv} and Radyushkin
\cite{Radyushkin:1997ki}, while $G$ was later discovered by Polyakov and Weiss \cite{Polyakov:1999gs}.
 They are a natural solution of the polynomiality constraint.
Since then, DDs have been used to model GPDs based on extracted PDFs in the framework of the Radyushkin Double Distribution Ansatz (RDDA) \cite{Musatov:1999xp}. 
The Goloskokov-Kroll model \cite{Goloskokov:2005sd,
Goloskokov:2007nt, Goloskokov:2009ia} is a popular example of the family of models following this approach.
As a consequence of the success of these phenomenological models, DDs have been
remembered by most as a convenient way to implement polynomiality in GPD models.

It has become a common misconception to believe that DDs appear only in this
restricted subset of GPD parameterizations.
On the contrary, our reasoning above shows that DDs are the \emph{essence} of
the polynomiality property.
Obeying the polynomiality property is \emph{exactly equivalent} to being
constructed from a DD, and this is a key argument of our approach.

Three main representations --or \emph{schemes}-- for DDs, all of them related to each other by the transformations \refeqs{eq:gauge-transform-F}{eq:gauge-transform-G}, have been so far employed:
\begin{description}
\item[PW] One DD $\Fcpw$ and an extra one-variable function $\Dcpw$
\cite{Polyakov:1999gs} such that:
\begin{equation}
\Gcpw(\beta, \alpha) = \delta(\beta) \Dcpw(\alpha) \;,  
\label{eq:polyakov-weiss-gauge-G}
\end{equation}
which, precisely, correspond to the DDs $F_D(\beta,\alpha)$ and $\delta(\beta)\, D(\alpha)$ used above in
\eqref{eq:radon-transform-in-dd-plus-d-gauge}.
\item[BMKS] One function $\hbmks$ \cite{Belitsky:2000vk} such that:
\begin{eqnarray}
\Fcbmks(\beta, \alpha) 
& = & 
\beta \, \hbmks(\beta, \alpha) \;, \label{eq:bmks-gauge-F} \\
G_{\textrm{BMKS}}(\beta, \alpha) 
& = & 
\alpha \, \hbmks(\beta, \alpha) \;. \label{eq:bmks-gauge-G} 
\end{eqnarray}
\item[P] One function $\hp$ \cite{Pobylitsa:2002vi} such that:
\begin{eqnarray}
\Fcp(\beta, \alpha) 
& = & 
(1-\left|\beta\right|) \, \hp(\beta, \alpha) \;, \label{eq:pobylitsa-gauge-F} \\
G_{\textrm{P}}(\beta, \alpha) 
& = & 
-\sgn(\beta) \, \alpha \, \hp(\beta, \alpha) \;. \label{eq:pobylitsa-gauge-G} 
\end{eqnarray}
\end{description}
By an abuse of terminology, $\hbmks$ and $\hp$ are often called DDs.
Explicit formulas for $\chi$ are given in the literature:
\begin{itemize}
  \item to convert a general $(F, G)$ scheme to a PW scheme
  \cite{Teryaev:2001qm, Tiburzi:2004qr},
  \item to convert a general $(F, G)$ scheme to a BMKS scheme
  \cite{Tiburzi:2004qr, Belitsky:2005qn},
  \item and to convert a BMKS scheme to a P scheme
  \cite{Mueller:2014hsa}.
\end{itemize}
This last formula is
 given without proof in \refcite{Mueller:2014hsa}, and to the best of our
 knowledge, none has been published. In fact, the connection between the BMKS
 and P schemes has remained unclear until recently. Since both BMKS and
 P schemes will play a central role in the following, we add a derivation of the
 last converting formula in \refapp{sec:families-ocdd}, and discuss it in
 details.
 
 In a general $(F, G)$ scheme, following \refcite{Teryaev:2001qm}, we
 define the D-term \cite{Polyakov:1999gs} by:
 \begin{equation}
 \label{eq:def-dterm} 
 D(\alpha) = \int_{-1+|\alpha|}^{+1-|\alpha|} \mathrm{d}\beta \, G(\beta,
 \alpha) \;.
 \end{equation}
 This terms is not modified by the transformations 
 \refeqs{eq:gauge-transform-F}{eq:gauge-transform-G} with functions $\chi$
 vanishing on the border of $\Omega$. Through
 \refeq{eq:D-term-as-radon-transform} it contributes only to the ERBL region.

The PW scheme has been the one mainly used in phenomenology through the
aforementioned RDDA, complemented with various assumptions concerning the
D-term. Attempts to model DDs in the BMKS scheme have been undertaken in
\refcite{Mezrag:2013mya} following a new regularization procedure
\cite{Radyushkin:2011dh, Radyushkin:2013hca}.
It is worth highlighting that none of these schemes can guarantee 
\apriori the
positivity property.
The additional P scheme has been designed in an attempt to fulfill
both polynomiality and positivity at the same time. 
It has not been used in the
building of phenomenological GPD models to the best of our knowledge.

\subsubsection{Quark and anti-quark GPDs}
\label{subsec:quark-GPDs}

We will use here a general notation $h$ for the DDs in all schemes ($\hbmks$, $\hp$ and $\hpw=\Fcpw$, when omitting an extra D-term).
Classically denoting $\theta$ the Heaviside function, the DD $h$ can be
decomposed as follows:
\begin{equation}
h(\beta, \alpha) = h^{>}(\beta, \alpha) \theta(\beta) 
+ h^{<}(\beta, \alpha) \theta(-\beta) \;, \label{eq:quark_antiquark_dd}
\end{equation}
where $h^{>}$ and $h^{<}$ are called respectively ``quark'' (with support on $\Omega^{>}=\Omega \cap \left\{\beta>0\right\}$), and ``anti-quark'' (with support on  $\Omega^{<}=\Omega \cap \left\{\beta<0\right\}$) distributions~\citep{Diehl:2003ny}; and which yield the following ``quark''  and ``anti-quark'' GPDs (corresponding to Radyushkin's original GPDs~\citep{Radyushkin:1998bz}):
\begin{equation}
H^{>}(x, \xi) = C^{>}(x, \xi) \int_{\Omega^{>}} \mathrm{d}\beta\mathrm{d}\alpha 
\, h^{>}(\beta, \alpha) \delta(x-\beta-\alpha\xi) \;, \label{eq:quark_gpd}
\end{equation}
with support $x \in [-\xi, +1]$, and
\begin{equation}
H^{<}(x, \xi) = C^{<}(x, \xi) \int_{\Omega^{<}} \mathrm{d}\beta\mathrm{d}\alpha
\, h^{<}(\beta, \alpha) \delta(x-\beta-\alpha\xi) \;, \label{eq:antiquark_gpd}
\end{equation}
with support on $x \in [-1, +\xi]$, the total GPD being of course $H=H^{>}+H^{<}$.
The factors $C^{>}$ and $C^{<}$ can be respectively:
\begin{itemize}
  \item both equal to 1 in the PW scheme when the GPD follows a degree $m$ polynomiality property~(\ref{eq:def-polynomiality}) or when we consider the GPD minus its D-term contribution \refeq{eq:polyakov-weiss-gauge-G};
  \item both equal to $x$ in the BMKS scheme \refeqs{eq:bmks-gauge-F}{eq:bmks-gauge-G};
  \item $1-x$ and $1+x$ in the P scheme \refeqs{eq:pobylitsa-gauge-F}{eq:pobylitsa-gauge-G}.
\end{itemize}

In the following, we will consider only quark GPDs (\ie $\beta>0$), unless explicitly stated otherwise.
It should be therefore understood that $H$ stands for $H^{>}$.

\subsubsection{Polynomiality at work: a simple example}
\label{sec:example-polynomiality-toy-model}

We give a practical illustration of the main assertions of the previous
subsection, in particular about how the polynomiality condition is 
implemented by the representation of a GPD as the Radon transform of a DD in
different schemes. 
Adopting an overall normalization for later convenience, we consider the
following DD in the BMKS scheme:
\begin{equation}
\label{eq:def-toy-model-dd-bmks}
\htoymodelbmks\left(\beta, \alpha\right) = 15 \left(\left(1-\beta\right)^2 - \alpha^2 \right)
\theta\left(\beta\right) \;.
\end{equation}
The restrictions $\HtoymodelDGLAP$ and $\HtoymodelERBL$ of the associated GPD:
\begin{equation}
\label{eq:toy-model-gpd-in-bmks-gauge}
\Htoymodel(x, \xi) = x \int_\Omega \mathrm{d}\beta \mathrm{d}\alpha \,
\htoymodelbmks(\beta, \alpha) \delta(x - \beta - \alpha \xi) \;,
\end{equation}
to the DGLAP and ERBL region read:
\begin{eqnarray}
\HtoymodelDGLAP(x, \xi)
& = &
\frac{20 \left(1-x\right)^3 x}{\left(1 - \xi^2\right)^2} \;, 
\label{eq:toy-model-gpd-dglap} \\
\HtoymodelERBL(x, \xi)
& = &
-\frac{5 x (\xi +x)^2 (-\xi  (\xi +2)+2 \xi  x+x)}{\xi ^3 (\xi +1)^2} \;, 
\label{eq:toy-model-gpd-erbl}
\end{eqnarray}
with support $x \in [-\xi, +1]$.

The computation of its Mellin moments can be readily performed, and the first 11
of them are presented in \reftab{tab:mellin-moments-gpd-example-toy-model-dd} in \refapp{app:gpd-toy-model-vanish-rhombus-border}.
The DGLAP and ERBL contributions to the Mellin moments are 
rational fractions, not polynomials! 
One needs to integrate over both DGLAP and ERBL regions to obtain a polynomial
in the variable $\xi$.
This is true at all order as can be seen exactly from the expressions of
$\int_{-\xi}^{+\xi}
\mathrm{d}x \, \Htoymodel(x, \xi)$ and $\int_{+\xi}^{+1}
\mathrm{d}x \, \Htoymodel(x, \xi)$ in
\refeqs{eq:gpd-toy-model-mellin-moment-erbl}{eq:gpd-toy-model-mellin-moment-dglap}.
The contribution of the DGLAP region is a rational fraction with apparent double
poles at $\xi = \pm 1$. Since both the numerator and its derivative vanish when
$\xi = 1$, this fraction actually possesses only a double pole at $\xi = -1$.
The residues can be straightforwardly computed, exhibiting the general structure
of both contributions to the Mellin moments:
\begin{equation}
\label{eq:example-mellin-moment-dglap-contribution}
\int_{+\xi}^{+1} \mathrm{d}x \, x^m \Htoymodel(x,\xi) = P^{\rm DGLAP}_{m+1}(\xi)
+ \frac{R_{-1}(m)}{1+\xi} + \frac{R_{-2}(m)}{(1+\xi)^2} \;,
\end{equation} 
and
\begin{equation}
\label{eq:example-mellin-moment-erbl-contribution}
\int_{-\xi}^{+\xi} \mathrm{d}x \, x^m \Htoymodel(x,\xi) = P^{\rm
ERBL}_{m+1}(\xi) - \frac{R_{-1}(m)}{1+\xi} - \frac{R_{-2}(m)}{(1+\xi)^2} \;.
\end{equation} 
$P^{\rm DGLAP}_{m}$ and $P^{\rm ERBL}_{m}$ are polynomial of
degree $m$ and:
\begin{eqnarray}\label{eq:residue-R1}
R_{-1}(m) = 20 (-1)^{m+1} \frac{ 7+2 \left(m+(-1)^m\right)}{\left(5+2m+(-1)^m\right)\left(9+2m+(-1)^m\right)} \; ,\\
\label{eq:residue-R2}
R_{-2}(m) = 20 (-1)^{m} \frac{ 2 m^2 + \left( 12+2 (-1)^m\right) m +16 + 5 (-1)^m}{\left(5+2m+(-1)^m\right)\left(9+2m+(-1)^m\right)} \; . 
\end{eqnarray}

We observe that even with a very simple GPD model, the expressions of the DGLAP 
and ERBL contributions to the Mellin moments of the GPD are intricate; 
their $\xi$-dependencies are not trivially related but add up to yield a
polynomial of degree $m+1$: 
\begin{equation}
\int_{-\xi}^1 \mathrm{d}x \, x^m \Htoymodel(x,\xi) = P^{\rm DGLAP}_{m+1}(\xi) +
P^{\rm ERBL}_{m+1}(\xi) \ .
\end{equation}
Such a delicate cancellation, which suppresses order by order the poles of
two rational fractions to produce a polynomial, cannot be the result of a coincidence. 
It is difficult to imagine that more complex GPD models will not exhibit such a feature.

\subsubsection{A necessary and sufficient condition for the BMKS and P schemes}
\label{sec:condition-bmks-p-schemes}

As a final but not less important result of this section, we carefully 
examine the implications of expressing as a Radon transform the formal 
link between a GPD and the one-component DD schemes BMKS and P.

A GPD expressed as a Radon transform of a DD in the BMKS scheme
(see \refeqs{eq:bmks-gauge-F}{eq:bmks-gauge-G}) reads:
\begin{equation}\label{eq:dictionnary-H-Radon-f-bmks}
\frac{H( x, \xi)}{x} = \int \mathrm{d}\beta\mathrm{d}\alpha \, \hbmks(\beta, \alpha) \delta(x-\beta-\alpha\xi) \;.
\end{equation}
The $l$-order Mellin moment ($l \geq 0$) of the \lhs is a $l$-degree polynomial of $\xi$, as can be checked easily.
Specializing to the case $l=0$\footnote{This would correspond to $m=-1$ in \refeq{eq:def-polynomiality}.} yields:
\begin{equation}\label{eq:LH-BMKS-m-equal-0}
\int_{-1}^{+1} \mathrm{d}x  \,
\frac{H( x, \xi)}{x} = \ C_{\textrm{BMKS}} \;, 
\end{equation}
where $C_{\textrm{BMKS}}$ is independent of $\xi$.

Similarly, a quark GPD expressed as a Radon transform of a DD in the P scheme (see \refeqs{eq:pobylitsa-gauge-F}{eq:pobylitsa-gauge-G}) writes\footnote{In the case of an anti-quark GPD, the \lhs would be of course modified with a $1+x$ denominator.\label{fn:anti-quark-P}}:
\begin{equation}\label{eq:dictionnary-H-Radon-f-pobylitsa}
\frac{H(x, \xi)}{1-x} =
\int \mathrm{d}\beta\mathrm{d}\alpha \, \hp(\beta, \alpha) \delta(x-\beta-\alpha\xi) \;.
\end{equation}
The integral of the \lhs (its $l = 0$ Mellin moment) fulfills\footref{fn:anti-quark-P}:
\begin{equation}\label{eq:LH-Pobylitsa-m-equal-0}
\int_{-1}^{1} \mathrm{d}x \, 
\frac{H(x, \xi)}{1-x} = \ C_{\textrm{P}} \;, 
\end{equation}
where $C_{\textrm{P}}$ is independent of $\xi$.

On the technical side, we assumed in the derivation of
\refeq{eq:LH-BMKS-m-equal-0} and \refeq{eq:LH-Pobylitsa-m-equal-0} that we could
swap the order of integrations in $\int_{-1}^{+1} \mathrm{d}x \, \int_\Omega
\mathrm{d}\beta \mathrm{d}\alpha$.
This is justified only if the integrated function is summable over $[-1, +1]
\times \Omega$.
However the polynomiality condition (\ref{eq:def-polynomiality}) implies the
existence of all moments $\int_\Omega \mathrm{d}\beta \mathrm{d}\alpha
\beta^m \alpha^n F(\beta, \alpha)$ and $\int_\Omega \mathrm{d}\beta
\mathrm{d}\alpha \beta^m \alpha^n G(\beta, \alpha)$ for $n, m \geq 0$ but does
not say anything about $\int_\Omega \mathrm{d}\beta \mathrm{d}\alpha
\hbmks(\beta, \alpha)$ or $\int_\Omega \mathrm{d}\beta \mathrm{d}\alpha
\hp(\beta, \alpha)$ which may even not exist. 
Both equations
 (\ref{eq:LH-BMKS-m-equal-0}) and (\ref{eq:LH-Pobylitsa-m-equal-0}) precisely
rely on the finiteness of these last two integrals. 
This summability assumption,
supplemented by the polynomiality condition, makes possible the application of
the Ludwig-Helgason consistency condition
(\ref{eq:ludwig-helgason-consistency-condition}) to specify 
a sufficient and necessary condition for $H(x, \xi)/x$
or $H(x, \xi)/(1-x)$ to be in the range of the Radon transform of summable functions or compactly-supported distributions.

From a strict mathematical point of view, there seems to be no reason to prefer
one DD scheme or  the other. 
From the physical point of view, we know that GPDs
are continuous functions of $x$, with support in $[-1, +1]$.
According to perturbative QCD \cite{Yuan:2003fs}, one should expect
$H(x, \xi)$ to exhibit a smooth quadratic\footnote{It would be a smooth \emph{cubic} decrease $\simeq (1-x)^3/(1-\xi^2)^2$ if we were considering
the nucleon instead of the pion.} decrease at
large $x$: $H(x, \xi) \simeq (1-x)^2/(1-\xi^2)$.
Hence $H(x, \xi)/(1-x)$ should be as singular as the GPD $H(x, \xi)$ itself, and
we may expect $\int_{-1}^{+1} \mathrm{d}x \,H(x, \xi)/(1-x)$ to be finite. 
On the contrary, PDF extractions indicate that $H(x, \xi = 0)$ has a power-law
divergence at small $x$.
Consequently we expect the Regge singularity of the PDF to be aggravated in
$\int_{-1}^{+1} \mathrm{d}x \, H(x, \xi)/x$, which may even not exist 
for phenomenological valence quark GPD models based on the RDDA.
Although we have employed the same notation $H(x,\xi)$ for the GPDs represented 
either by $\hbmks$ in \refeq{eq:dictionnary-H-Radon-f-bmks} and 
satisfying (\ref{eq:LH-BMKS-m-equal-0}), or by $\hp$ in
\refeq{eq:dictionnary-H-Radon-f-pobylitsa} and obeying
(\ref{eq:LH-Pobylitsa-m-equal-0}), it is not self-evident that 
the same GPD $H$ can be represented with both DD schemes. 
We will illustrate this point with an example in the following section.
At last, irrespective of their values, the integrals $\int_{-1}^{+1} \mathrm{d}x
\, H(x, \xi)/x$ and $\int_{-1}^{+1} \mathrm{d}x \,H(x, \xi)/(1-x)$ would still
have to be independent of $\xi$.
We may already anticipate the P scheme to be more tractable for computing
purposes.

\subsubsection{One GPD, several DD schemes: a simple example}
\label{sec:example-dd-schemes-toy-model}

We consider again the GPD $\Htoymodel$ of \refeqs{eq:toy-model-gpd-dglap}{eq:toy-model-gpd-erbl}. 
This model is built to be algebraically simple, not phenomenologically realistic, and the considerations above
about GPD phenomenology are temporarily left aside.
This model is defined in the BMKS scheme by the DD
$\htoymodelbmks$ (\ref{eq:def-toy-model-dd-bmks}).
The condition (\ref{eq:LH-BMKS-m-equal-0}) writes:
\begin{equation}
\label{eq:sumrule-bmks-gauge-toymodel}
\int_{-1}^{+1} \mathrm{d}x \,\frac{\Htoymodel(x, \xi)}{x} = 5 \;,
\end{equation} 
which is indeed independent of $\xi$.

The scheme transform (\ref{eq:toy-model-gauge-transform-bmks-to-pobylitsa})
converts this BMKS representation to the P representation
\refeq{eq:toy-model-dd-pobylitsa}. The condition
(\ref{eq:LH-Pobylitsa-m-equal-0}) writes:
\begin{equation}
\label{eq:sumrule-pobylitsa-gauge-toymodel}
\int_{-1}^{+1} \mathrm{d}x \,\frac{\Htoymodel(x, \xi)}{1-x} = -\frac{5 \left(\xi 
\left(\xi ^2-2\right)-2 \left(\xi ^2-1\right) \tanh ^{-1}(\xi )\right)}{\xi ^3} \;,
\end{equation} 
which has a non trivial $\xi$-dependence.
There is no contradiction with the previous statements, because the underlying
DD in the P scheme $\htoymodelp$ (\ref{eq:toy-model-gpd-in-pobylitsa-gauge}) is
not summable over $\Omega$ as testified by 
\refeq{eq:toy-model-dd-pobylitsa-gauge-non-summability}.
The integrals over all lines of $\htoymodelp$ are well-defined and finite
(their values are obtained from the GPD $\Htoymodel$) but $\htoymodelp$ is
too singular to be summable over $\Omega$.

We can pick up the singularities (see
\refeq{eq:toy-model-dd-pobylitsa-gauge-singular-part}) 
of this insightful model in the P scheme and get rid of them. 
Doing so we build a regularized DD $\htoymodelpreg$ and a regularized GPD 
$\Htoymodelreg$ which differs from the original one $\Htoymodel$ only by the
introduction of a D-term (see
\refeqs{eq:toy-model-regularized-gpd-dglap}{eq:toy-model-regularized-gpd-erbl}).
This D-term modifies the condition (\ref{eq:LH-Pobylitsa-m-equal-0}) by adding
an extra contribution:
\begin{equation}
\label{eq:extra-contribution-in-sumrule-brought-by-dterm}
\int_{-1}^{+1} \frac{\mathrm{d}x}{1-x} \frac{5 \left(x^3-\xi ^2 x\right)}{\xi
^3} = \frac{10 \left(\xi  \left(2 \xi ^2-3\right)-3 \left(\xi ^2-1\right) \tanh
   ^{-1}(\xi )\right)}{3 \xi ^3} \;.
\end{equation}
The original value of the sum rule (\ref{eq:sumrule-pobylitsa-gauge-toymodel})
supplemented by this contribution yields:
\begin{equation}
\label{eq:sumrule-regularized-pobylitsa-gauge-toymodel}
\int_{-1}^{+1} \mathrm{d}x \,\frac{\Htoymodelreg(x, \xi)}{1-x} =
\frac{5}{3} \;,
\end{equation}
which is now independent of $\xi$.

What did we learn? 
Condition (\ref{eq:LH-Pobylitsa-m-equal-0}) did not 
natively hold because the underlying DD was not summable over $\Omega$.
In principle a GPD can be expressed in any DD scheme, but there is no reasons
for DDs to be smooth simple functions in every scheme. 
In our example the DD $\htoymodelp$ in the P scheme can even not be identified
with a compactly-supported distribution.
Adding a D-term produced a GPD model obeying condition
(\ref{eq:LH-Pobylitsa-m-equal-0}) and deriving from a smooth DD, summable over
$\Omega$.  
It is a different GPD model but a model as theoretically consistent as the
initial one.
It fulfills polynomiality and has the required behavior under discrete
symmetries.
It has the same forward limit and satisfies the same positivity 
inequalities. 
Since the D-term is immaterial to the $m = 1$ Mellin moment, the form factor sum
rules stays the same too. 
They only differ by the realization of the additional sum rule
\eqref{eq:LH-Pobylitsa-m-equal-0} relying on the $m=0$ Mellin moment, yielding
either \eqref{eq:sumrule-pobylitsa-gauge-toymodel} or \eqref{eq:sumrule-regularized-pobylitsa-gauge-toymodel}.
To summarize, adding this D-term produced a consistent GPD model which is
indistinguishable from the original one based on the sole knowledge of the DGLAP region.

\section{Covariant extension to the ERBL region \label{sec:radon-inverse}}

The way Lorentz covariance binds the DGLAP and ERBL regions together has been 
questioned for many years.
To the best of our knowledge, the first systematic discussion of this link is
the work of M\"uller and Sch\"afer \cite{Mueller:2005ed}.
Under analyticity assumptions, they argued that an
extension of the GPD from the ERBL to the DGLAP region exists and is unique.
They also mentioned that the converse does not hold because of the D-term
ambiguity.
However this study does not solve the problem of \emph{actually} extending the
GPD from one region to the other. 
Indeed analytically continuing a GPD is a daunting task only if we cannot start
with GPD expressions in simple closed forms.
In the following, we present an original discussion based on Radon
transform properties of covariant
extensions of GPDs from the DGLAP to the ERBL region. 
We then deduce a general procedure to construct these
covariant extensions starting from the knowledge of the values of the GPDs in
the DGLAP region.

\subsection{Intuitive picture}
\label{sec:intuitive-picture}

\begin{figure*}
\begin{tabular}{m{0.4\textwidth}m{0.1\textwidth}m{0.4\textwidth}}
\includegraphics[width=0.4\textwidth]{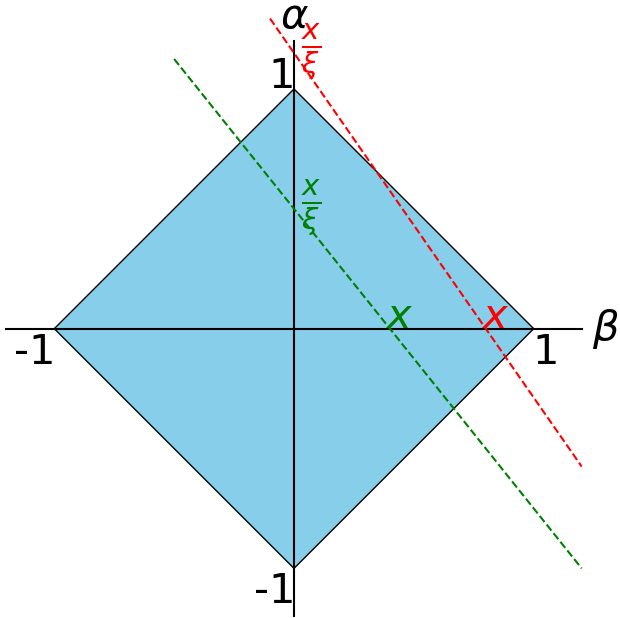} & 
\begin{eqnarray*}
 & \mathcal{R}\\
 & \longrightarrow
\end{eqnarray*}
 & \includegraphics[width=0.4\textwidth]{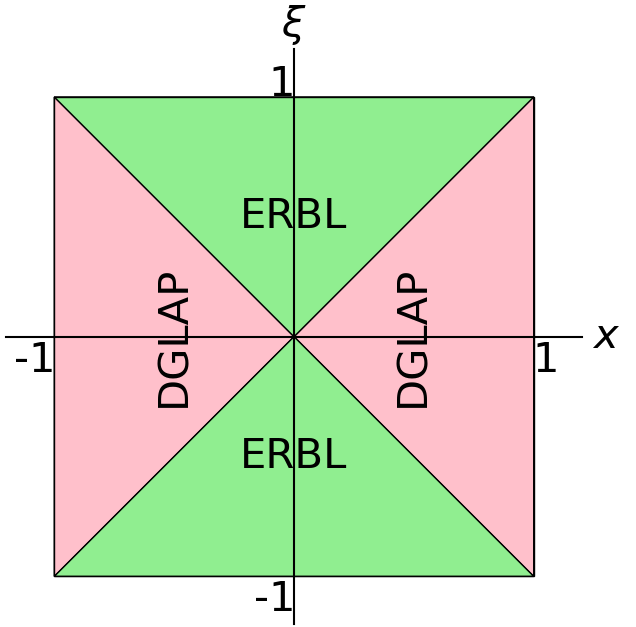}\tabularnewline
\end{tabular}
\caption{The domains $\Omega^{<}$ and $\Omega^{>}$ of the DD (resp. DGLAP
and ERBL of the GPD) on the left (resp. right). The Radon transform
$\fRadon{h}{x}{\xi} \propto \int \mathrm{d}\beta\mathrm{d}\alpha \, h(\beta, \alpha) \delta (x-\beta-\alpha\xi )$, 
which is an integration of $h$ on a line parameterized by the couple
$(x, \xi)$, is the operation that sends one domain to the other. The goal of
the inversion of the Radon transform is to rely only on DGLAP information,
meaning that we have only access in DD space to integration lines that cross 
the $\alpha$ axis on $x/\xi > 1$ (red lines). In this example, both $x$ and
$\xi$ are positive. \label{fig:radon_transform}}
\end{figure*}

We explained in \refsec{sec:double-distribution-and-radon-transform} that the
polynomiality property requires the existence of compactly-supported
distributions $\hbmks(\beta, \alpha)$, $\hp(\beta, \alpha)$, $\Fcpw(\beta,
\alpha)$, and $\Dcpw(\alpha)$ such that:
\begin{eqnarray}
\sqrt{1+\xi^2} \, \frac{H(x, \xi)}{x} 
& = & 
\Radon{\hbmks} \;, \label{eq:relation-gpd-dd-bmks-scheme} \\
\sqrt{1+\xi^2} \, \frac{H(x, \xi)}{1-x} 
& = & 
\Radon{\hp} \;, \label{eq:relation-gpd-dd-pobylitsa-scheme} \\
\sqrt{1+\xi^2} \left[ H(x, \xi) - \Dcpw\left(\frac{x}{\xi}\right) \right]
& = & 
\Radon{\Fcpw} \;, \label{eq:relation-gpd-dd-pw-scheme}
\end{eqnarray}
where writing the first two relations is
subject to the conditions mentioned in \refsec{sec:condition-bmks-p-schemes}.
The geometrical content of both equations is the same: the \lhs is the integral over lines in the plane of the function appearing in the \rhs.
Going back to
\refeq{eq:gpd-in-general-F-and-G-gauge}, one realizes that DDs are integrated in the $(\beta, \alpha)$-plane along the lines:
\begin{equation}
  \label{eq:Lines}
  \alpha = \frac{1}{\xi}(x-\beta) \;,
\end{equation}
which cross the $\alpha$-axis at $x/\xi$ and the $\beta$-axis at $x$ (see
\reffig{fig:radon_transform}).

Consider now a real $z$ such that $|z| > 1$, \ie $(0, z) \notin \Omega$.
The families of straight lines joining $(0, z)$ to $(x, 0)$ with $x \in ]0, 1]$
cover the whole domain $\Omega^{>}$. 
Thus any point in $\Omega^{>}$ contributes to DGLAP kinematics.
If $|z| < 1$, all points in the cone of apex $(0, z)$ and base delimited by
$(0, 0)$ and $(1-|z|, 0)$ brings a contribution to the ERBL region.
Proceeding with the limit $|z| \rightarrow 1^-$ we see that all points in $\Omega^{>}$ contribute to
ERBL kinematics. 
The same discussion can of course be carried for $\Omega^{<}$.
Therefore the entire DD support but the $\beta=0$ line
is integrated over when $(x, \xi)$ covers the DGLAP or ERBL region.
At this stage the question of the covariant extension of GPDs from the
DGLAP to the ERBL region becomes to know whether it is possible to unravel an
underlying DD $h(\beta, \alpha)$ or $\Fcpw(\beta, \alpha)$ on $\Omega$
\emph{uniquely} from the DGLAP region.

\subsection{Formalization}
\label{sec:formalization}

At the GPD level, a D-term $\delta(\beta) D(\alpha)$ is visible only in the ERBL
region. 
At the DD level, a D-term is irrelevant in the geometric
construction of the previous paragraph. 
There is a minimal ambiguity in the
extension from the DGLAP to the ERBL region. 
Is it the only one?

In other words, assume there are two GPDs $H_1$ and $H_2$ which are equal over
the DGLAP region. 
Are they equal over the ERBL region, up to a D-term?
Without loss of generality, we can consider $H = H_1 - H_2$ by linearity of the
Radon transform. 
This $H$ is zero in the DGLAP region. 
Since this reasoning holds up to a D-term, we can express $H$ in the PW scheme
assuming that $\Dcpw = 0$. 
We want to show that $\Fcpw = 0$. 
We will rely on the theorem of Boman and Todd Quinto mentioned in
\refsec{app:Radon}.
Switching to Radon transform canonical variables with 
\refeqs{eq:dictionnary-s-cos-phi-x}{eq:dictionnary-tan-phi-xi}, the membership
to the DGLAP and ERBL regions becomes:
\begin{eqnarray}
(x, \xi) \in \textrm{ DGLAP} & \Leftrightarrow & |s| \geq |\sin \phi| \;, \label{eq:def-dglap-canonical-variables} \\
(x, \xi) \in \textrm{ ERBL} & \Leftrightarrow & |s| \leq |\sin \phi| \;. \label{eq:def-dglap-canonical-variables}
\end{eqnarray}
Note that $|\xi| \leq 1$ means $\phi \in [0, \pi/4] \cup [3\pi/4, 5\pi/4] \cup [7\pi/4, 2\pi]$.
The complementary range in $[0, 2\pi]$ corresponds to the physical domain of
Generalized Distribution Amplitudes (GDAs) \cite{Mueller:1998fv, Diehl:1998dk,
Diehl:2000uv} which are the crossed-channel analog of GPDs.
In the DGLAP region, the GPD is zero if $|x| > 1$, which is equivalent to $|s| >
|\cos \phi|$, which is larger than $|\sin \phi|$ for $|\xi| < 1$, \ie in the
GPD physical domain.
We therefore assume here that:
\begin{equation}
\label{eq:uniqueness-from-boman-todd-quinto-canonical-variables}
H\left(\frac{s}{\cos \phi}, \tan \phi\right) = 0 \quad \textrm{ for } s > \sin
\phi \textrm{ and } \phi \in [0, \pi/4] \;.
\end{equation}
Let us choose $\xi_0 \in [0, 1]$ and $x_0 \in ]\xi_0, +\infty[$. 
We write $\tan \phi_0 = \xi_0$ with $\phi_0 \in [0, \pi/4]$
and take $s_0$ verifying $x_0 \cos \phi_0 > s_0 > \sin \phi_0.$
By continuity of the $\sin$ function, there exists $\epsilon > 0$ such that
$s_0 > \sin \phi$ for $|\phi - \phi_0| < \epsilon$.
It follows that $\fRadon{\Fcpw}{s}{\phi} = 0$ for $s > s_0$ and $\phi \in
]\phi_0-\epsilon, \phi_0+\epsilon[$. 
The aforementioned theorem of Boman and Todd Quinto asserts that:
\begin{equation}
\label{eq:conclusion-from-boman-todd-quinto-canonical-variables}
\Fcpw(\beta, \alpha) = 0 \quad \textrm{ for all } (\beta, \alpha) \in \Omega^{>}
\textrm{ such that } \beta \cos \phi_0 + \alpha \sin \phi_0 = s > s_0 \;.
\end{equation}
Selecting $s = x_0 \cos \phi_0$, this last condition implies:
\begin{equation}
\label{eq:conclusion-from-boman-todd-quinto-gpd-variables}
\Fcpw(\beta, \alpha) = 0 \quad \textrm{ for all } (\beta, \alpha) \in \Omega^{>}
\textrm{ such that } \beta + \alpha \xi_0 = x_0 \;.
\end{equation}
Thus $\Fcpw$ vanishes on all lines contributing to the DGLAP region, \ie $\Fcpw =0$ on $\Omega^{>}$.
Therefore the knowledge of a GPD in the DGLAP region fully determines it over the whole physical domain up to terms inherited from DDs with support on the line $\beta = 0$. 
This is completely consistent with the discussion about the D-term at the beginning of this section.

What is the manifestation at the GPD level of DDs with support on the line $\beta=  0$? 
Assume we started from a function $H_{\textrm{source}}$ defined in the DGLAP region, and that we were able to identify one DD $\Fcpw$ in the PW scheme such that $H_{\textrm{source}} = \Radon{\Fcpw}$. 
Let us further assume that $\Fcpw$ is properly normalized, \ie that it yields the correct valence quark number (from the PDF) or the correct electric charge (from the form factor at vanishing momentum transfer).
If $\Fcpw$ is a function, its values along the line $\beta = 0$ do not
contribute to line integrals over $\Omega$ since this subset has measure 0.
In this case there is no remaining freedom.
If $\Fcpw$ is not a function, we already saw the D-term contribution. 
We cannot exclude \apriori contributions like $\delta(\beta) D^+(\alpha)$ where
$D^+$ is an even function.
For $\xi > 0$ this modifies $H$ by the addition of a term:
\begin{equation}
\label{eq:def-ambiguity-F-delta-beta}
\delta_F H(x, \xi) = \frac{1}{\xi} D^+\left( \frac{x}{\xi} \right) \;.
\end{equation}
If $\xi = 0$, this new term manifests itself in the PDF through:
\begin{equation}
\label{eq:def-ambiguity-F-delta-beta-pdf}
\delta_F H(x, 0) = 2 \delta(x) \int_0^1 \mathrm{d}\alpha D^+(\alpha) \;.
\end{equation}
Such a term would violate quark number conservation if $\int_0^1
\mathrm{d}\alpha D^+(\alpha) \neq 0$, but there does not seem to be any
first principle reason to forbid such terms if this integral does indeed vanish.
Such terms have already been considered \eg in \refcite{Radyushkin:1998bz}.
Generally speaking, we end up with two families of ambiguities at the DD level:
\begin{enumerate}
\item Modification of the DD $F$ on the line $\beta$ = 0: terms like
$\delta(\beta) D^+(\alpha)$ where $D^+$ is even, has support in $[-1, +1]$ and
has vanishing integral over $[0, 1]$.
\item Modification of the DD $G$ on the line $\beta$ = 0: terms like
$\delta(\beta) D^-(\alpha)$ where $D^-$ is odd and has support in $[-1, +1]$.
\end{enumerate}
In principle, there could also exist ambiguities involving derivatives of the Dirac distribution, \ie $\delta^{(n)}(\beta) D^{\pm}_n(\alpha)$. 
Line integrals of such terms contribute to the GPD $H$ with $1/(|\xi| \xi^n) D_n^{(n)}(x/\xi)$ up to a factor $\xi$ depending on whether this term is attached to the DD $F$ (with $D^+_n$) or to the DD $G$ (with $D^-_n$).
The presence of such terms does not change the argument of our discussion. Their phenomenological relevance will be discussed elsewhere, and we will stick to the $\delta(\beta) D^{\pm}(\alpha)$ ambiguities in the following.

Let us summarize: the knowledge of a GPD in the DGLAP region is enough to
constrain it in the ERBL region up to additional terms like (for $\xi \in [-1,
+1]$):
\begin{equation}
\label{eq:ambiguity-on-H-from-dglap}
\delta H(x, \xi) = \frac{1}{|\xi|} D^+\left( \frac{x}{\xi} \right) +
\sgn(\xi) D^-\left( \frac{x}{\xi} \right) \;.
\end{equation}
These terms contribute only to the ERBL region.
In other words, the inverse problem of reconstructing a GPD from its restriction to the DGLAP region admits infinitely many solutions. 
Two distinct solutions differ by $D^+$ and $D^-$ terms as in \refeq{eq:ambiguity-on-H-from-dglap}.
This statement is independent of the choice of an underlying DD scheme.

The $D^+$ and $D^-$ terms modify the polynomiality relation (\ref{eq:def-polynomiality}) by acting on
the two highest degree coefficients:
\begin{equation}
\label{eq:def-polynomiality-modification}
\int_{-1}^{+1} \mathrm{d}x x^m \delta H(x, \xi) = \left( \xi^m
\int_{-1}^{+1} \mathrm{d}z z^m D^+(z) + \xi^{m+1}
\int_{-1}^{+1} \mathrm{d}z z^m D^-(z) \right) \;.
\end{equation}
Since $D^+$ and $D^-$ have opposite parities, the $F$- and $G$-terms do not
simultaneously modify the polynomiality relation. 
To the best of our knowledge, the ambiguity linked to $F$-terms has not been
discussed so far in this context.

\subsection{Problem reduction}
\label{sec:problem-reduction}

We have so far discussed that the overlap representation, as expressed by Eq.~\eqref{eq:OverlapDGLAP}, makes sure the positivity of the GPD;
that there is no simple way to exploit the overlap in the ERBL region and, for the same price,
respect the polynomiality condition at any finite truncation order in Fock space;
and, finally, that the DD representation ensures this polynomiality.
Thus, an immediate and natural approach to model a GPD,
by fulfilling both positivity and polynomiality conditions and exploiting 
the physical information encoded in the LFWFs would
result from:
\begin{enumerate}
  \item the computation of the overlap DGLAP GPD, symmetric in
  Fock space (overlap of LFWFs with the same number of constituents),
  \item the derivation of a DD from this DGLAP GPD by an inverse problem,
  \item the extension of the GPD to its full kinematic domain by means of
  the DD representation.
\end{enumerate}
This is the program we will apply in the following.

In particular, given a GPD $H(x, \xi)$ with support
$x \in [-1, +1]$ and with available non-trivial information
only in the DGLAP region $0 \leq \xi \leq |x| \leq 1$, our goal is to find a 
DD $h$ such that:
\begin{eqnarray}
H(x, \xi) & = &
C^{>}(x,\xi) \int_{\Omega^{>}} \mathrm{d}\beta\mathrm{d}\alpha \, h(\beta,
\alpha) \, \delta(x-\beta-\alpha\xi) \nonumber \\
& + &
C^{<}(x, \xi) \int_{\Omega^{<}} \mathrm{d}\beta \mathrm{d}\alpha \,
h(\beta, \alpha) \, \delta(x-\beta-\alpha\xi) \;, \label{eq:1CDD}
\end{eqnarray}
where the factors $C^{>}$ and $C^{<}$ were defined in \refsec{subsec:quark-GPDs}.

In the next section, we describe a well-established numerical procedure to do it.
But before, we make the following remarks:
\begin{itemize}
  \item The quark and anti-quark GPDs are not correlated in the DGLAP region. 
  In positive DGLAP ($0 \leq \xi \leq x$), only $H^{>}$ is present, while in negative DGLAP ($x \leq -\xi \leq 0$), only $H^{<}$ is. 
  The two parts interfere in the ERBL region  ($-\xi < x < \xi$) where \mbox{$H=H^{>}+H^{<}$}.
  Therefore, the task to accomplish is an independent inversion of \refeq{eq:quark_gpd} and \refeq{eq:antiquark_gpd}. 
  \item As mentioned in \refsec{subsec:double-distributions}, $h$ is $\alpha$-odd as the consequence of time reversal invariance.
\end{itemize}

\begin{figure*}
\begin{tabular}{m{0.4\textwidth}m{0.1\textwidth}m{0.4\textwidth}}
\includegraphics[width=0.4\textwidth]{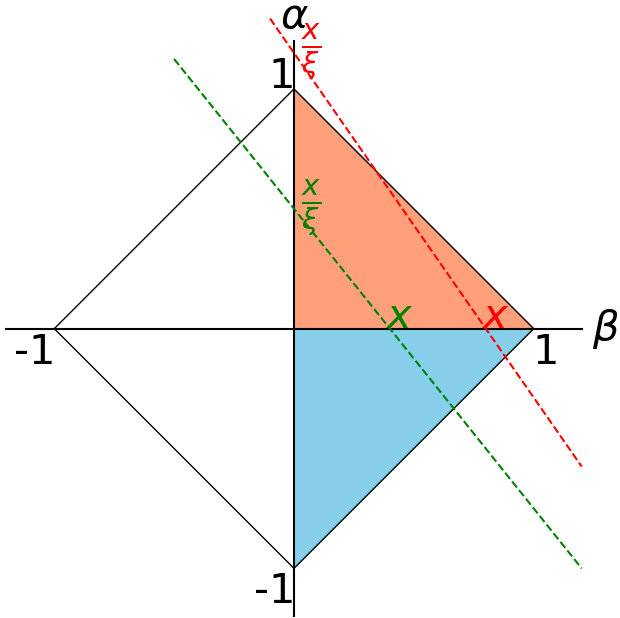} & 
\begin{eqnarray*}
 & \mathcal{R}\\
 & \longrightarrow
\end{eqnarray*}
 & \includegraphics[width=0.4\textwidth]{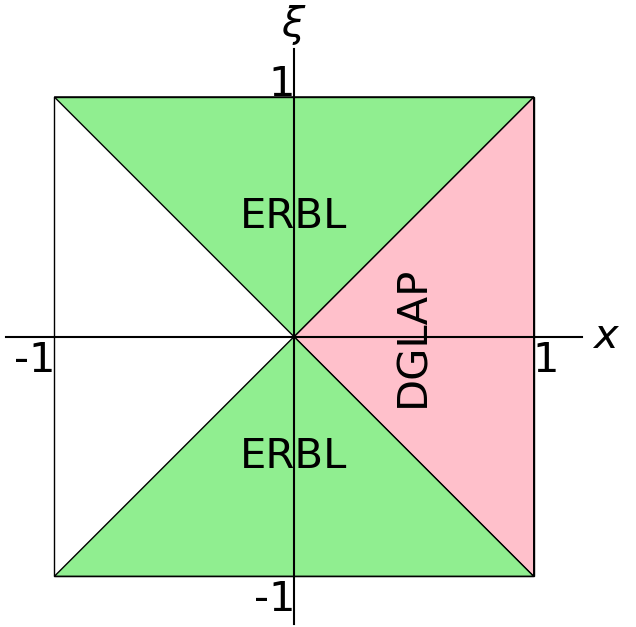}\tabularnewline
\end{tabular}
\caption{DD and GPD supports when only $\Omega^{>}$ is considered. Only the salmon red domain is used. The blue one is
deduced by parity. And the white one, \ie $\Omega^{<}$, is not correlated in DGLAP, and therefore
can be dealt with separately.\label{fig:inversion_domain}}
\end{figure*}

These two properties together reduce the size of a numerical problem by $4$ by comparison of a direct numerical inversion of \refeq{eq:1CDD}.
Indeed, we can separate a general GPD into two distinct problems $H^{>}$ and $H^{<}$, by virtue of linearity of \refeq{eq:1CDD} and the non-correlation in the DGLAP region.
This limits us to half the DD domain (either $\Omega^{>}$ or $\Omega^{<}$) without loss of generality.
And by parity, we reduce again the problem by half.
Figure~\ref{fig:inversion_domain} summarizes this. 
This significantly decreases the computing cost of the numerical inversion\footnote{Given an algorithm with polynomial complexity $\mathcal{O}\left(N^p\right)$ where $N$ is the size of the problem, solving two equal-size independent problems would have a $\mathcal{O}\left(2\,N^p\right)$ complexity, which is much better than a joint problem of complexity $\mathcal{O}\left(\left(2N\right)^p\right)$.} and further constrains the target solution.

As stated in \refsec{subsec:quark-GPDs}, we will discuss only the case of quark DDs and GPDs. 
The treatment of anti-quark DDs and GPDs is essentially the same. 

\section{Numerical implementation \label{sec:numerical-implementation}}

This section presents the considered numerical implementation of the
covariant extension of a GPD from the DGLAP to the ERBL region with its
challenges and results.
In \refsec{sec:radon-inverse} we explained that we can solve our physics problem
by inverting the Radon transform. 
This may seem straightforward since the
Radon transform is linear, but this task is in fact much more difficult. 
Indeed
the inverse Radon transform may not be continuous (see \refapp{app:Radon}) and,
in a loose sense, two ``close'' GPDs may be obtained as
Radon transforms of very ``different'' DDs.
Since we are facing an incomplete data problem (we know the GPD only in the
DGLAP region), the sensitivity to noise is expected to be even stronger than
in the complete data problem, where we search a DD from the complete knowledge
of a GPD and a GDA over their whole kinematic domains.
In this respect, we note that reconstruction artifacts have already been
reported \cite{Teryaev:2001qm, Mezrag:2015mka} for the latter situation.

One key remark is in order here. 
We do not know any closed-form formula for the
inverse of the Radon transform restricted to the DGLAP region. 
It is not even clear that such a formula even exists.
However we do not need it, and it would be of limited practical interest:
the potential amplification of noise is related to the discontinuous nature of
the inverse Radon transform.
It is not the manifestation of a poor numerical scheme or of a badly-designed
computing code. 
It is the inescapable consequence of a precise and general mathematical
statement.
Even if we had at our disposal a closed-form expression of the inverse Radon
transform, we should expect this phenomenon of noise amplification except in the
lucky but rare situations where all computations can be performed analytically. 
As soon as approximations enter the game, the discontinuous nature of the
inverse Radon transform may generate some artifacts in the sought-after DD.

The way to go is well-known in the mathematical literature
(see \eg \refcite{Natterer:2001m}). 
Assuming that the underlying DD is smooth enough, it is possible to numerically
invert the Radon transform while keeping noise under control. 
This is called regularization.

This sections falls into three parts. We first discretize our problem to reduce
it to the computation of the pseudo inverse of a rectangular matrix. 
Then we select adequate linear solver and regularization procedure. 
At last we
validate our computing chain with simple but relevant test case scenarios.

\subsection{Discretization} \label{sec:discretization}

The goal now is to obtain a discrete problem
from the integral equation~\eqref{eq:quark_gpd}, and we will use the usual notation:
\begin{equation}
AX=B\,,\label{eq:linear_problem}
\end{equation}
where $A$ is a $m \times n$ matrix, $X$ a vector of dimension $n$, and $B$ 
a vector of dimension $m$.

\subsubsection{Mesh}

To obtain this finite-dimensional linear problem,
the DD space should first be discretized.
In an abstract way, we use a set of basis functions $\left\{v_{j}\right\}$ for the decomposition:
\begin{equation}
h\left(\beta, \alpha\right)
= \sum_{j}h_{j}\,v_{j}\left(\beta, \alpha\right)\,,\label{eq:DD_decomposition}
\end{equation}
where the index $j$ labels the set of basis functions, and therefore the 
degrees of freedom.
Adopting a formalism close to the one of Finite Element Methods~(FEM)~\cite{LanMar16}, these basis functions are in one-to-one correspondence to given nodes in the DD domain. Indexing these nodes means indexing the basis functions.
Applying this to a given mesh which is a set of vertices (or corners) and 
edges defining its elements, we can be more explicit.
A basis function is non-zero only on elements adjacent to its corresponding 
node, and the restriction of a basis function to one such element is the Lagrange interpolation with respect to this node,
\ie the polynomial that is equal to $1$ on the said node, and $0$ on all others.
See \reffig{fig:basis-function-P1} for an example of such a basis function.

Following the conventions of FEM~\cite{Arnold2014}, we will consider the
following classification:
\begin{description}
  \item[$P_n$-Lagrange] Used for triangular meshes, where the restriction of a
  basis function to a triangular element is an interpolating polynomial of total degree at most $n$.
  For example, for $P_1$, it would be a polynomial of the form $a+b\,\beta+c\,\alpha$. 
  \item[$Q_n$-Lagrange] Used for quadrilateral meshes, where the restriction
  of the basis function to a mesh element is an interpolating polynomial of partial degree at most $n$.
  For example, for $Q_1$, it would be a polynomial of the form $a+b\,\beta+c\,\alpha+d\,\beta\,\alpha$.
\end{description}

In the case of linear piece-wise functions ($P_1$ or $Q_1$), the considered nodes of the basis functions are the vertices of the mesh.
For higher orders, the nodes also include other points (such as the middle of the edges for $P_2$ and $Q_2$).
We will also consider constant piece-wise functions and we will call those elements $P_0$
(which corresponds to $dP_0$ in FEM notations).
In this case, each basis function corresponds to one element (or any node in the interior of the element,
\eg the center of gravity, to keep the same correspondence between nodes and basis functions).
\reftab{tab:basis-functions} summarizes this.

Our unknowns $\left\{ h_{j}\right\}$ of \refeq{eq:DD_decomposition} correspond to the values of the DD $h$ on the nodes $j$,
and will be recast into the vector $X$ of the discrete problem \eqref{eq:linear_problem}.

\begin{table}
\begin{centering}
\begin{tabular}{|c|c|c|}
\hline 
Order $n$ & Basis function & Node\tabularnewline
\hline 
\hline 
0 & Piece-wise constant & Center of gravity of an element\tabularnewline
\hline 
1 & Piece-wise linear & Vertex of the mesh\tabularnewline
\hline 
2 & Piece-wise quadratic & Vertex or middle of an edge\tabularnewline
\hline 
\end{tabular}
\par\end{centering}
\caption{Summary of the different $P_n$ elements. Each basis function has
support on the elements surrounding the corresponding node. The restriction
to an element is a Lagrange interpolation: it takes a value of $1$
on the said node, and $0$ on all the other nodes.\label{tab:basis-functions}}
\end{table}

For the work presented here, we will consider only a triangular mesh, since the domain is a triangle anyway
(see \reffig{fig:inversion_domain}), with $P_1$ or $P_0$ elements.

We will always use the index $j$ in the following to label the basis functions, \ie the nodes,
and the index $k$ for labelling the triangular elements.
Of course, in the case of $P_0$, the indices will be interchangeable, since a basis function is defined by an element.

\subsubsection{Basis functions}

\begin{figure}[h]
  \includegraphics[width=0.7\columnwidth]{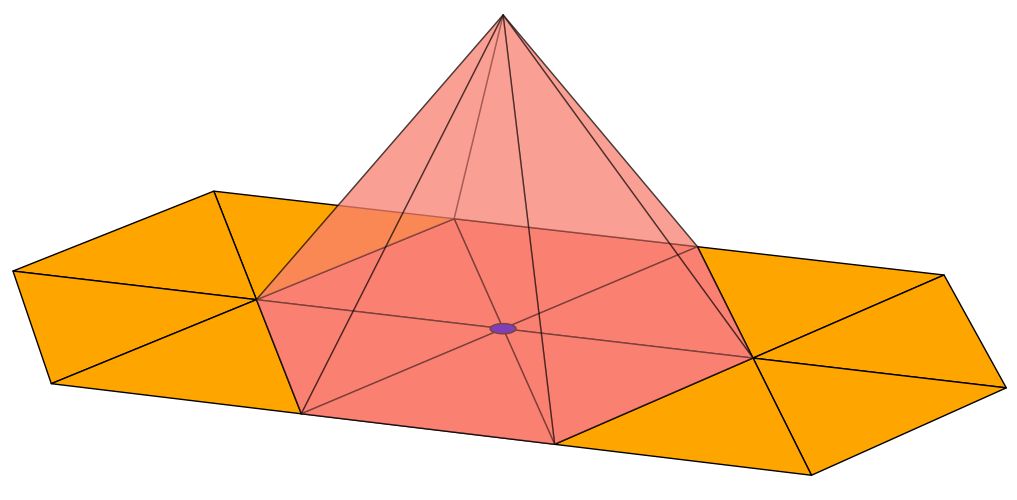} 
  \caption{Example of a $P_1$ basis function. The corresponding node (\ie a vertex in this case) is represented in blue. The value of the basis function on this node is 1, and 0 on the others. The support is limited to the adjacent triangles (in salmon red color).}
  \label{fig:basis-function-P1}
\end{figure}

In the case of a triangular mesh, it is natural to use barycentric coordinates to define the basis functions (instead of the Cartesian coordinates).
For a given triangle $k$, we will denote by \{$\lambda_k^1\left(\beta,\alpha\right)$, $\lambda_k^2\left(\beta,\alpha\right)$,
$\lambda_k^3\left(\beta,\alpha\right)$\} the barycentric coordinates with respect to the three vertices. Note that the number of degrees of freedom is still $2$, since ${\lambda_k^1+\lambda_k^2+\lambda_k^3=1}$.
A given point $\left(\beta,\alpha\right)$ belongs to the triangle $k$ if all three barycentric coordinates are positive. Moreover, for $P_1$ elements, they provide natural restrictions for the basis functions,
since they are exactly the linear Lagrange interpolations at the vertices.
If we denote by \mbox{$\left(\beta_i,\alpha_i\right)$, $i=1\ldots3$}, the three vertices of a triangle, then the barycentric coefficient with respect to the first vertex can be written as:
\begin{equation} \label{eq:barycentric-coeff}
\lambda^1\left(\beta,\alpha\right) =
\frac {\beta_3 \, \alpha_2 - \beta_2 \, \alpha_3 + \left(\alpha_3 - \alpha_2\right) \beta + \left(\beta_2 - \beta_3\right) \alpha}
{\beta_3 \, \alpha_2 - \beta_2 \, \alpha_3 + \left(\alpha_3 - \alpha_2\right) \beta_1 + \left(\beta_2 - \beta_3\right) \alpha_1} \, ,
\end{equation}
and the others similarly by cycling indices.

The matrix of the linear problem is determined by the linear operator 
that transforms a DD into a GPD in \refeq{eq:quark_gpd}.
To build this matrix, we only need to know the Radon transform\footnote{To simplify the notations, we redefined the operator $\mathcal{R}$ without the $\xi$-dependent factor.} of a basis function:
\begin{eqnarray}
\mathcal{R}v_j\left(x,\xi\right) & = &
\int v_j\left(\beta,\alpha\right)\delta\left(x-\beta-\alpha\xi\right)\mathrm{d}\beta\,\mathrm{d}\alpha\,.
\label{eq:transform_basis-function}
\end{eqnarray}
Let us first express this basis function in the $P_0$ and $P_1$ cases 
(superscripts 0 and 1 respectively):
\begin{eqnarray}
v_j^0\left(\beta,\alpha\right) & = &
\theta\left(\lambda_j^1\left(\beta,\alpha\right)\right)\,\theta\left(\lambda_j^2\left(\beta,\alpha\right)\right)\,
\theta\left(\lambda_j^3\left(\beta,\alpha\right)\right) \, , \label{eq:basis-function-P0} \\
v_j^1\left(\beta,\alpha\right) & = & \sum_{\substack{k \, \in \, \text{elements} \\ \text{adjacent to vertex } j}}
\theta\left(\lambda_k^1\left(\beta,\alpha\right)\right)\, \theta\left(\lambda_k^2\left(\beta,\alpha\right)\right)\,
\theta\left(\lambda_k^3\left(\beta,\alpha\right)\right) \, \lambda_k^{\bar{j}}\left(\beta,\alpha\right) \, ,
\label{eq:basis-function-P1}
\end{eqnarray}
where $\bar{j}$ is the vertex $j$ recast to the limited set $\left\{1,2,3\right\}$ of vertices of the element $k$.
The $P_1$ basis function is also represented in \reffig{fig:basis-function-P1}.
Applying the Radon transform on these basis functions yields:
\begin{eqnarray}
\mathcal{R}v_j^0\left(x,\xi\right) & = & 
\theta\left(\alpha^j_{\text{max}}-\alpha^j_{\text{min}}\right)
\left(\alpha^j_{\text{max}}-\alpha^j_{\text{min}}\right) \, ,
\label{eq:transform_basis-function-P0} \\
\mathcal{R}v_j^1\left(x,\xi\right) & = & \sum_{\substack{k \, \in \, \text{elements} \\ \text{adjacent to vertex } j}}
\theta\left(\alpha^k_{\text{max}}-\alpha^k_{\text{min}}\right)
\int_{\alpha^k_{\text{min}}}^{\alpha^k_{\text{max}}} \mathrm{d}\alpha \,
\lambda_k^{\bar{j}}\left(x-\alpha \xi,\alpha\right) \, ,
\label{eq:transform_basis-function-P1}
\end{eqnarray}
where the bounds of the integration $\left\{\alpha^k_{\text{min}},\alpha^k_{\text{max}}\right\}$ are determined with the
three inequalities given by the Heaviside functions (positive barycentric coordinates).
For higher order elements, the idea is the same, only the integrated function will change.

\subsubsection{Sampling}

The next step is then to discretize the GPD variables $\left(x,\xi\right)$, 
\ie to sample the set of straight lines intersecting the domain $\Omega$.
Given that we have only access to DGLAP kinematics,
we will use the couples $(x, y)\in [-1, +1]^2$ with
$y=\xi/x$.
The choice of $(x, y)$ will determine a line of the matrix.
More precisely, the matrix $A$ will have the coefficients:
\begin{equation}
A_{ij}=C^{>}\left(x_i,\xi_i\right)\,\fRadon{v_j}{x_i}{\xi_i} \, ,
\label{eq:matrix-coefficients}
\end{equation}
where $1 \leq i \leq m$ indexes the lines of the matrix, 
and $1 \leq j \leq n$ indexes the columns, \ie the nodes in DD space.
The $C^{>}$ factor was introduced in \refeq{eq:quark_gpd}.

The size of the matrix is chosen such that we maximize the information, 
\ie we need to integrate over lines that cross all the elements of the DD mesh.
A value of $m\sim4\,n$ is empirically satisfying.
The matrix can be therefore built by picking random couples $\left(x,y\right)$ until we attain the desired size.
The results will of course depend on the matrix used and it is interesting to consider this as a source of ``statistical error'',
whereas the regularization procedure (see the following section) would be the source of ``systematic error''.
The statistical error can be managed quite easily and reduced considerably by picking as many samples as we want,
whereas the systematic error remains a challenge to estimate.

Once the matrix is built, we use the set of chosen couples $\left(x,y\right)$
to build the vector $B$ \rhs of \refeq{eq:linear_problem} with simply:
\begin{equation}
B_{i}=H^{>}\left(x_{i},\xi_{i}\right) \, . \label{eq:rhs_linear_problem}
\end{equation}

In summary, $A$ is a matrix $m\times n$ where $n$ is the number of 
mesh elements for $P_0$ (or number of vertices for $P_1$) and $m$ the number
of straight lines intersecting $\Omega$. 
Each line will typically cross $\mathcal{O}(\sqrt{n})$ mesh elements, which means that
only $\mathcal{O}(\sqrt{n})$ coefficients on a matrix line are non-zero 
and $A$ is a sparse matrix. 
We need more constraints
than parameters ($m > n$) and we usually use $m = 4n$, making the rank 
of $A$ $\lesssim n$ (\ie close to full-rank).
$B$ is a vector of dimension $m$, and $X$ of dimension $n$.

\subsection{Linear solver and regularization\label{subsec:Linear-solver}}

An additional complexity arises in the selection of the matrix inversion
routine.
In \refsec{sec:formalization} we assumed the existence 
of a covariant extension of a DGLAP-restricted GPD $\HDGLAP$ and
showed its uniqueness up to the manifestations of ambiguities on the line $\beta
= 0$.
The key question is to know whether such an extension ever exist?
Or, stated differently, given a function defined in the DGLAP region 
(a putative overlap of LFWFs), is it possible to extend it to the 
ERBL region in a way satisfying polynomiality? 
Existing criterions as in \refeq{eq:LH-BMKS-m-equal-0} and
\refeq{eq:LH-Pobylitsa-m-equal-0}, complemented by the Ludwig-Helgason
consistency conditions, deal with the GPD known over its whole physical domain,
not its restriction to the DGLAP region. 
A numerical solver may have to handle a linear system as in 
\refeq{eq:linear_problem} but without one and only one solution.
This is common in computerized tomography, not because the 
solution does not exist 
(there was one object inserted inside the scanning
device), but because the experimental signal comes with noise which
may apparently modify the original situation to an \emph{inconsistent data
problem}.
In the framework of the Radon transform, causes may be multiple: 
the integration lines may not cross the same domain (no solutions), or they may
be parallel and close to one another and bring redundant information
(infinitely many solutions).
One efficient way to ensure that the solution always exists and is unique is to
turn to a least-square formulation: 
\begin{equation}
\label{eq:least-square-problem}
\textrm{Search } X \in \mathbb{R}^n \textrm{ such that } ||A X - B||^2 \textrm{
is minimum,}
\end{equation}
where $||.||$ generically denotes a norm in a finite-dimensional vector space.

In the present work, we use a recent iterative conjugate-gradient type algorithm
for sparse least-squares problems: LSMR \cite{LSMR}.
For inconsistent problems (where the least-square formulation is favored), 
it is
equivalent to a Minimum Residual algorithm for the problem:
\begin{equation}
^{t}AA\,X={}^{t}AB\,,\label{eq:least-squares_normal_eq}
\end{equation}
but it can also solve directly the problem (\ref{eq:linear_problem}) when it is consistent,
\ie when the numerical approximation of the target solution is equal to the exact function.
In the $P_0$ case, it means functions that are already piece-wise constant,
whereas a $P_1$ approximation can reproduce exactly a (piece-wise) linear polynomial.

\begin{figure}
\begin{centering}
\includegraphics[width=0.8\columnwidth]{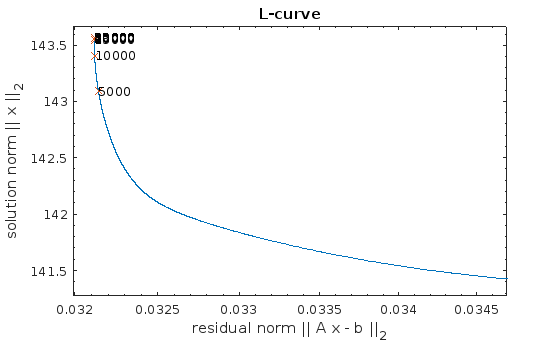}
\par\end{centering}
\caption{L-curve obtained with the number
of iterations as regularization parameter for the case of \refsec{sec:AlgebraicModel}.\label{fig:L_curve}}
\end{figure}

This type of algorithms applies naturally its own regularization process,
with the number of iterations being the regularization parameter.
To illustrate this, we can use the so-called L-curve \cite{Hansen2007}, which is a curve
following a regularization parameter (which is in our case the number of
iterations) and shows the compromise between the norm of the solution $\left\Vert X\right\Vert $
(the larger the norm, the larger the impact of noise) and the residual
norm $\left\Vert r\right\Vert$, where $r=AX-B$ (which we desire to be small enough to converge to the real solution).
This procedure gives the optimal regularization factor to choose for each problem,
as the point of maximum curvature of the ``L'', as shown in \reffig{fig:L_curve}.

In practice, as illustrated on \reffig{fig:L_curve},
it is very difficult to determine this optimal regularization parameter for the considered problems.
A better way to stop the iterations is to consider the stopping criteria used by these algorithms such as LSMR:
\begin{itemize}
\item For a consistent problem: $\left\Vert r\right\Vert \leq\texttt{atol}\,\left\Vert A\right\Vert \,\left\Vert X\right\Vert +\texttt{btol}\,\left\Vert B\right\Vert $~;
\item For a least-squares problem: $\left\Vert A\,r\right\Vert \leq \texttt{atol}\,\left\Vert A\right\Vert \,\left\Vert r\right\Vert $,
\end{itemize}
where $\texttt{atol}$ and $\texttt{btol}$ are the input tolerances.

An empirical value of $10^{-5}$ for the least-square tolerances gives a good
compromise between noise and convergence, and this has the benefit of being valid for all considered cases,
assuming that the considered DD is smooth and not one of the problematic cases with singularities such as \refsec{subsec:Regge} (for which a specific workaround is needed and explained therein).
A $10^{-5}$ value for $\texttt{atol}$ (resp. $\texttt{btol}$) means that the
matrix $A$ (resp. the right hand side $B$) is known exactly up to the fifth decimal, while the rest is numerical noise.
Of course, in practice,
we can compute analytically $A$
(if the chosen basis functions and mesh allow us to compute the Radon Transform without numerical integration,
as it is the case for the method presented here) and $B$ (if the GPD is known analytically),
and they are therefore known exactly, up to machine precision.
But in the inconsistent (\ie least-squares) case,
the considered vector $B$ is different from the one due to a GPD calculated from the discrete numerical DD.
This difference is the finite limit of the residual, in contrast with the consistent case where the residual has a zero limit.
Even though small, when allied with the ill-posed character of the inversion, it can have a large impact on the solution.
This is why we consider in practice that $B$ (or equivalently $A$) is not known exactly and neglect higher decimals~;
we apply a regularization procedure by doing so.

\subsection{Test and validation of the numerics} \label{subsec:test}

The first immediate check we can perform to validate 
 the numerical implementation described above, consists in the following.
We first take a simple Ansatz for the DD, irrespective of the considered 
value of $t$ (\eg $t=0$).
We then compute the associated analytical GPD by applying the Radon transform, and use only its DGLAP part to apply our numerical inversion and obtain a numerical estimate of the DD.
Finally, we compare this result with the original Ansatz.
We will apply this testing procedure to the following 
three quark GPDs, each one of them deriving from a DD in the 
P scheme: 
\begin{itemize} 
\item[(i)] A constant DD $h_p^{\rm cst}(\beta,\alpha)$ on the half-domain
$\Omega^{>}$:
\begin{equation}
\label{eq:DDconstant}
\hp^{\rm cst}(\beta, \alpha) = \frac{3}{2} \theta(\beta) 
 \ \ \ \Rightarrow \ \ \ 
H(x, \xi) = \frac{3 (1-x)^2}{1-\xi^2} \;,
\end{equation}
\item[(ii)] the example defined by
\refeq{eq:toy-model-dd-regularized-pobylitsa-gauge} and already 
introduced in
\refsec{sec:example-polynomiality-toy-model},
\item[(iii)] and a simplified case of the RDDA:
\begin{equation}
\label{eq:DDRDDA}
\hp^{\rm RDDA}(\beta, \alpha) = \frac{\Gamma(N+3/2)}{\sqrt{\pi} \, \Gamma(N+1)} 
\frac{\left[(1-\beta)^2-\alpha^2\right)^N}{\left(1-\beta\right)^{2N+1}} 
\frac{q(\beta)}{1-\beta} \;, 
\end{equation}
where $q(\beta)$ is the associated PDF. 
In particular, we specialize for the case $N=1$ and 
take\footnote{This is a very good practical approximation of the 
result for the valence-quark pion's PDF obtained in
\refscite{Chang:2014lva, Mezrag:2014jka}, within a Bethe-Salpeter 
and Dyson-Schwinger approach and by including the appropriate 
correction to the impulse-approximation. It also results directly 
from  the overlap of the LFWF derived from the same Bethe-Salpeter
 wave function~\cite{Mezrag:2016hnp}.} 
 $q(\beta)=30 \, \beta^2 (1-\beta)^2$.
We thus obtain a closed algebraic formula for the GPD which, 
in turn, can be numerically inverted with the 
procedure described above.
\end{itemize}

\begin{figure*}
 \begin{tabular}{cc} 
  \includegraphics[width=0.5\textwidth]{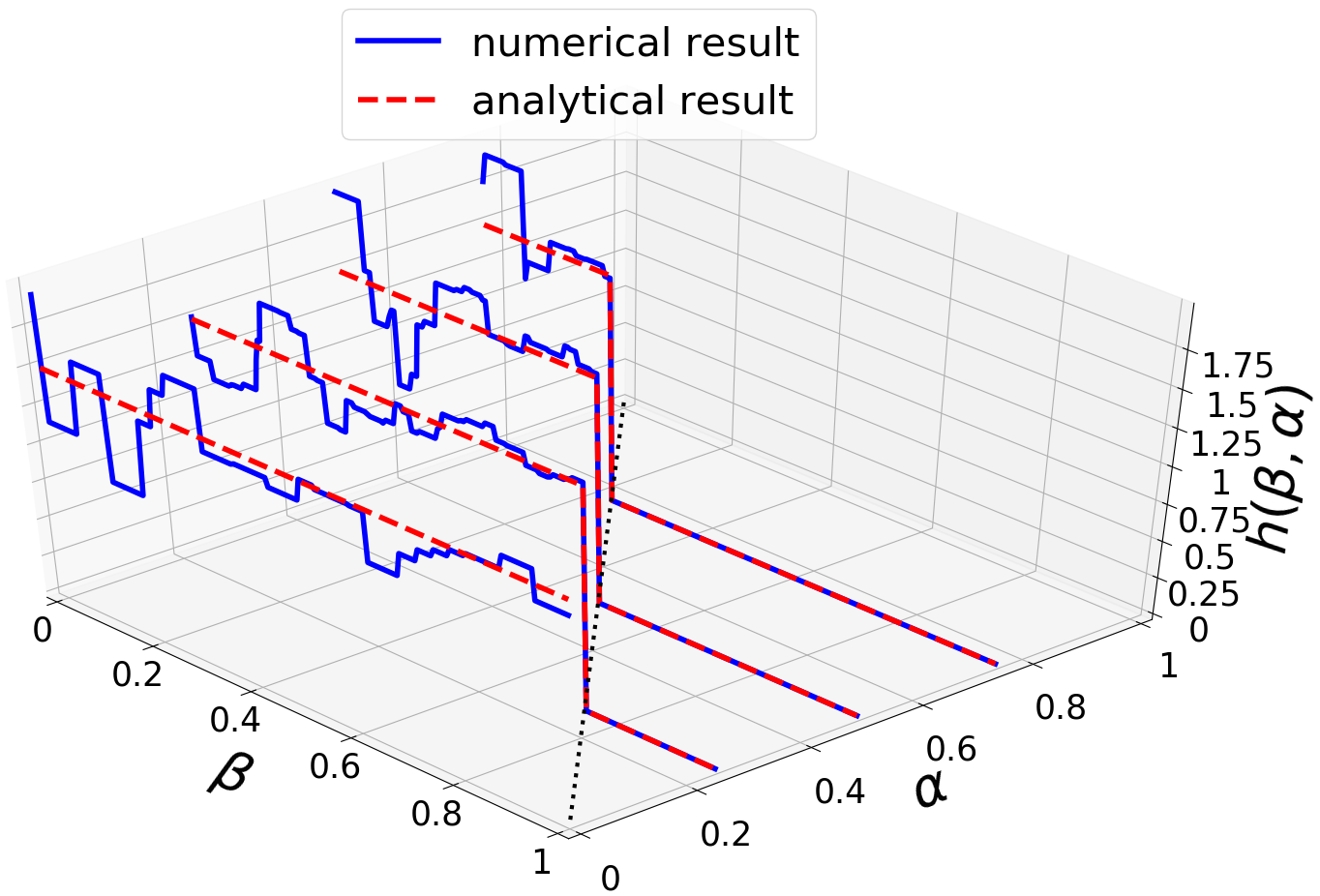} & 
   \includegraphics[width=0.5\textwidth]{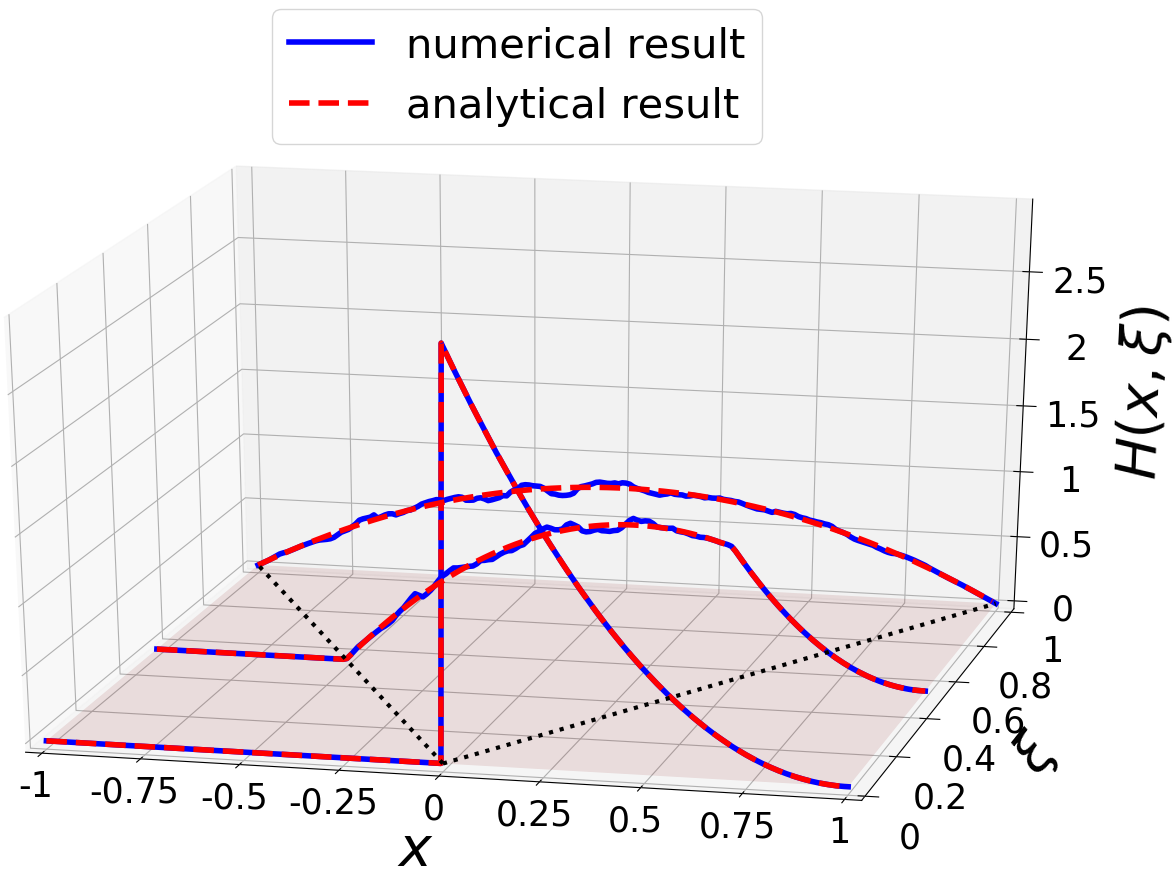} \\
     \includegraphics[width=0.5\textwidth]{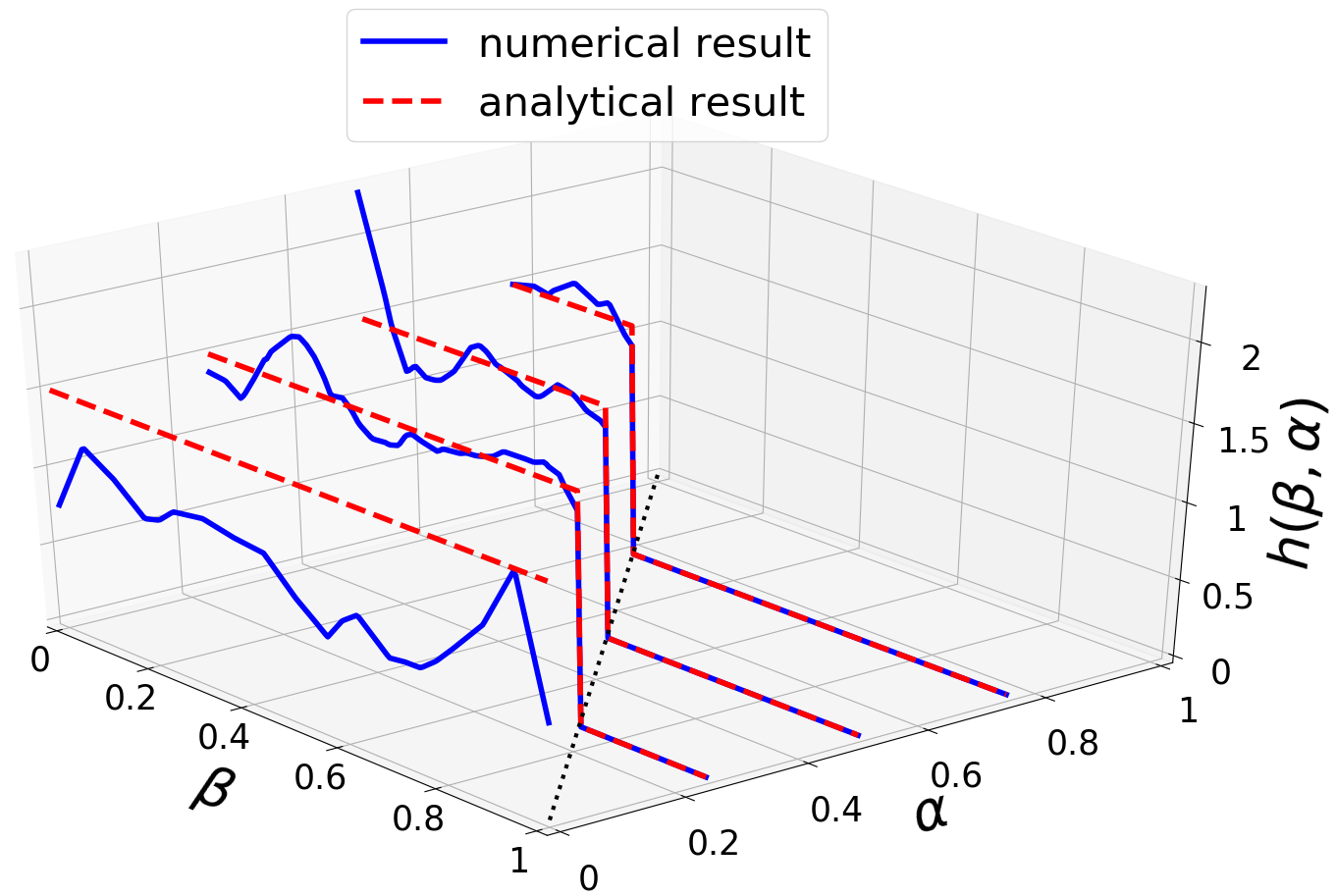} & 
   \includegraphics[width=0.5\textwidth]{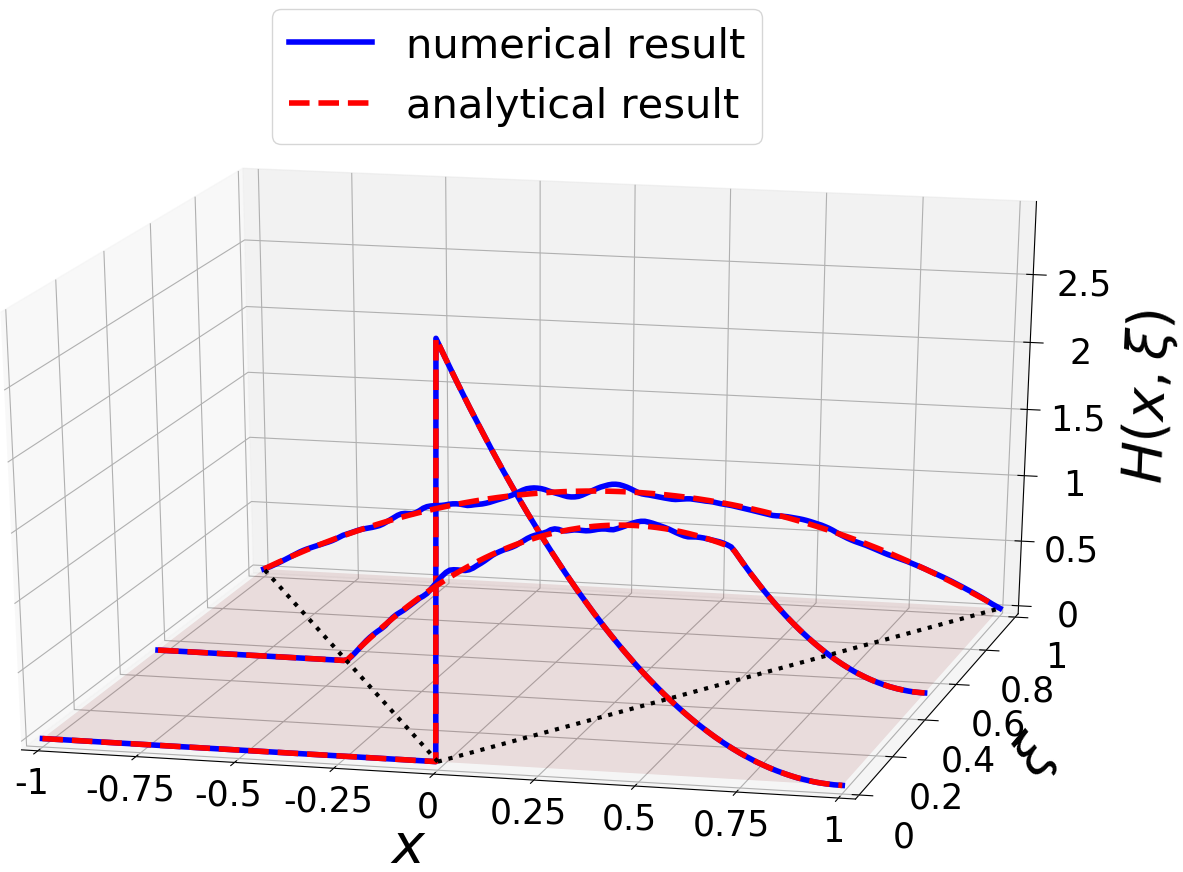}
   \end{tabular}
        \caption{Comparison between algebraic, 
        given by \refeq{eq:DDconstant}, and numerical results 
        for $\hp^{\rm cst}(\beta,\alpha)$ at fixed values of $\alpha=\left[0,0.25,0.5,0.75\right]$ (left panel) 
        and the corresponding GPD at fixed values of $\xi=\left[0,0.5,1\right]$ (right panel) for the case (i).
        The blue solid curves display the numerical results
         while the red dashed ones show the algebraic 
         results. The black dotted curve indicate
         either the line $\beta+\alpha=1$ (left panel) or $x \pm \xi = 0$
         (right panel). The upper panel stands for a discretization 
         obtained with
         $P_0$ elements, while the lower panel displays the 
         results with $P_1$ elements. }
    \label{fig:DDconstant}
\end{figure*}

\begin{figure*}
 \begin{tabular}{cc} 
  \includegraphics[width=0.5\textwidth]{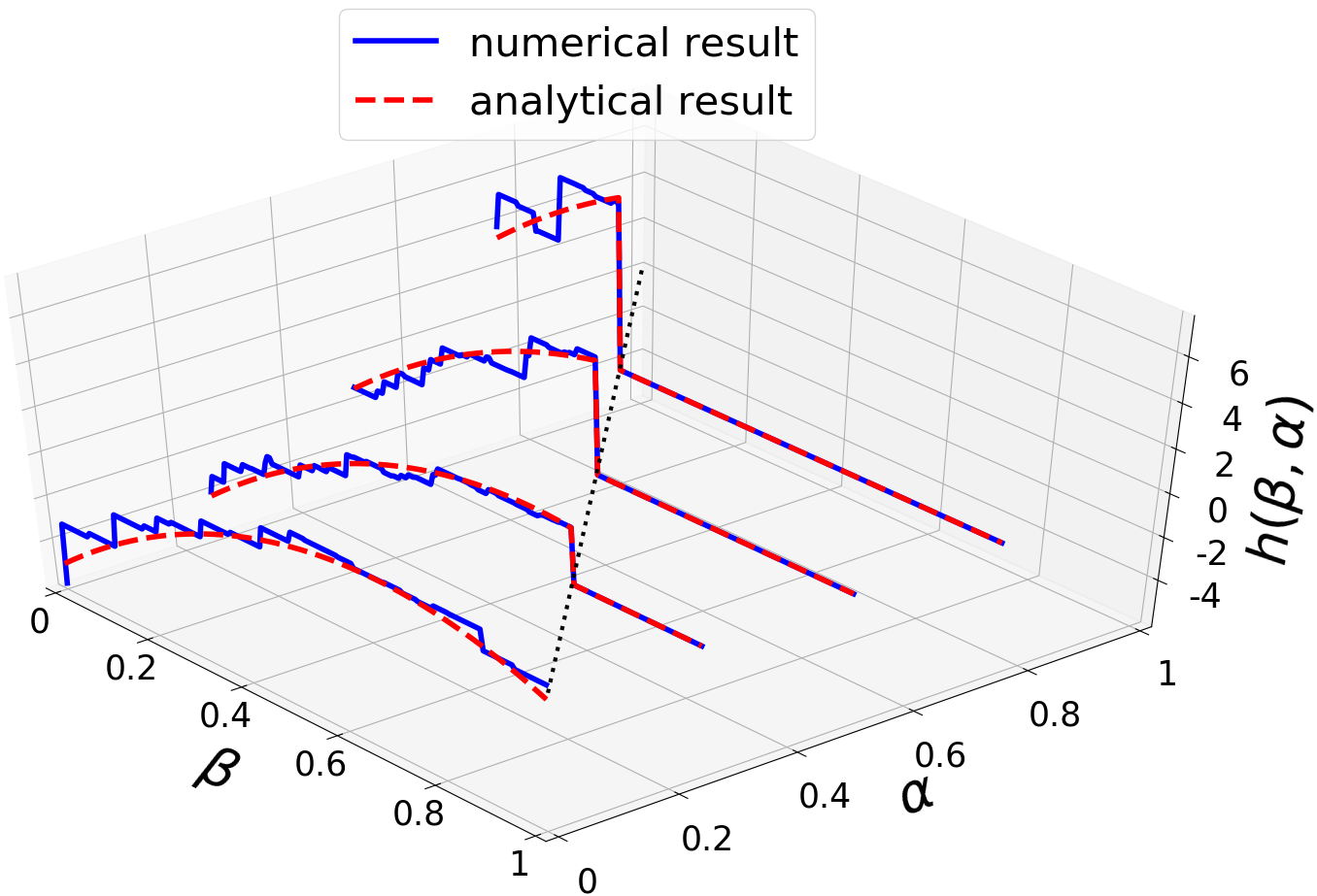} & 
   \includegraphics[width=0.5\textwidth]{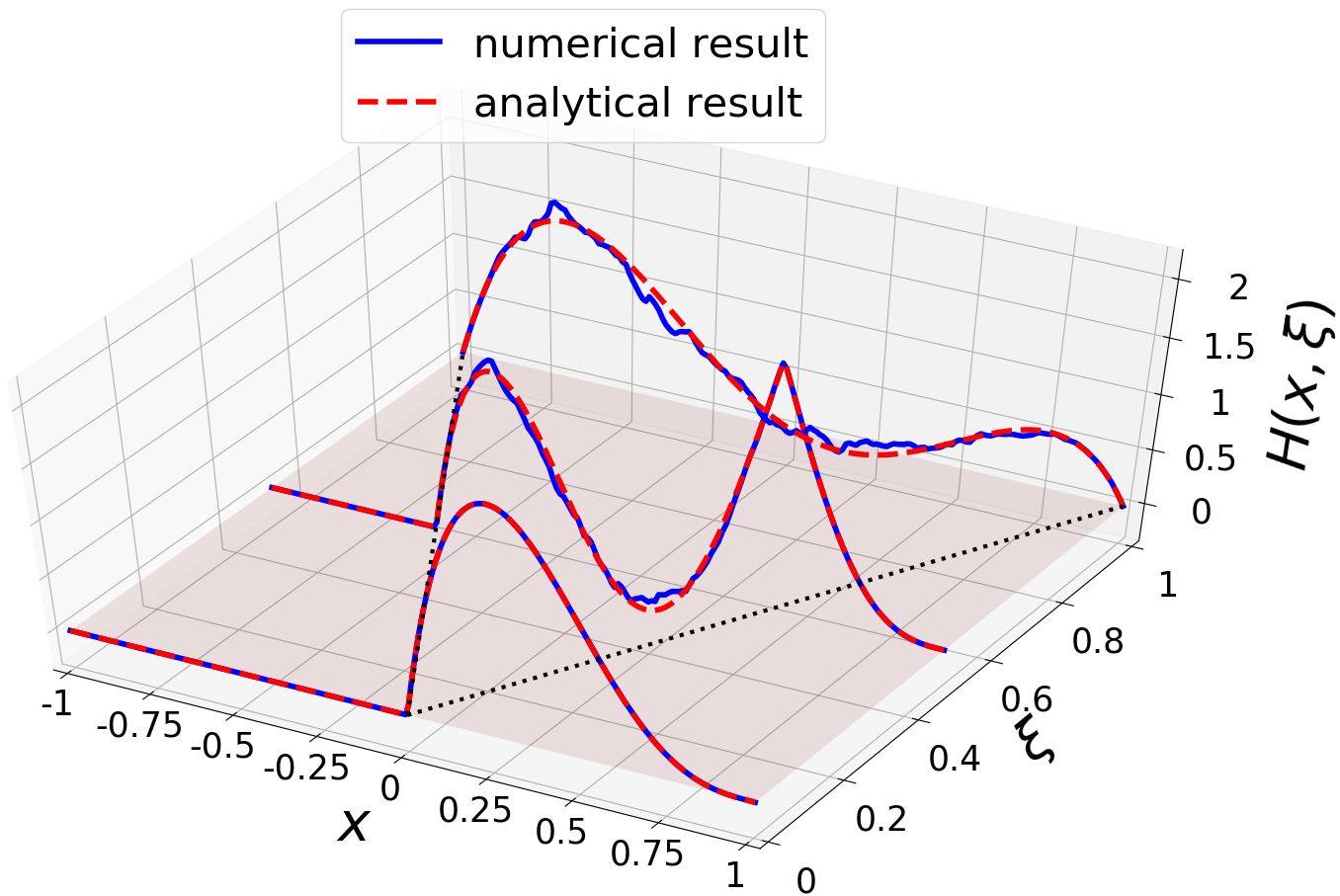} \\
     \includegraphics[width=0.5\textwidth]{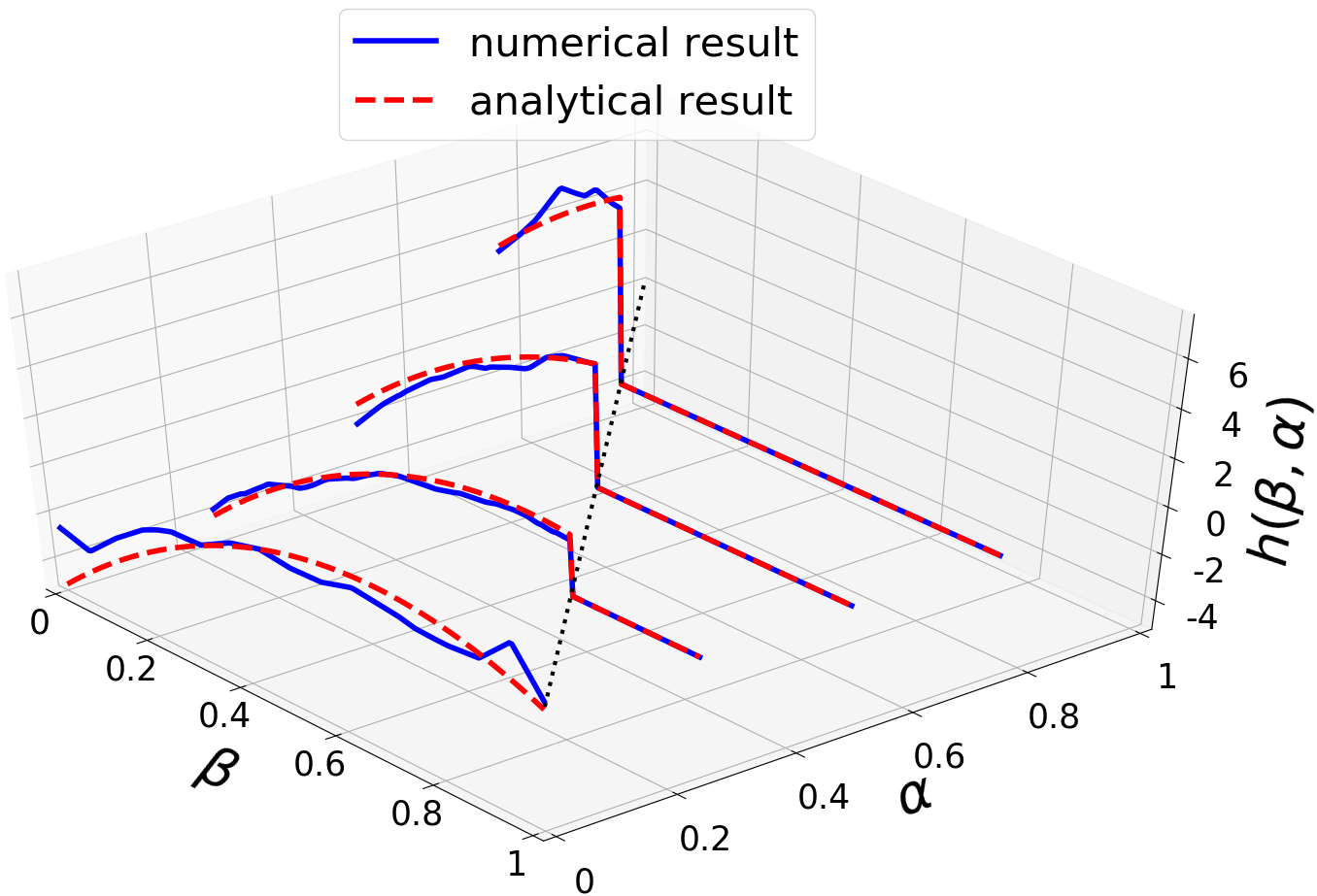} & 
   \includegraphics[width=0.5\textwidth]{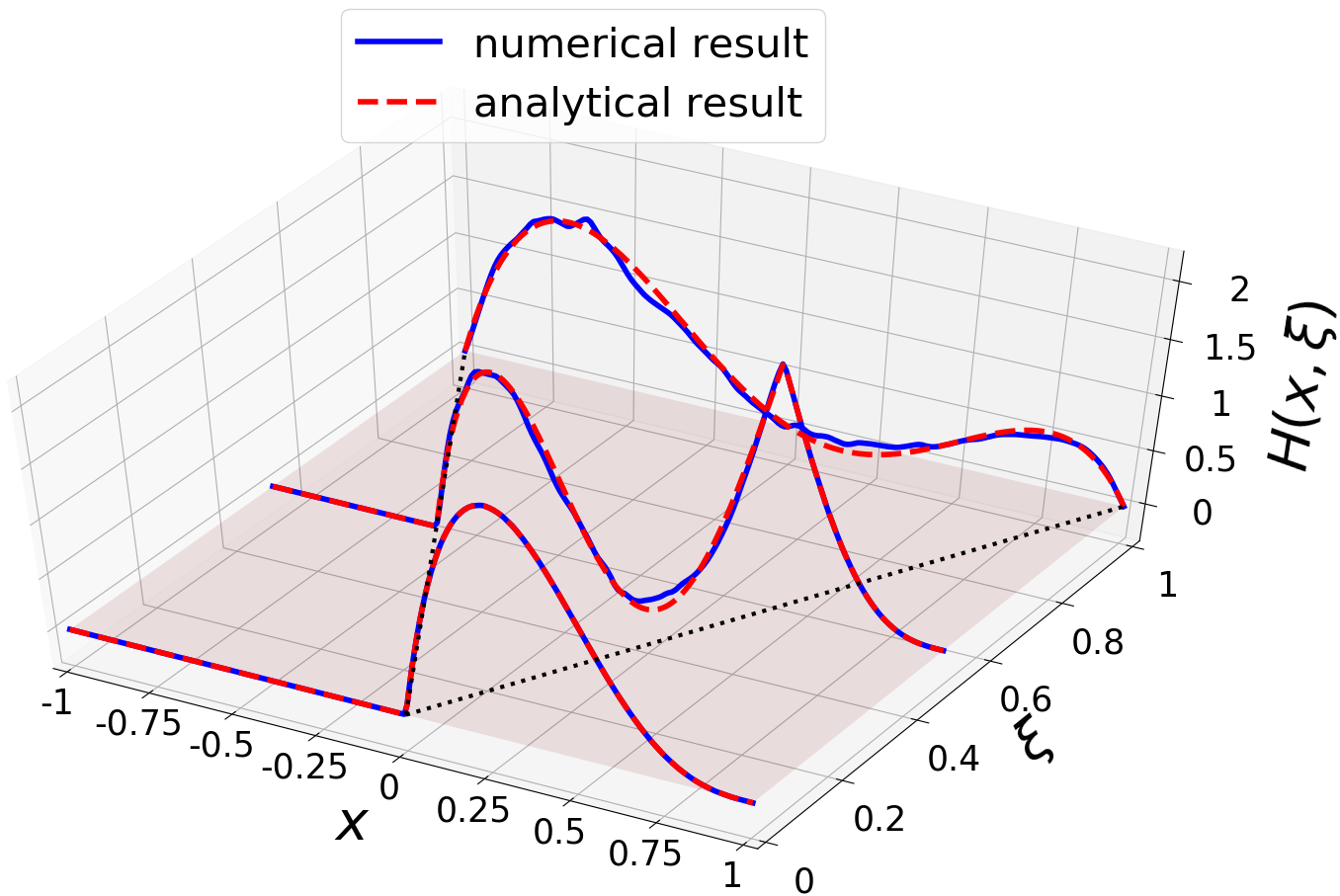}
   \end{tabular}
        \caption{Comparison between algebraic, given by 
        \refeq{eq:toy-model-dd-regularized-pobylitsa-gauge}, and
        numerical results for $\htoymodelp(\beta, \alpha)$ 
        (left panel)
        and the corresponding GPD (right panel) for the case (ii).
          Same conventions as in \reffig{fig:DDconstant}.}
    \label{fig:DDtoy}
\end{figure*}

\begin{figure*}
 \begin{tabular}{cc} 
  \includegraphics[width=0.5\textwidth]{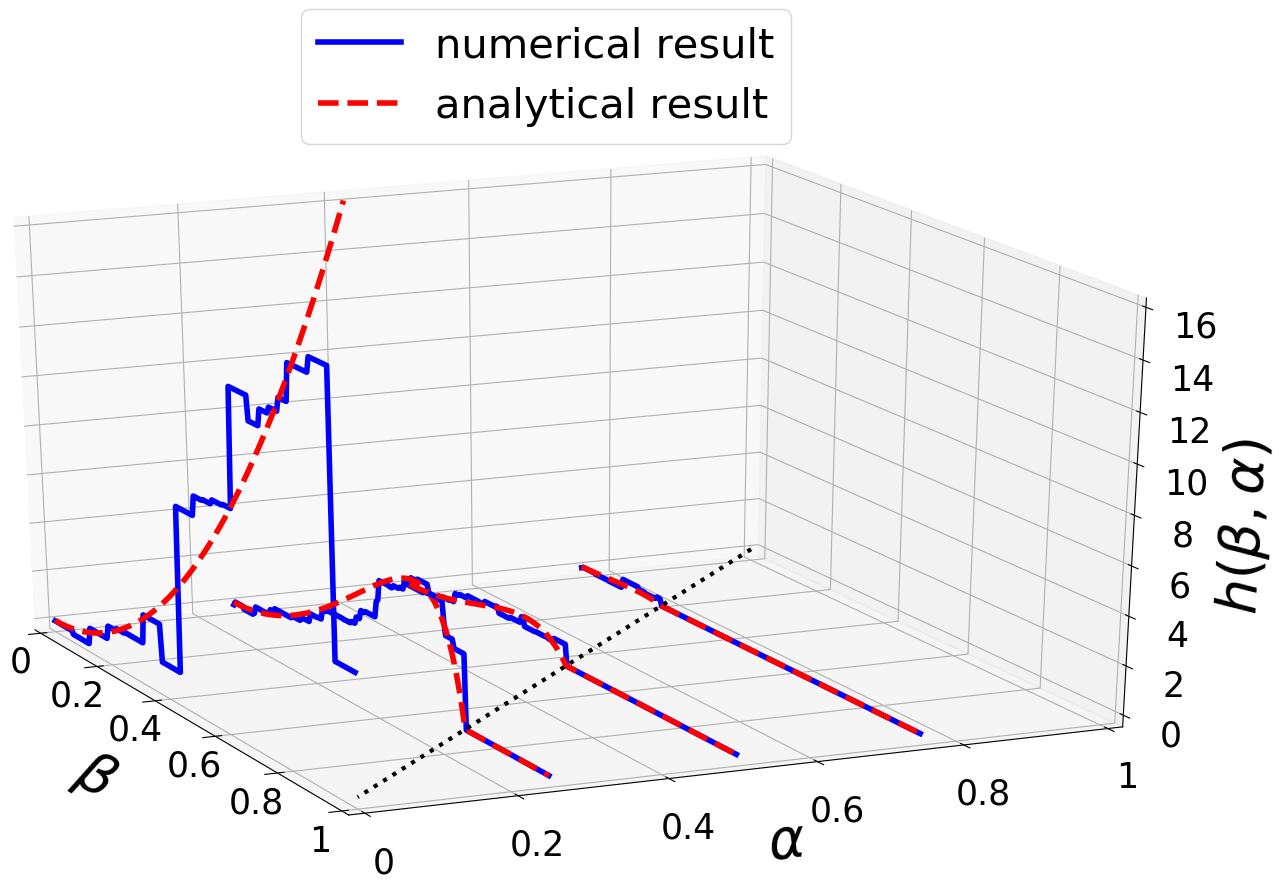} & 
   \includegraphics[width=0.5\textwidth]{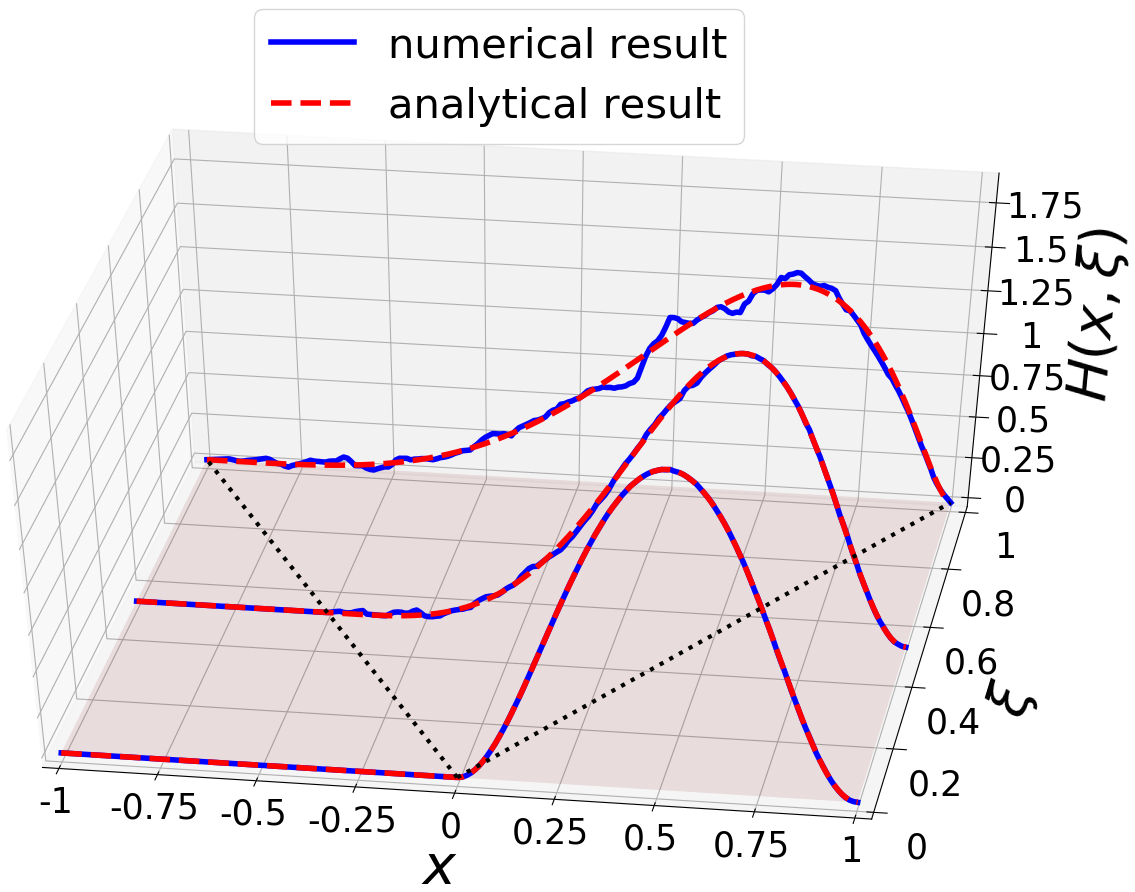} \\
     \includegraphics[width=0.5\textwidth]{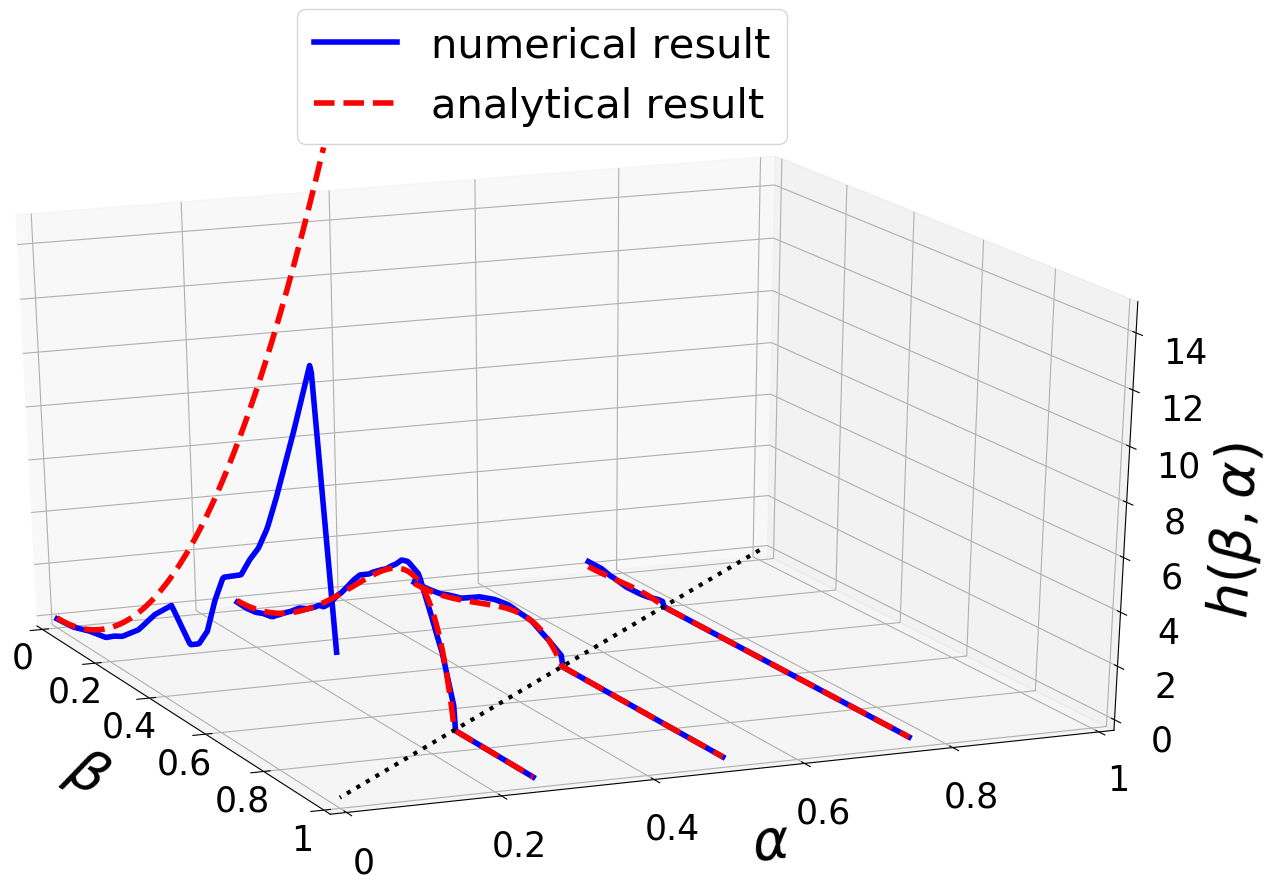} & 
   \includegraphics[width=0.5\textwidth]{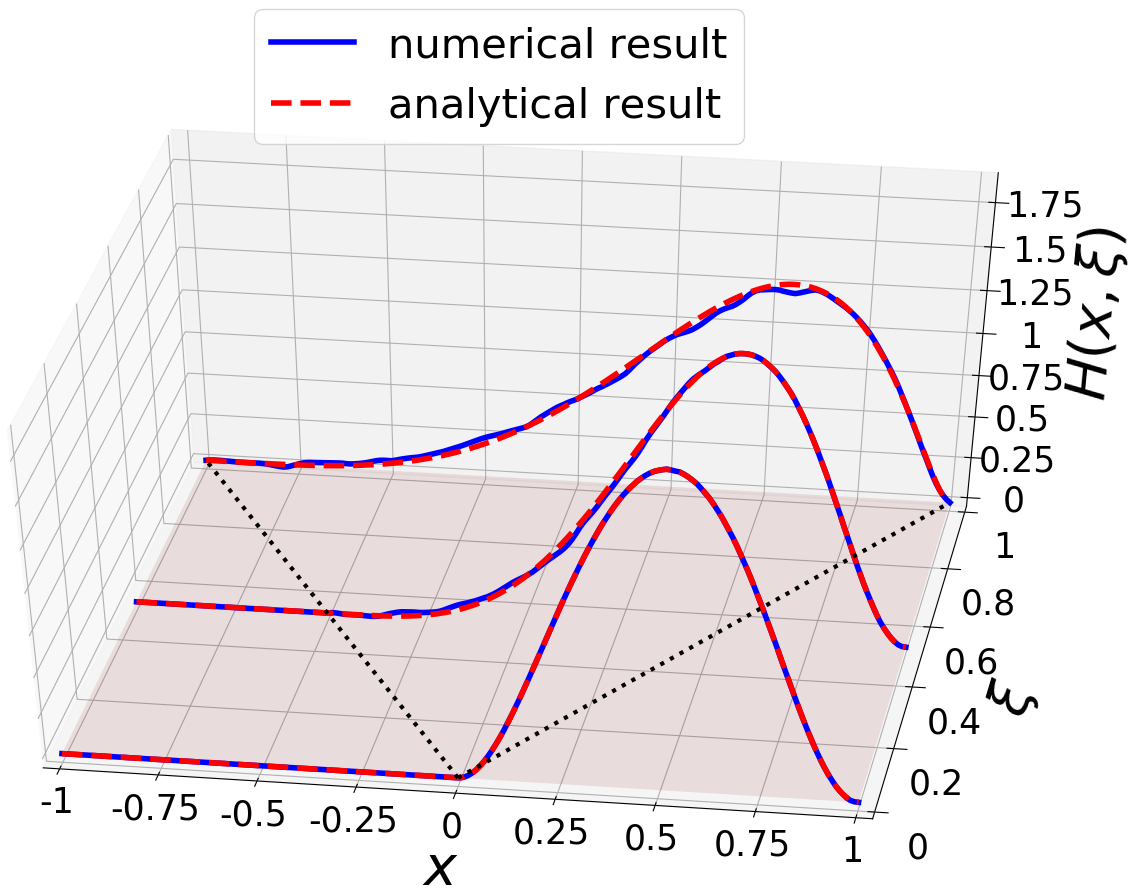}
   \end{tabular}
        \caption{Comparison between algebraic, given by 
        \refeq{eq:DDRDDA}, and numerical results for
        $\hp^{\rm{RDDA}}(\beta, \alpha)$ (left panel) and the 
        corresponding
        GPD (right panel) for the case (iii).  
        Same conventions as in \reffig{fig:DDconstant}.}
    \label{fig:DDRDDA}
\end{figure*}

In all cases, we adopted the P scheme as it appears to be a 
convenient representation for GPD models derived through a 
covariant extension of an overlap of LFWFs, 
as discussed in \refsec{sec:gpd-theory-modeling}. 
Therefore it 
will systematically be employed for the analysis of the 
more practical examples scrutinized in the next section.
In the leftmost plots of \reffig{fig:DDconstant}, we display the 
comparison of the exact algebraic and numerically approximated 
$\hp^{\rm cst}(\beta, \alpha)$ for the case (i), 
while the rightmost ones stand for the comparison of the 
GPDs directly obtained from both DDs.
The upper and lower plots have been obtained, respectively, 
with piecewise constant (\ie $P_0$ elements) and piecewise linear 
(\ie $P_1$)
basis functions when discretizing the DD space for the reduction to
 a finite-dimensional problem, as explained 
 in \refsec{sec:discretization}.
Analogous plots, and similarly arranged, are
 displayed in \reffig{fig:DDtoy} for
case (ii) and \reffig{fig:DDRDDA} for case (iii).

The mesh was generated using the Triangle software~\cite{shewchuk96b} with a requirement of maximal area for the triangular elements equal to $0.001$, which produced a mesh of $427$ vertices and $780$ elements.
The linear solver used is described in \refsec{subsec:Linear-solver}.
As justified before, we used a tolerance of $10^{-5}$ as a regularization
procedure.

All in all, as an overall conclusion we can assert that 
the numerical inversion approximates very well the three known 
GPDs as they
appear not to differ significantly in all the cases. 
Several important points should be however stressed out:
\begin{itemize}
  \item The numerical inversion 
  relies only on the knowledge of the GPD within the DGLAP region 
  and its extension to the ERBL is our main goal, wherefore the 
  examination of algebraic and numeric GPDs over the ERBL 
  region is the main outcome of Figs.~\ref{fig:DDconstant}-\ref{fig:DDRDDA}.
  \item The numerical reconstruction of the DDs may seem 
  quite noisy or far off in some cases, but these discrepancies 
  do not hinder the reconstruction of the GPD, for which the 
  convolution helps smooth these defaults\footnote{This should not come as a
  surprise. The Radon transform is a smoothing operator, since it
  integrates a DD over lines. Conversely, the inverse Radon operator has to
  undo this smoothing to reconstruct the DD, hence provoking noise
  amplification.}.
  The physical object of interest is the GPD, not the DD, 
  therefore these discrepancies are not an issue.
  \item It should be noted however that the constant DD can be 
  reconstructed numerically exactly (up to machine precision).
  Indeed, the regularization in that case is not needed, 
  since there is no distinction between the analytical DD and 
  its discretized version.
  We could therefore directly invert the discrete problem and 
  recover the exact DD (which would be equivalent to using a 
  tolerance of $0$ instead of $10^{-5}$), but for the sake of 
  homogeneity,
  we decided to employ the same method for all shown examples.
  \item It may seem from the plots that the $P_1$ discretization 
  does not improve on the $P_0$ result, but this is not true.
  We chose here to use the same mesh for both $P_0$ and $P_1$ 
  elements.
  This particular mesh has $427$ vertices and $780$ triangular 
  elements.
  Hence, the number of degrees of freedom for $P_1$ (\ie the number
   of vertices) is half the one for $P_0$ (\ie number of elements).
  In other words, we attain with $P_1$ a similar result to the 
  $P_0$ one but with half the degrees of freedom, \ie at a much 
  lower cost.
  It is therefore a significant improvement.
  In \refsec{sec:applications}, we will keep only the $P_1$ method 
  for non-trivial applications with popular LFWFs occurring 
  in realistic descriptions of hadron structure.
\end{itemize}

\section{Examples of Applications \label{sec:applications}}

The approach described in \refsec{sec:gpd-theory-modeling} and
\refsec{sec:radon-inverse} can be either applied to numerous 
existing LFWF-based GPD models to covariantly extend them from 
the DGLAP region
to the ERBL one, or used to build a covariant GPD model, 
reliable on both DGLAP
and ERBL regions, from the knowledge of the LFWF.
Although in some particular cases, an analytical derivation of the DD is possible and a full GPD in both DGLAP and ERBL regions can be thereupon obtained, one can only proceed systematically by applying the numerical technique that has been 
introduced in section \ref{sec:numerical-implementation}. 
In this section, aiming to illustrate the procedure without the intention of being exhaustive, we provide four examples of GPD models, three of which can be extended to the ERBL region both analytically and numerically, allowing us to benchmark our algorithm. 

\subsection{Algebraic Bethe-Salpeter model}
\label{sec:AlgebraicModel}

We consider first a specific pion GPD model described in
\refcite{Mezrag:2016hnp}, based on the following helicity-0:
\begin{equation}\label{eq:LCWF}
\Psi_{l=0}\left(x,\vperp{k}\right) = 8 \, \sqrt{15} \, \pi \, \frac{M^3}{\left(\vperp{k}^2 + M^2\right)^2} \, \left(1-x\right) x \, , 
\end{equation}
and helicity-1: 
\begin{equation}\label{eq:LCWFh1}
i \, k_{\perp}^j \, \Psi_{l=1}\left(x,\vperp{k}\right) = 8 \, \sqrt{15} \, \pi \, \frac{k_{\perp}^j \, M^2}{\left(\vperp{k}^2 + M^2\right)^2} \, \left(1-x\right) x \, , \qquad j=1,2 \, ,
\end{equation}
contributions to the LFWF, obtained by integrating and properly projecting the pion Bethe-Salpeter wave function resulting from the algebraic model described in \cite{Chang:2013pq} where $M$ is a model mass parameter introduced at the level of the quark propagator.
A value of $M\sim0.318~\GeV$ allows to recover the pion charge
radius~\cite{Amendolia:1986wj}.
\refeqs{eq:LCWF}{eq:LCWFh1} correspond to specializing the helicity-0
and helicity-1 components given, respectively, by Eqs.~(154) and~(155) in \refcite{Mezrag:2016hnp} for the asymptotic case $\nu=1$ therein.
Then, exploiting the GPD overlap representation as described in \refsec{subsec:overlap} and extending \refeq{eq:overlap_ud} with  the two contributions:
\begin{equation}
\label{eq:overlap_ud_2contributions}
H_{\pi^{+}}^{u}\left(x,\xi,t\right)=\int\frac{\mathrm{d}^{2}\mathbf{k}_{\bot}}{16\,\pi^{3}}
\left[
\Psi_{l=0}^{*}\left(\hat{x}^{'},\hat{\mathbf{k}}_{\bot}^{'}\right)
\Psi_{l=0}\left(\tilde{x},\tilde{\mathbf{k}}_{\bot}\right)
+ \hat{\mathbf{k}}_{\bot}^{'} \cdot \tilde{\mathbf{k}}_{\bot}
\Psi_{l=1}^{*}\left(\hat{x}^{'},\hat{\mathbf{k}}_{\bot}^{'}\right)
\Psi_{l=1}\left(\tilde{x},\tilde{\mathbf{k}}_{\bot}\right)
\right]\,,
\end{equation}
one is left 
with\footnote{There is a normalization mismatch 
between Eq.~(168) and Eq.~(169) in \refcite{Mezrag:2016hnp}, 
which has been corrected here.}:
\begin{equation}
\label{eq:GPDAlgebraicDGLAP}
\left. H_{\pi^+}^{u}(x,\xi,t)\right|_{\xi \le x} \ = \
\frac{15}{2} \ \frac{(1-x)^2 (x^2-\xi^2)}{(1-\xi^2)^2} \ \frac 1 {(1 + \zeta)^2} \left( 3 + \frac{1-2 \zeta}{1+\zeta} 
\ \frac{\displaystyle \arctanh{\left(\sqrt{ \frac{\zeta}{1+\zeta}}\right)}}{ \displaystyle \sqrt{\frac{\zeta}{1+\zeta}}} \right) \ ,
\end{equation}
as a fully algebraic result for the DGLAP region, where:
\begin{equation}
\zeta \ = \ \frac{-t}{4M^2} \frac{(1-x)^2}{1-\xi^2} \;, 
\end{equation}
encodes the correlated dependence of the kinematical variables $x$ and $t$.

 \begin{figure*}[t]
 \begin{tabular}{cc} 
  \includegraphics[width=0.5\textwidth]{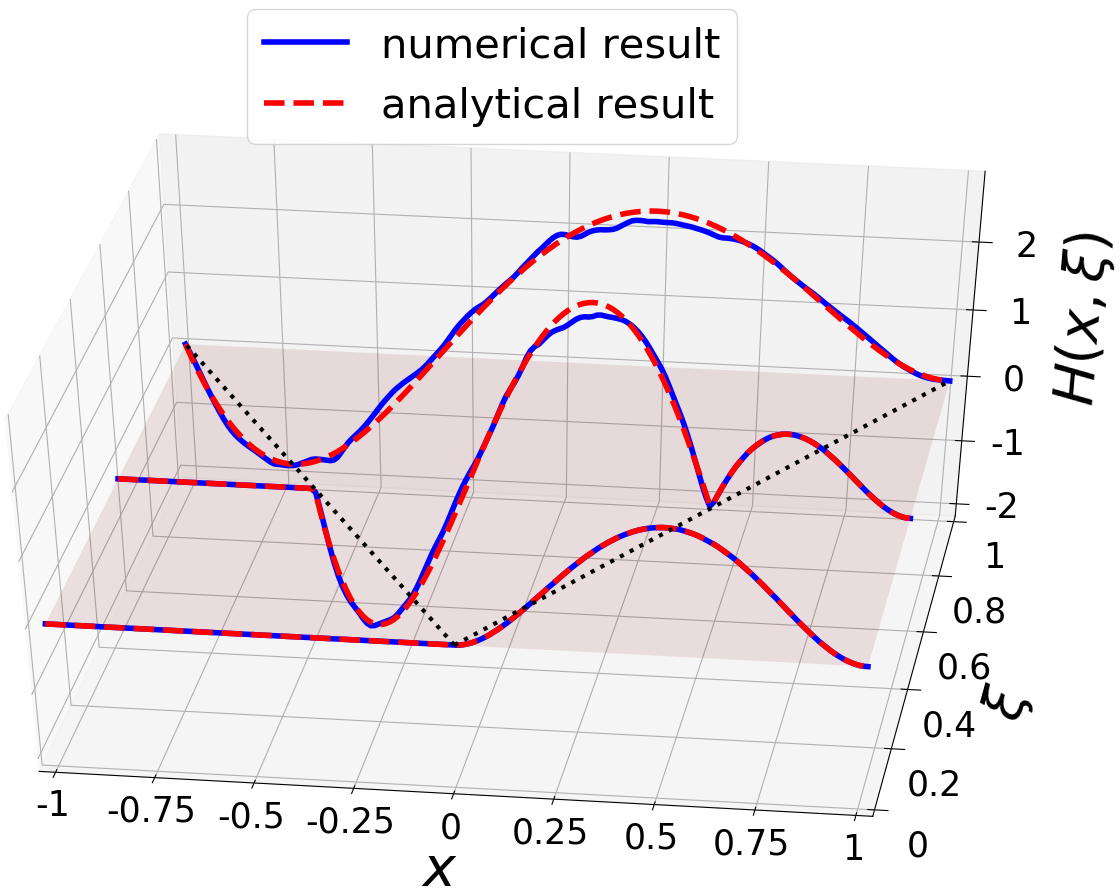} & 
   \includegraphics[width=0.5\textwidth]{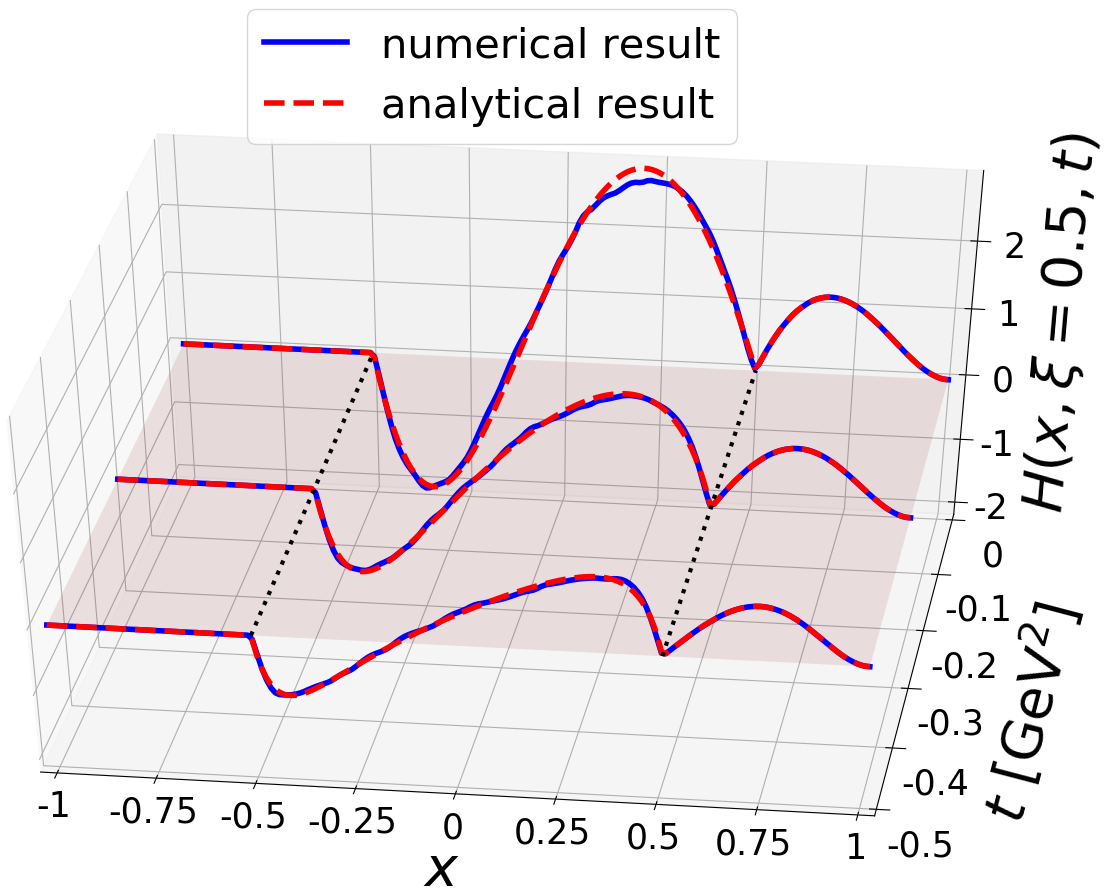}
   \end{tabular}
     \caption{\small Comparison between algebraic and numerical 
     results for the pion GPD modeled in \refcite{Mezrag:2016hnp}. 
      As it was the case in \reffig{fig:DDconstant}, 
      the blue solid curves 
      display the numerical results while the red dashed ones 
      show the
      results algebraically derived and given by
      \refeqs{eq:GPDAlgebraicDGLAP}{eq:GPDAlgebraicERBL}. 
      The left panel
      stands for the case $t=0$ for fixed values of $\xi=\left[0,0.5,1\right]$ and the right one shows 
      the $t$-behavior for fixed values $\left[0,-0.25,-0.5\right]$ at $\xi=0.5$. 
      We retain only $P_1$ elements. 
      For more details, see \refsec{subsec:test}.}
    \label{fig:AlgebraicModel}
 \end{figure*}

A careful computation allows for the derivation of the 
following closed expression for the DD in the P scheme 
given by \refeqs{eq:pobylitsa-gauge-F}{eq:pobylitsa-gauge-G}:
\begin{equation}
\label{eq:DDAlgebraic}
\hp(\beta, \alpha, t) = \frac {15}{2} 
\frac{1 -3 \alpha^2 - 2 \beta + 3 \beta^2 
+ \frac{-t}{4 M^2} \left(1 -\alpha^4+2\alpha^2\beta^2 - \beta^4 + 4 \beta^2 - 4 \beta \right)}
{\left(1 + \frac{-t}{4 M^2} \left(
\left(1-\beta\right)^2-\alpha^2\right)\right)^3} \theta(\beta) \;.
\end{equation}
This last result then can be applied, as explained in
\refsec{sec:radon-inverse}, to provide us with a covariant 
extension of the result given in \refeq{eq:GPDAlgebraicDGLAP} 
to the ERBL kinematic domain.
A thorough scrutiny of the main features resulting from the 
pion GPD model of \refcite{Mezrag:2016hnp} will be the object of 
a further work, in particular the study of the evolution in $t$ 
and its implications.
However, the main purpose of bringing here this model 
computation is to benchmark the numerical technique 
developed in \refsec{sec:numerical-implementation}. 
This can be easily done,
without any loss of generality, by focusing on the case $t=0$, 
for which one can write down, invoking \refeq{eq:quark_gpd}, 
the following simple expression for the GPD within 
the ERBL domain:
\begin{equation}
\label{eq:GPDAlgebraicERBL}
\left. H_{\pi^+}^{u}(x,\xi,0)\right|_{|x| \le \xi} \ = \
\frac {15} 2 \frac{(1-x) (\xi^2-x^2)}{\xi^3 (1+\xi)^2} \left( x + 2 x \xi + \xi^2 \right) \ .
\end{equation}
Thus, we compare the results given by \refeq{eq:GPDAlgebraicDGLAP} and
\refeq{eq:GPDAlgebraicERBL}, respectively derived for the DGLAP and ERBL
kinematics,  with those obtained by the numerical inversion of the linear problem described in section~\ref{sec:numerical-implementation}; and display the outcome in the leftmost panels of \reffig{fig:AlgebraicModel}.
As can be seen, both numerical and algebraic results compare strikingly well, not only at a qualitative but also a quantitative level.
Albeit the successful comparison we obtain for the case $t=0$ is satisfactorily enough, aiming at the practical validation of the numerical technique at other values of $t$, we have also displayed in the rightmost panel of \reffig{fig:AlgebraicModel} the results for the GPD evolved in $t$ at a constant value of $\xi=0.5$.
The algebraic expression for the DGLAP region is given in
\refeq{eq:GPDAlgebraicDGLAP}, while that for ERBL can be obtained by the
integration of \refeq{eq:DDAlgebraic} through \refeq{eq:quark_gpd}.
The algebraic expression for the DGLAP region is given in
\refeq{eq:GPDAlgebraicDGLAP}, while that for ERBL can be obtained by the
integration of \refeq{eq:DDAlgebraic}.

\subsection{Algebraic spectator model}

If LFWFs have been widely used in attempts to model the pion, the case of the nucleon has also been treated previously, in particular in the pioneering paper of Hwang and M\"uller~\cite{Hwang:2007tb}.
They have developed an algebraic parametrization for two-body LFWFs of
the nucleon (described by a constituent quark and a spectator scalar diquark):
\begin{eqnarray}
  \label{eq:MuellerLFWFPara}
  \Psi_{+1/2}^\uparrow\left(x,\vperp{k} \right) & = & \Psi_{-1/2}^\downarrow\left(x,\vperp{k} \right) = \left(M + \frac{m}{X} \right)\varphi(x, \vperp{k}),  \\
  \label{eq:MuellerLFWFanti}
  \Psi_{-1/2}^\uparrow\left(x,\vperp{k} \right) & = & -\frac{k^1+ik^2}{x}\varphi(x,\vperp{k}),  \quad \Psi_{+1/2}^\downarrow\left(x,\vperp{k} \right)  =  \frac{k^1-ik^2}{x}\varphi(x,\vperp{k}),\\
  \label{eq:MuellerLFWFScalar}
  \varphi(x,\vperp{k}) & = & \frac{gM^{2p}}{\sqrt{1-x}}x^{-p}\left(M^2-\frac{\vperp{k}^2+m^2}{x}-\frac{\vperp{k}^2+\lambda^2}{1-x} \right)^{-p-1},
\end{eqnarray}
with $M$, $m$ and $\lambda$ being respectively the nucleon, quark and spectator masses, $p$ being a free parameter.
The authors have shown that with such a model, after calculating the overlap of
wave functions (see Eqs.~(13-14) of \refcite{Hwang:2007tb}), one can write the GPD $E$ in the P scheme:
\begin{equation}
  \label{eq:GPDEMueller}
  E(x,\xi,t) = (1-x) \int_0^1 \textrm{d}\beta \int_{-1+\beta}^{1-\beta}\textrm{d}\alpha\ e(\beta,\alpha,t) \, \delta(x-\beta-\xi \alpha) ,
\end{equation}
and they have been able to extract analytically:
\begin{equation}
\label{eq:DDeMueller}
e\left(\beta,\alpha,t\right) = \frac{N \, \left(\beta +\frac{m}{M}\right) \left((1-\beta )^2-\alpha ^2\right)^p }
{\left( 2 \left(\frac{(1-\beta ) m^2}{M^2}+\frac{\beta  \lambda ^2}{M^2}-(1-\beta ) \beta -\frac{t \left((1-\beta )^2-\alpha ^2\right)}{4 M^2}\right)\right)^{2 p + 1} }
\end{equation}
where $N$ is a constant determined by the usual PDF normalization (obtained with the GPD H, and not E).
Therefore, the authors managed to extend their specific model in the ERBL region.
Consequently, we use this model as an additional benchmark for our numerical technique.

\begin{figure*}[t]
 \begin{tabular}{cc} 
  \includegraphics[width=0.5\textwidth]{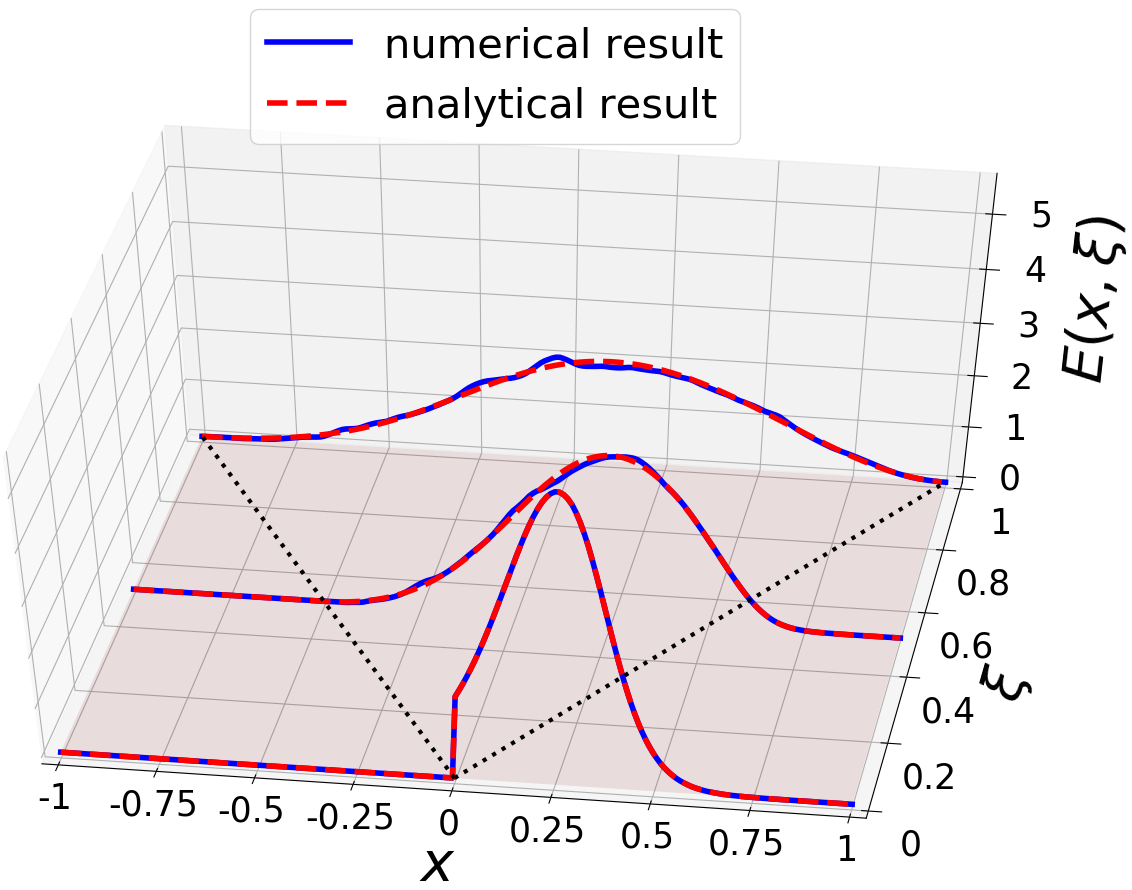} & 
   \includegraphics[width=0.5\textwidth]{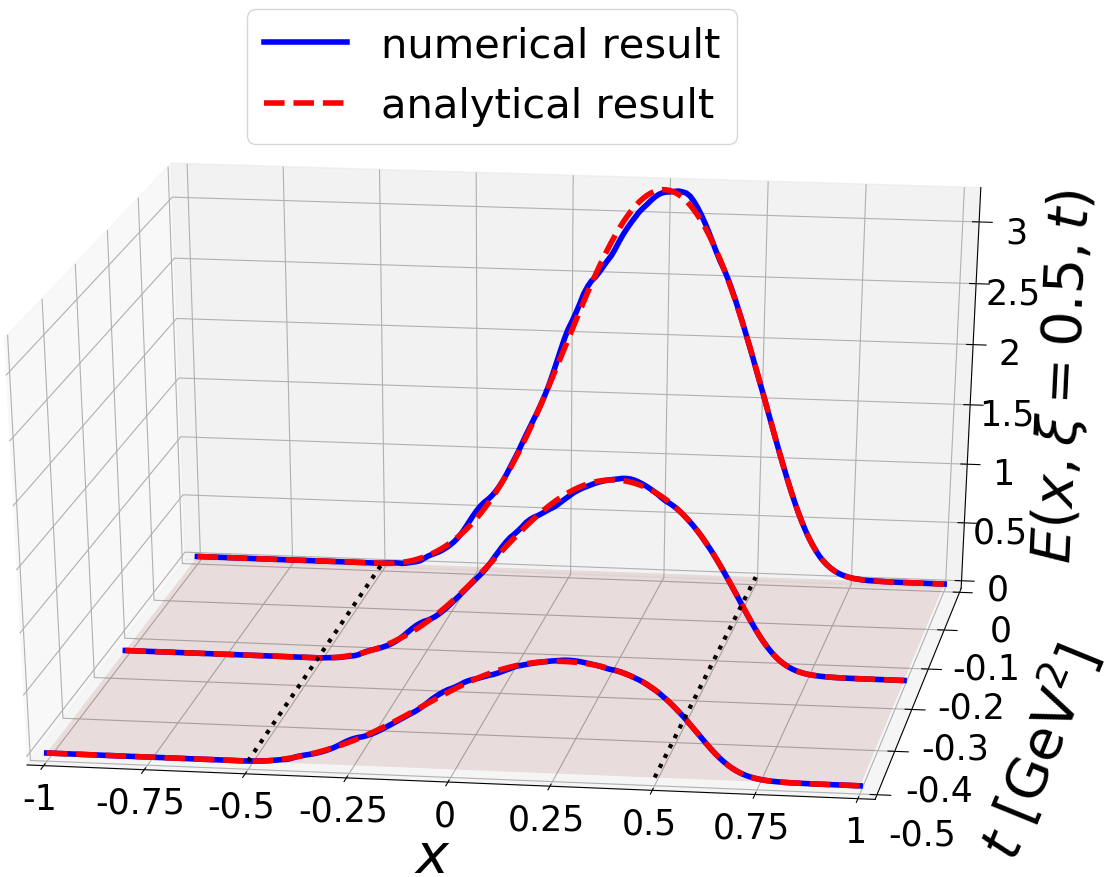}
   \end{tabular}
  \caption{Comparison for the GPD E between our numerical 
  algorithm and the algebraic result of \refcite{Hwang:2007tb}. 
  Same conventions as in \reffig{fig:AlgebraicModel}.}
  \label{fig:MuellerComparison}
\end{figure*}

The comparison between our numerical reconstruction and the algebraic 
result is shown on \reffig{fig:MuellerComparison}.
We use the same parameters values as in \refcite{Hwang:2007tb}, 
\ie $M=1~\GeV$, $m=0.45~\GeV$, $\lambda=0.75~\GeV$ and $p=1$.
In this case, the normalization constant for the DD is $N\sim0.176$.
We stress that in \reffig{fig:MuellerComparison}, the qualifying terms 
``analytical'' and ''numerical'' refer to the DDs. 
In other words, ``analytical result'' means that the GPD is calculated 
through an integration of the ``analytical'' DD \eqref{eq:DDeMueller}
in the ERBL region (or through the integration of Eq.~(15) in
\refcite{Hwang:2007tb} in the DGLAP region), whereas ``numerical result'' means
that the GPD is calculated from the ''numerically'' reconstructed DD
(from DGLAP information only), through \eg Eqs.~\eqref{eq:DD_decomposition}, 
\eqref{eq:transform_basis-function-P1} and \eqref{eq:GPDEMueller}.

\subsection{Parametrization with Regge behabior \label{subsec:Regge}}

All the previous examples dealt with GPDs smoothly behaving at $x=0$.
However, phenomenological models for valence quark GPDs often exhibit 
an integrable singularity, typically a $1/\sqrt{x}$ behavior.
If such a GPD is related to a DD $\hp$ in the P scheme, then
it can be shown that $H(x, 0) \sim \int_{-1}^{+1} \mathrm{d}\alpha
\, \hp(x, \alpha)$ at small $x$. 
Therefore $\hp$ itself may also exhibit an integrable singularity
$1/\sqrt{\beta}$.
The numerical method presented up to now approximates the target DD by
piecewise constant or piecewise linear functions on $\Omega$.
In particular all approximations are bounded, even if in principle they
can be more and more peaked when the mesh gets thinner.
As discussed in \refsec{sec:numerical-implementation}, 
careful choices of the size of the mesh and of the number of iterations
are essential for the resolution of the inverse problem with adequate control of
the numerical noise.
With a naive discretization, it is difficult when dealing with singularities to make sense of a solution~;
it would probably require a number of iterations that is not attainable due to the necessary truncation of the regularization.
We thus adopt a more educated discretization~; knowing that we 
are dealing with Regge-type singularities, we can adapt our method accordingly,
by discretizing:
\begin{equation} \label{eq:DDsqrtbeta}
\hp'(\beta, \alpha) = \sqrt{\beta} \ \hp(\beta, \alpha)
\end{equation}
which will be less singular that $\hp$, possibly even free of singularities. 
This change of target function only modifies the kernel of \refeq{eq:quark_gpd} (it is not a Radon transform anymore), otherwise everything readily follows the same procedure.

\begin{figure}[t]
  \includegraphics[width=0.5\textwidth]{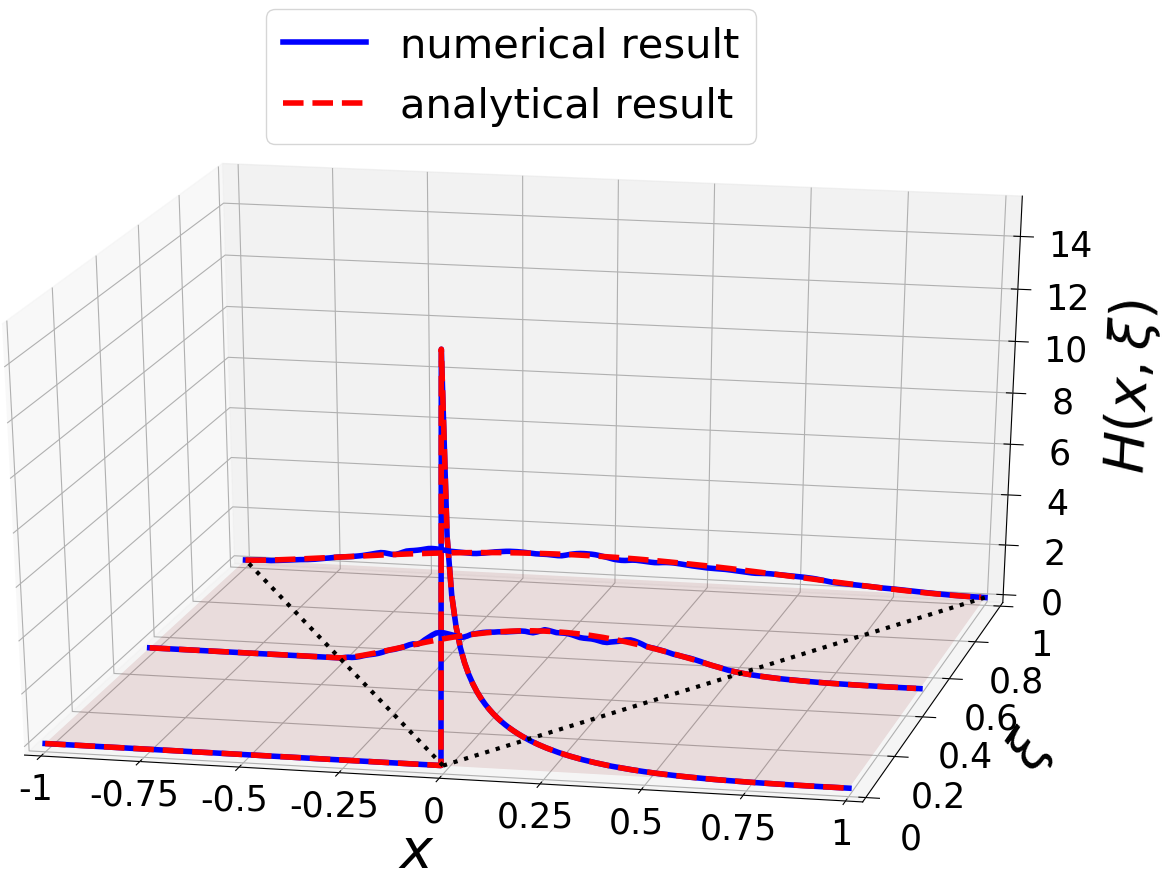}
  \caption{Comparison between algebraic and numerical results for the GPD 
  corresponding to the model of \refeq{eq:DDregge}. 
  Same conventions as in \reffig{fig:AlgebraicModel} (left panel).}
  \label{fig:ReggeComparison}
\end{figure}

Let us exemplify this technique on a simple parameterization of nucleon DD for
which, once again, the expression is already known. 
We use the RDDA as presented in \refsec{subsec:test}, \refeq{eq:DDRDDA} for $N=1$, 
but this time with a singular PDF\footnote{This simple model is similar in
spirit to the parameterizations of the nucleon GPDs $H$, $E$,
$\tilde{H}$ and $\tilde{E}$ used in popular phenomenological models, see \eg
the review \refcite{Guidal:2013rya} and refs therein.}:
\begin{equation} \label{eq:PDFregge}
q_{\textrm{Regge}}\left(x\right) = \frac{35 \ (1-x)^3}{32 \ \sqrt{x}} \ ,
\end{equation}
which would give the following DD (with a $1/\sqrt{\beta}$ singularity) on
$\Omega^{>}$:
\begin{equation} \label{eq:DDregge}
h_p^{\textrm{Regge}}\left(\beta,\alpha\right) = \frac{105 \left(\alpha ^2-\left(\beta -1\right)^2\right)}{128 \ \left(\beta -1\right) \sqrt{\beta }} \ .
\end{equation}
The comparison between the algebraic GPD and the result obtained through the numerical reconstruction of the DD with DGLAP information is shown on \reffig{fig:ReggeComparison}.

\subsection{Gaussian wave functions}

\begin{figure*}
  \centering
   \begin{tabular}{cc} 
  \includegraphics[width=0.5\textwidth]{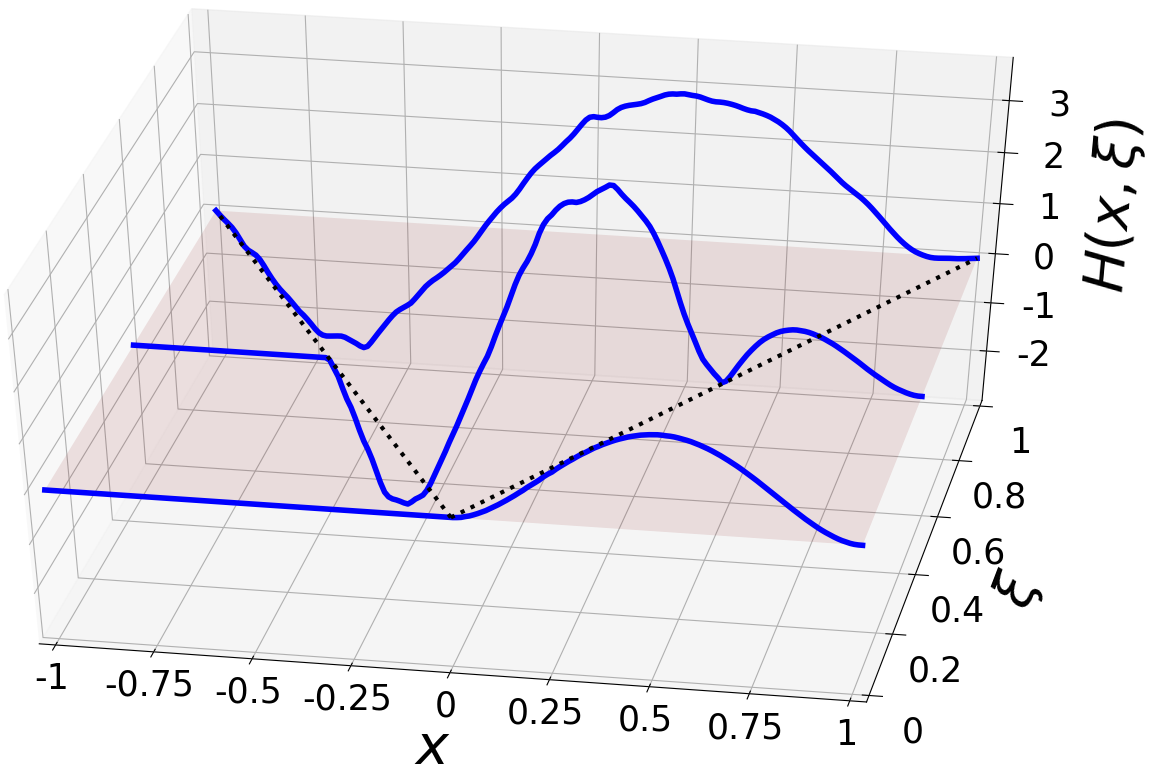} & 
   \includegraphics[width=0.5\textwidth]{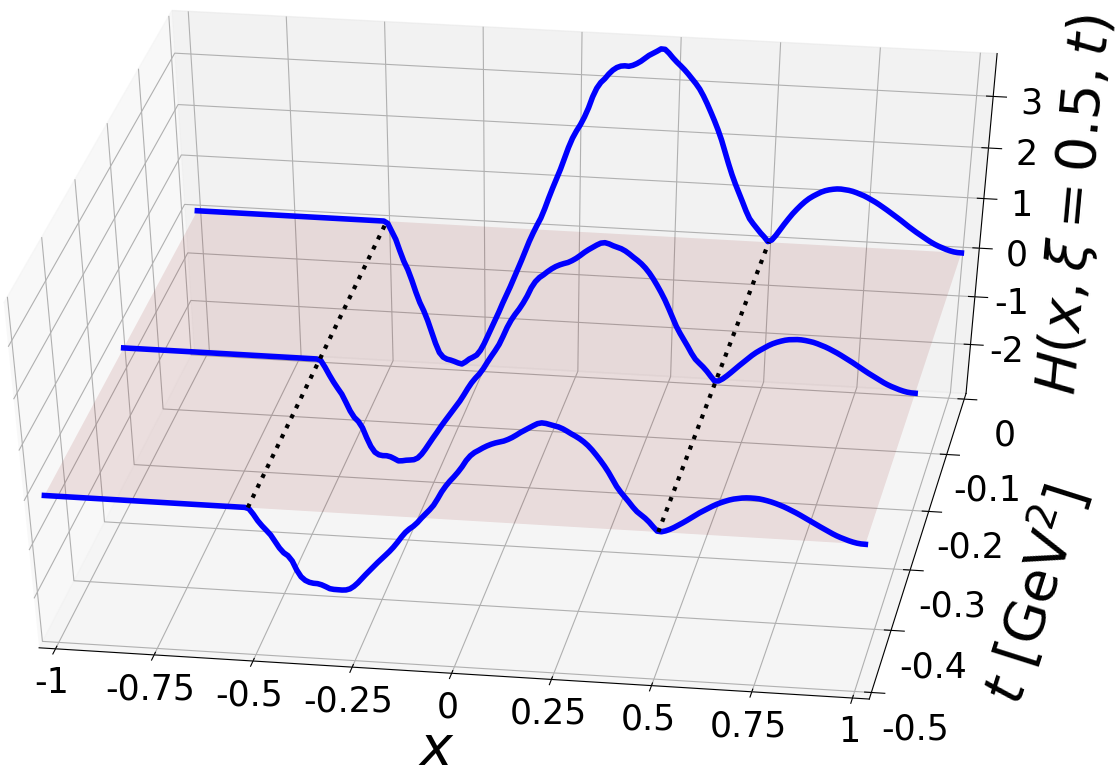}
   \end{tabular}
  \caption{Results obtained for the GPD $H$ in the case of a Gaussian LFWF.
  Same conventions as in \reffig{fig:AlgebraicModel}, but without analytical benchmark.}
  \label{fig:GaussianLFWF}
\end{figure*}

Last, on top of the Bethe-Salpeter wave function already mentioned above in the
pion case, one can compute a LFWF with a Gaussian Ansatz (such
Ansätze have been used for instance in ADS/QCD computations, see
\refcite{Brodsky:2014yha} and refs therein).
We choose to work with the following one:
\begin{equation}
  \label{eq:GaussianLFWF}
  \Psi\left(x,\mathbf{k}_{\perp}^{2}\right)=\frac{4\sqrt{15}\,\pi}{M}\,\sqrt{x\,(1-x)}\,e^{-\frac{\mathbf{k}_{\perp}^{2}}{4M^{2}(1-x)x}}\,.
\end{equation}
This wave function yields in the DGLAP region (using \refeq{eq:overlap_ud}):
\begin{equation}
  \label{eq:DGLAPFromGaussian}
H\left(x,\xi,t\right)=\frac{30(x-1)^{2}\left(x^{2}-\xi^{2}\right)^{3/2}\exp\left(\frac{\left(1-\xi^{2}\right)t(1-x)}{8M^{2}\left(\xi^{2}(x-2)+x\right)}\right)}{\left(1-\xi^{2}\right)\left(-2\xi^{2}+\xi^{2}x+x\right)}\,.
\end{equation}
To the best of our knowledge, an algebraic expression for the associated DD (it
it exists) has not been published yet. 
We anyhow examine this
case for illustrative purposes and display in \reffig{fig:GaussianLFWF} the
results for the GPD in both DGLAP and ERBL derived from the DDs obtained with our numerical inversion.
Indeed, some qualitative similarities in the shape can be noticed when compared to the results previously obtained with the Bethe-Salpeter LFWF and depicted in \reffig{fig:AlgebraicModel}. 
The parameter value of $M\sim0.315~\GeV$ has been used here.

\section{Discussion}
\label{sec:discussion}
 
\subsection{Chiral symmetry and the soft pion theorem}
\label{sec:soft-pion-theorem}

Pion GPDs should satisfy one extra first principle constraint, tied to the specific role of the pion with respect to chiral symmetry. 
The isosinglet and isovector pion GPDs $H^{I=0}$ and $H^{I=1}$ are defined in terms of the following matrix elements:
\begin{eqnarray}
  \label{eq:def-isoscalar-isovector-gpds}
\left\{ \begin{array}{c} 
 \delta^{ab} H^{I=0}(x,\xi,t) \\
   i\epsilon^{abc} H^{I=1}(x,\xi,t)
\end{array}
\right.
& = & 
\frac{1}{2} \int \frac{\mathrm{d}z^-}{2\pi} \, e^{i x P^+ z^-} 
  \\ 
 & \times & 
 \bra{\pi^b,P+\frac{\Delta}{2}} \bar{\Psi}\left(-\frac{z}{2}\right)
\left\{ \begin{array}{c}
\mathcal{I} \gamma^+ \\ 
\tau^c\gamma^+
\end{array} 
 \right\}
\wl{-\frac{z}{2}}{\frac{z}{2}}\Psi\left(\frac{z}{2}\right) \ket{\pi^a,P-\frac{\Delta}{2}}_{z^+=0, z_\perp=0} \ , 
\nonumber
\end{eqnarray}
where $\Psi=(\psi^u, \psi^d)$ denotes the doublet of $u$ and $d$ quark fields, $\mathcal{I}$ is the identity, $\tau^c$ the Pauli matrices, $\tau^{\pm} = ( \tau^1 \pm i \tau^2 ) / \sqrt{2}$, and $\ket{\pi^1}$, $\ket{\pi^2}$ and $\ket{\pi^3}$ is a basis of the adjoint representation of the Lie algebra $\mathfrak{su}(2)$. Expressing theses vectors in terms of charge eigenstates:
\begin{eqnarray}
  \label{eq:isospin_basis_charged_pions}
  \ket{\pi^\pm} & = & \frac{1}{\sqrt{2}}\left(\ket{\pi^1}\pm i\ket{\pi^2}\right) \;, \\
 \label{eq:isospin_basis_neutral_pion}
 \ket{\pi^0} & = & \ket{\pi^3} \;,
\end{eqnarray}
yields in particular:
\begin{eqnarray}
H^{I=0}(x,\xi,t) 
& = & 
H^u_{\pi^+}(x,\xi,t) + H^d_{\pi^+}(x,\xi,t) \;, \label{eq:relation-isoscalar-flavor-gpds-piplus} \\
& = & 
H^u_{\pi^-}(x,\xi,t) + H^d_{\pi^-}(x,\xi,t) \;, \label{eq:relation-isoscalar-flavor-gpds-piminus} \\
H^{I=1}(x,\xi,t) 
& = & 
H^u_{\pi^+}(x,\xi,t) - H^d_{\pi^+}(x,\xi,t) \;.\label{eq:relation-isovector-gpds-piplus} \\
& = & -H^u_{\pi^-}(x,\xi,t) + H^d_{\pi^-}(x,\xi,t) \;, \label{eq:relation-isovector-gpds-piminus}
\end{eqnarray}
The additional charge conjugation constraint, $H^q_{\pi^+}(x,\xi,t) =  - H^q_{\pi^-}(-x,\xi,t)$ for $q=u,d$ dictates:
\begin{equation}
H^u_{\pi^+}(x,\xi,t) = - H^d_{\pi^+}(-x,\xi,t) \;.
\label{eq:G-parity-GPD}
\end{equation}
Therefore the information for the whole system of pion GPDs can be equivalently encoded in $H^{I=0}$ and $H^{I=1}$, or $H^u_{\pi^+}$ and $H^d_{\pi^+}$, which makes $H^{I=0}$ (resp. $H^{I=1}$) an odd (resp. even) function of $x$: 
\begin{equation}
\label{eq:parity-isovector-isoscalar}
H^I(-x,\xi,t) = (-1)^{1-I} H^I(x,\xi,t) \quad \textrm{for} \quad I = 0, 1 \;.
\end{equation}

In the chiral limit, the soft pion theorem \cite{Polyakov:1998ze} states that:
\begin{eqnarray}
H^{I=0}(x, \xi=1, t=0) 
& = & 0 \;, \label{eq:soft_pion_theorem_isoscalar} \\
H^{I=1}(x, \xi=1, t=0)
& = & 
\varphi_{\pi}\left(\frac{1+x}{2}\right) \;, \label{eq:soft_pion_theorem_isovector} 
\end{eqnarray}
where $\varphi_{\pi}$ is the leading-twist pion DA. 
This theorem holds in any framework where chiral symmetry is properly implemented, as clearly exemplified in a recent independent derivation in the \ds-\bs framework \cite{Mezrag:2014jka}. 

We saw in \refsec{sec:formalization} that an overlap of LFWFs in the DGLAP region characterizes a set of GPDs of the form:
\begin{equation}
H(x, \xi) = H_{\textrm{LFWF}}(x, \xi) + \delta H(x, \xi) \;,
\end{equation}
with $\delta H$ defined in \refeq{eq:ambiguity-on-H-from-dglap} and $H_{\textrm{LFWF}}(x, \xi)$ obtained from a numerical inversion of the Radon transform in a particular DD scheme. 
By linearity, this relation holds in a flavor basis or in an isospin basis. 
In particular, the $x$-parity of $D^+$ and $D^-$ induces two different contributions to the isoscalar and isovector GPDs, and hence to the soft pion theorem:
\begin{eqnarray}
\big(D^{u -}+D^{d -}\big)(x, t=0) 
& = &
- H_{\textrm{LFWF}}^{I=0}(x, \xi=1, t=0) \;, \label{eq:soft-pion-theorem-isoscalar-dminus} \\
\big(D^{u +}-D^{d +}\big)(x, t=0) 
& = &
 - H_{\textrm{LFWF}}^{I=1}(x, \xi=1, t=0) + \varphi_{\pi}\left(\frac{1+x}{2}\right) \;. \label{eq:soft-pion-theorem-isovector-dplus} 
\end{eqnarray}
The requirement $\int \mathrm{d}z D^+(z)=0$ is manifest in this last equation. 
The normalization of the LFWFs used in the overlap leading to $H_{\textrm{LFWF}}$ imposes the equality of the form factor at vanishing momentum transfer and of the integral of the leading twist DA.
Therefore the remaining freedom in the covariant extension of a GPD from the DGLAP to the ERBL region is flexible enough to satisfy the soft pion theorem. The latter is explicitly shown by a more detailed analysis of the algebraic Bethe-Salpeter model in \refcite{Chouika:2017rzs}.

We already emphasized that, for any hadron, our model building strategy systematically produces GPDs satisfying \apriori polynomiality, positivity, and the correct consequences of discrete symmetries.
We have just shown that, in the pion case, the soft pion theorem can always be fulfilled too.
Whatever the input overlap is, we can construct a model obeying \apriori all the first principles constraints. The reduction formulas to obtain PDFs or form factors impose further restrictions on the choice of the LFWFs --which can be read from the overlap in the DGLAP region-- and not on the covariant extension in itself.

At last, to the best of our knowledge, nothing prevents the soft pion theorem to be satisfied in a model with an inadequate chiral symmetry behavior. 
Fulfilling the soft pion theorem may not be equivalent to enjoying a proper implementation of chiral symmetry breaking.
However it is tantalizing to think that the soft pion theorem can only be met if the underlying LFWFs do possess a correct chiral behavior, in which case we may speculate about the physics content of the $D^-$ and $D^+$ terms restoring this chiral property. 
These considerations go beyond the scope of the present paper, and are left to a later study.

\subsection{Double Distributions: smoothness assumptions and schemes}
\label{sec:smothness-assumptions}

The discussion of \refsec{sec:formalization} relied on the PW scheme, while the numerical examples of \refsec{sec:applications} involved the P scheme. 
From \refeqs{eq:relation-gpd-dd-bmks-scheme}{eq:relation-gpd-dd-pw-scheme}, it is manifest that the numerical resolution can be straightforwardly adapted from one DD scheme to the other. 
However, even if the underlying physics (the GPD itself) is invariant under the change of DD schemes, choosing one particular DD scheme may have a practical impact in actual GPD computations.
This is reminiscent of what is well-known about Green functions and Feynman
diagrams in gauge field theories or kinematics in special relativity.
Therefore we may wonder what is the impact of a DD scheme choice in the actual covariant extension of a GPD from the DGLAP to the ERBL region.
Due to the prefactors relating $H$ and the Radon transforms of DDs in \refeqs{eq:relation-gpd-dd-bmks-scheme}{eq:relation-gpd-dd-pw-scheme}, we may expect that different DD schemes will yield different GPD covariant extensions starting from the same GPD in the DGLAP region.
This may give a feeling of arbitrariness of the result.
On the contrary we will show below that the set of solutions of the inverse problem does not change. 
Whatever the choice of the DD scheme underlying the inverse of the Radon transform, given an input function in the DGLAP region, the set of GPDs whose restriction to the DGLAP region is this input function stays the same.

To prove this, we only need to study the influence of a DD scheme transformation on the set of solutions described in \refsec{sec:formalization} in the PW scheme, and in particular exhibit the manifestation of the $D^+$ and $D^-$ ambiguities in the BMKS and P schemes.
Since DD scheme transformations are linear, we only have to adapt the general expressions given in \refapp{sec:families-ocdd} to the particular case:
\begin{eqnarray}
F(\beta, \alpha) & = & \delta(\beta) D^+(\alpha) \;, \\
G(\beta, \alpha) & = & \delta(\beta) D^-(\alpha) \;.
\end{eqnarray}
Using \refeq{eq:gauge-transform-generalized-mueller-formula}, we see that the $D^+$ and $D^-$ ambiguities in the PW scheme bring a contribution $\delta \hp$ to the DD $\hp$ we would obtain by inverting the Radon transform in the P scheme:
\begin{equation}
\label{eq:propagation-ambiguities-from-pw-to-p-scheme}
\delta \hp(\beta, \alpha) = \delta(\beta) D^+(\alpha) + \frac{\theta(\beta)}{(1-\beta)^2} \bar{D}'\left(\frac{\alpha}{1-\beta}\right) \;,
\end{equation}
where $\bar{D}(z) = z D^+(z) + D^-(z)$.
The corresponding exercise in the BMKS scheme yields:
\begin{equation}
\label{eq:propagation-ambiguities-from-pw-to-bmks-scheme}
\delta \hbmks(\beta, \alpha) = \delta(\beta) \frac{D^-(\alpha)}{\alpha} - \delta'(\beta) \int_1^\alpha \mathrm{d}\sigma \frac{D^+(\sigma)}{\sigma} \;.
\end{equation}
Direct computations of the associated GPD contribution $\delta H$ establish that we actually recover the expression \eqref{eq:ambiguity-on-H-from-dglap} derived in the PW scheme. The manifestation of the ambiguities related to $D^+$ and $D^-$ nevertheless presents different forms in different DD schemes.

We will now explain with the simple example $\Htoymodel$ of \refeqs{eq:toy-model-gpd-dglap}{eq:toy-model-gpd-erbl} the impact of the different DD schemes we have considered so far.
This GPD model was defined in the BMKS scheme from \refeq{eq:def-toy-model-dd-bmks}. 
As mentioned in \refsec{sec:example-dd-schemes-toy-model} it satisfies the necessary and sufficient condition \eqref{eq:LH-BMKS-m-equal-0} expressing its membership to the range of the Radon transform of summable functions in the BMKS scheme.
We can express this GPD in the P scheme by means of \refeq{eq:gauge-transform-mueller-formula} to obtain a DD $\htoymodelp$ which is not summable over $\Omega$ and violates the analogous necessary and sufficient condition \eqref{eq:LH-Pobylitsa-m-equal-0} in the P scheme.

At the same time, we can numerically invert the Radon transform in the P scheme with the restriction $\HtoymodelDGLAP$ to the DGLAP region as an input function. 
The numerical procedure described in \refsec{sec:numerical-implementation} will approximate the DD in the P scheme by a piecewise linear function, and will naturally pick up a (reasonably) smooth solution among all the solutions of this inverse problem.  
This is indeed what has been successfully checked and showed in \reffig{fig:DDtoy} with the second test case of \refsec{subsec:test}. 
The DD $\htoymodelpreg$ \eqref{eq:toy-model-dd-regularized-pobylitsa-gauge} in the P scheme is smooth (it is a polynomial over $\Omega^>$), and is associated to a GPD equal in the DGLAP region to $\HtoymodelDGLAP$, which was derived from the original DD $\htoymodelbmks$ in the BMKS scheme or its scheme-transformed partner $\htoymodelp$ (see \refeqs{eq:toy-model-regularized-gpd-dglap}{eq:toy-model-regularized-gpd-erbl}).
The P scheme DD $\htoymodelpreg$ defines a GPD model $\Htoymodelreg$ which is smooth enough to fulfill the necessary and sufficient condition \eqref{eq:LH-Pobylitsa-m-equal-0} as explicitly evaluated in \refeq{eq:sumrule-regularized-pobylitsa-gauge-toymodel}.

The two GPD models $\Htoymodel$ and $\Htoymodelreg$ differ only in the ERBL region by a $D^-$-type contribution\footnote{We add the extra $\sgn$ function because all computations in \refapp{app:gpd-toy-model-vanish-rhombus-border} were performed under the assumption $\xi >0$. It is the presence of $\sgn(\xi)$ or $|\xi|$ that distinguishes the $D^-$ and $D^+$ contributions since $\xi$-parity reflects the time invariance of QCD.} \eqref{eq:toy-model-regularized-gpd-erbl}:
\begin{equation}
\Htoymodel(x, \xi) - \Htoymodelreg(x, \xi) = \sgn(\xi) D^-\left(\frac{x}{\xi}\right) \;,
\end{equation}
where:
\begin{equation}
\label{eq:toy-model-d-minus}
D^-(z) = 5 z (1 - z^2) \theta(1-|z|) \;.
\end{equation} 
A direct inspection of \refeq{eq:propagation-ambiguities-from-pw-to-p-scheme} reveals that, in the P scheme, the DDs $\htoymodelp$ and $\htoymodelpreg$, respectively underlying $\Htoymodel$ and $\Htoymodelreg$, should differ by a term generated by $\bar{D} = D^-$ (and $D^+ = 0$):
\begin{equation}
\label{eq:propagation-ambiguities-from-pw-to-p-scheme-toy-model}
\delta \hp(\beta, \alpha) = -15 \frac{\theta(\beta)}{(1-\beta)^2} \left[\left(\frac{\alpha}{1-\beta}\right)^2 - \frac{1}{3}\right] \;,
\end{equation}
which is exactly $\htoymodelpsing$ \eqref{eq:toy-model-dd-pobylitsa-gauge-singular-part}, the part of $\htoymodelp$ which had to be subtracted to yield the summable DD $\htoymodelpreg$ in the P scheme. 

We are therefore left with a clear picture of the inversion procedure. 
The input function $\HDGLAP$ does not know anything about DD schemes. 
In principle we may choose any DD scheme to perform the numerical inversion.
However, as pointed out here and in \refsec{sec:example-dd-schemes-toy-model}, a GPD may be related to a smooth DD in one DD scheme, and to a singular one in another scheme.
There is a difference between DD schemes when actually inverting the Radon transform, and the algorithm of \refsec{sec:numerical-implementation} will naturally select one of the smoothest DDs among the set of solutions of the inverse problem.
Consequently, the DD and the GPD provided by the inversion procedure in one DD scheme have no reason to be related by a simple DD scheme transformation to the DD and GPD obtained from the inversion procedure in another scheme.
This is also evident when looking at the necessary and sufficient conditions \refeq{eq:LH-BMKS-m-equal-0} and \refeq{eq:LH-Pobylitsa-m-equal-0} of \refsec{sec:condition-bmks-p-schemes} which seem barely compatible.
Thus for a fixed GPD model, we cannot generically expect $\hp$ and $\hbmks$ to be both reasonably smooth.
However, by properly keeping track of the $D^-$ and $D^+$ ambiguities in various DD schemes, we observe that the set of solutions to the inverse problem does not depend on the choice of the DD scheme, as it should be.

As a side remark, let us insist on the non-trivial structure of $\delta \hp$ in \refeq{eq:propagation-ambiguities-from-pw-to-p-scheme} and \refeq{eq:propagation-ambiguities-from-pw-to-p-scheme-toy-model}.
A contribution living on the $\beta = 0$ line in the PW scheme spans the whole domain $\Omega^>$ in the P scheme; the singularity structure in DDs may come under different guises when changing schemes: a $\delta(\beta)$ in the BMKS scheme manifests itself in the P scheme as a term potentially singular when $\beta$ is close to 1, and conversely. 
Forgetting that $\hp$ may not be summable, or even a compactly-supported distribution, a direct application of the support theorem of Boman and Todd Quinto quoted in \refapp{app:Radon} would have totally missed the $D^-$ contribution.
The difficulty arises from the fact that physics dictates hypothesis on the DDs $F$ and $G$, not on the DDs $\hp$ or $\hbmks$.
Regarding this, the PW scheme is a natural scheme to lead the discussion on the inversion of the Radon transform, and the propagation of the remaining freedom to fully characterize the set of solutions.
At last, the independence of the set of solutions to the inverse problem on the choice of the DD scheme implies in particular that we can select one DD scheme without loss of generality to invert the Radon transform.

We have been working with two one-component DD schemes (P and BMKS) and by an extension of the analysis of \refapp{app:gpd-toy-model-vanish-rhombus-border}, we may expect that some other one-component DD schemes exist. 
However, we already advocated for the advantages of the P scheme and the drawbacks of the BMKS scheme from a practical point of view.
We can exemplify the difference further by considering the algebraic model of \refsec{sec:AlgebraicModel}, whose DD $h_p$ is given in \refeq{eq:DDAlgebraic}.
At $t=0$, $h_p$ is simply a quadratic polynomial, and therefore can be easily handled by a computer.
Since we know the results algebraically, we can use the scheme change formulas for the BMKS (see \refapp{sec:families-ocdd}) and PW schemes (see \refeq{eq:toy-model-gauge-transform-bmks-to-polyakov-weiss-general-expression} or \refcite{Tiburzi:2004qr}). The scheme transform yields the following results:
\begin{eqnarray}
  \label{eq:AlgebraicTransformBMKSF}
  h_{\textrm{BMKS}}(\beta,\alpha)  & = & \frac{1}{4} \left(\alpha ^2
  \left(3-\frac{3}{(\left| \alpha \right| +\beta )^4}\right)+\beta ^2 \left(\frac{3}{(\left| \alpha \right|
   +\beta )^4}-3\right)\right. \nonumber \\
& & \left. +\beta  \left(4-\frac{4}{(\left| \alpha
   \right| +\beta )^3}\right)+\frac{2}{(\left| \alpha \right| +\beta
   )^2}-2 +\frac{1}{4} \left(1-\alpha ^2\right) \delta (\beta )\right ) \;,
\end{eqnarray}
in the BMKS scheme and: 
\begin{eqnarray}
  \label{eq:AlgebraicTransformPWF}
  F_{\textrm{PW}}(\beta,\alpha) & = & \frac{15}{2} \left(2 (1-\beta)
  \left(1-\beta +2\beta^2 -6 \alpha ^2\right)+\left| \alpha \right|  \left(9 \alpha ^2+4 |\alpha| -2\right) \right. \nonumber \\
    & & \left. +|\alpha|  \left(3 \alpha ^2-2\right)
    + \alpha^2 \left(2-3 |\alpha| \right)-\left| \alpha \right| ^3\right),\\
  \label{eq:AlgebraicTransformPWG}
  G_{\textrm{PW}}(\beta,\alpha) & = & -\frac{15}{2} \delta(\beta) \alpha 
  (\left| \alpha \right| -1) \left(3 \alpha ^2-\left| \alpha \right| ^2+\left| \alpha \right| -1\right) ,
\end{eqnarray}
in the PW scheme.

Several comments are in order. 
First, trading the P scheme for the BMKS one introduces singularities, making it significantly harder to be extracted numerically. 
If not singular, the PW remains nevertheless more complicated.
A similar statement holds with our toy model $\Htoymodel$ in the PW  scheme \refeqs{eq:toy-model-dd-F-polyakov-weiss}{eq:toy-model-dd-G-polyakov-weiss}. 
On top of this, the P scheme allows to extract a D-term contribution from the numerical inversion, while an inversion in the PW scheme will structurally miss a D-term and yield an incomplete polynomiality.
From the model-building perspective (and apart from the pion case and the soft pion theorem), a GPD model directly obtained from the numerical inversion of the Radon transform in the P scheme satisfies all required first principle constraints.
The need for $D^-$ and $D^+$ contributions may be dictated by phenomenology but not by first principles. 
Contrarily, a GPD model similarly produced as an output of the same algorithm but in the PW scheme will have to be complemented by a $D^-$ contribution. 
For all the reasons above, we think that there is a marked advantage in proceeding with the inversion of the Radon transform in the P scheme.

\section{Conclusion \label{sec:conclusion}}

GPDs computations, either from modeling or \textit{ab-initio} techniques, remain
a hot topic today in hadron physics, and the experiments at COMPASS, 
Jefferson Lab and on an EIC will certainly have a deep impact on our
understanding of hadron structure.
However, to fully exploit forthcoming experiments, the community needs
 models and \textit{ab-initio} techniques relying on a firm theoretical basis.

For the first time a systematic way to build GPD models
 such that \emph{all properties are \apriori fulfilled} is outlined. 
Using the method and algorithm presented here, all first-principle
theoretical constraints are met: not only the support property, but also both
positivity and polynomiality properties, the soft pion theorem in the case of pion GPDs, and, of course, the remaining, 
easier-to-implement constraints.

We showed that using of the Radon transform allows, 
 from the overlap representation of GPDs, 
 to covariantly extend results 
 obtained in the DGLAP region to the ERBL one. 
We emphasized the key role of the different DD schemes and 
how the choice of the latter impacts numerical reconstructions.
As a consequence, even knowing only a GPD in the DGLAP region, it is possible 
to obtain an extension to the ERBL region including a non-trivial D-term,
irrespective of potential additional "D-term-like" contributions. 
Such extra terms may be used for phenomenological purposes, but
are in principle not required as soon as the polynomiality property is obeyed up
to its highest degree.

We described the problems tied to the discontinuous nature of the inverse
Radon transform.
Our numerical results are in very good agreement with algebraic
evaluations when available, validating the method and 
giving confidence to its use in the vast majority of cases 
where no covariant extension of LFWF overlap is known. 
These robustness and flexibility emphasize the strength of 
our technique.

 In terms of interpretation however, one should keep in mind that 
 if the Fock state truncation is manifest in the DGLAP region, 
 it is no more visible in the ERBL one.
 The results we obtain in the ERBL region correspond to the 
 contribution needed to insure that, together with the DGLAP parts, 
 the GPD fulfills polynomiality.
 The number of Fock states implicitly used in the ERBL domain for that 
 covariant completion is so far unknown.
  This lack of interpretation in the ERBL
 region is compensated by the fact that LFWF model builders 
 and \textit{ab-initio} specialists will be able to extract 
 information from this region, and \textit{in fine} to compare 
 to experimental data.
 
 At last, on a longer time scale, the present work opens the 
 path to a phenomenology based on LFWFs, using observables 
 related not only to GPDs, but also to TMDs, PDFs, DAs or even 
 Wigner distributions.

\section*{Acknowledgments}
\label{sec:acknowledgments}

The authors would like to thank J.~Brémont, J.~Carbonell, R.~Clackdoyle,
M.~Defrise, L.~Desbat, P.~Lafitte, D.~M\"{u}ller, S.~Rit, C.D.~Roberts, C. Shi and O.~Teryaev for valuable
discussions and comments.
The authors are also thankful for the chance to participate in the workshops
“Non-Pertubative QCD 2014”, Punta Umbria, Spain, where the overlap
representation project started being discussed and “Non-Pertubative QCD 2016”,
Sevilla, Spain, where first phenomenologically relevant applications have been presented.
This work is partly supported by U.S. Department of Energy, Office of Science, Office of Nuclear
Physics, under contract no.~DE-AC02-06CH11357, 
by the Commissariat à l’Energie Atomique et aux Energies Alternatives, 
the GDR QCD “Chromodynamique Quantique”, the ANR-12-
MONU-0008-01 “PARTONS” and the Spanish ministry Research Project FPA2014-53631-C2-2-P.

\appendix

\section{The Radon Transform: selected results}
\label{app:Radon}

For $\phi \in [0, 2 \pi]$ and $s \in \mathbb{R}$, the \emph{Radon transform}
$\Radon{f}$ of the function of two real variables $f$ is defined by
\cite{Radon:1917tr}:
\begin{equation}
\label{eq:radon-transform-canonical-form}
\fRadon{f}{s}{\phi} = \int_{-\infty}^{+\infty} \mathrm{d}\beta \mathrm{d}\alpha \, f( \beta, \alpha ) \delta( s - \beta \cos \phi - \alpha \sin \phi ) \;.
\end{equation}

\noindent The variable ranges $\phi \in [0, 2 \pi]$ and $s \in \mathbb{R}$ cover
the real plane $\mathbb{R}^2$ twice. This redundancy is manifest through the
relation: 
\begin{eqnarray}
\fRadon{f}{-s}{\phi}
& = & 
\int_{-\infty}^{+\infty} \mathrm{d}\beta \mathrm{d}\alpha \, f( \beta, \alpha ) \delta( - s - \beta \cos \phi - \alpha \sin \phi ) \nonumber \\
& = & 
\int_{-\infty}^{+\infty} \mathrm{d}\beta \mathrm{d}\alpha \, f( \beta, \alpha ) \delta\big( s - \beta \cos ( \phi \pm \pi ) - \alpha \sin ( \phi \pm \pi ) \big) \nonumber \\
& = & 
\fRadon{f}{s}{\phi \pm \pi} \textrm{ for all } s \in \mathbb{R} \textrm{ and } \phi \in [0, 2\pi] \;. \label{eq:radon-transform-opposite-direction}
\end{eqnarray}

A theorem of Hertle \cite{Hertle:1983} says that a distribution $g(s, \phi)$
satisfying
\refeq{eq:radon-transform-opposite-direction} and the \emph{Ludwig-Helgason
consistency condition}\footnote{This condition is also coined
\emph{Cavalieri condition} in Teryaev's seminal paper \cite{Teryaev:2001qm}.}:
\begin{equation}
\label{eq:ludwig-helgason-consistency-condition}
\textrm{for all nonnegative integer } m \quad \int \mathrm{d}s \, s^m g(s, \phi)
= p_m(\cos \phi, \sin \phi) \,,
\end{equation}
where $p_m$ is a homogeneous polynomial of degree $m$,
is in the range of the Radon transform $\Radon$. In other words there exists a
distribution $f(\beta, \alpha)$ such that $g(s, \phi) = \fRadon{f}{s}{\phi}$. 
The converse is true from the same theorem: a function in the range of the Radon
transforms naturally satisfies \refeq{eq:radon-transform-opposite-direction} and
the Ludwig-Helgason consistency condition \ref{eq:ludwig-helgason-consistency-condition}. 

Discussing the inversion of the Radon transform goes through the determination
of its kernel. 
It has long been known (see \eg \refcite{Besicovitch:1958r}) that
a summable function which has a zero Radon transform vanishes over the whole
plane $\mathbb{R}^2$. 
Zalcman \cite{Zalcman:1982a} constructed a non-zero
non-summable continuous function admitting a zero Radon transform. 
The function $f(\beta,
\alpha) = (\beta^2 - \alpha^2)/(\beta^2 + \alpha^2)^2$ if $\alpha, \beta \neq
0$ and 0 otherwise is a simple example of a non-summable and discontinuous
function with a vanishing Radon transform.
Since we are interested here in DDs $f(\beta, \alpha)$
with support $\Omega = \left\{\left(\beta,\alpha\right) \in \mathbb{R}^2 / \,
|\beta|+|\alpha| \le 1\right\}$, we will restrict in the
following to compactly-supported functions 
where the situation dramatically simplifies.
Boman and Todd Quinto proved \cite{boman:1987rad} that if the Radon transform
$\Radon{f}$ of a compactly-supported distribution $f$ verifies
$\fRadon{f}{s}{\phi} = 0$ for $s > s_0$ and $|\phi - \phi_0| < \epsilon$ (where $s_0$ is real and $\epsilon$ real and positive) then:
\begin{equation}
\label{eq:boman-todd-quinto-theorem-made-simple}
f(\beta, \alpha) = 0 \quad \textrm{for all } \beta, \alpha \textrm{ such that }
\beta \cos \phi_0 + \alpha \sin \phi_0 > s_0 \,.
\end{equation}
In particular, remembering \refeq{eq:radon-transform-opposite-direction}, if the
Radon transform $\fRadon{f}{s}{\phi}$ vanishes for all $s$, then $f = 0$. 
By linearity, we see that a compactly-supported distribution is uniquely
determined by its Radon transform.

An important mathematical literature has been dedicated to the Radon transform,
notably because of its central role in the field of computerized tomography.
Inverting the Radon transform indeed allows to unravel the internal structure of
an object exposed to some kind of radiation propagating along straight lines. 
Sometimes it is not possible to scan a whole object; for example only directions
in an angular range smaller than $[0, 2\pi]$ may be accessible. Such cases are
referred to as incomplete data problems. 

The aforementioned theorem of Boman and Todd Quinto gives some examples of
situations where a function is uniquely determined by its Radon transform. This
allows to define the inverse Radon transform. However is has early been shown on
simple but general examples, that the inverse Radon transform
may not be continuous. In that case nothing prevents the artificial
amplification of noise (numerical or experimental when dealing with actual
measurements) when inverting the Radon transform.
This problem may be even more severe when dealing with incomplete data problems
(see \refcite{Natterer:2001m} and refs therein).

\section{Double Distributions in the BMKS and P schemes}
\label{sec:families-ocdd}

Consider a quark GPD $H$ represented by DDs in a general $(F, G)$-scheme as in \refeq{eq:gpd-in-general-F-and-G-gauge}. $H(x, \xi)$ has support $x \in [-\xi, +1]$ while $F(\beta, \alpha)$ and $G(\beta, \alpha)$ have support $\beta \geq 0$. This choice conveniently simplifies calculations without loss of generality since we will consider only linear relations. 

Is it possible to transform this (F, G)-representation following \refeqs{eq:gauge-transform-F}{eq:gauge-transform-G} 
in order to get a GPD representation in the P scheme:
\begin{equation}
\label{eq:def-pobylitsa-gauge}
\int_{\Omega} \textrm{d}\beta
\textrm{d}\alpha \big[F(\beta, \alpha) + \xi G(\beta, \alpha)\big] \delta(x - \beta - \alpha \xi) = (1-x)
\int_{\Omega} \textrm{d}\beta \textrm{d}\alpha \hp(\beta, \alpha) \delta(x - \beta - \alpha \xi)
\;?
\end{equation}

In other words, we look for an $\alpha$-odd function $\chip(\beta, \alpha)$,
vanishing on the border of $\Omega$: $\{(\beta, \alpha) \in
\mathbb{R}^2 / |\alpha|+|\beta|=1\}$, such that:
\begin{eqnarray}
F(\beta, \alpha) + \frac{\partial \chip}{\partial
\alpha}(\beta, \alpha) 
& = & 
(1 - \beta) \hp(\beta, \alpha) \;,
\label{eq:gauge-transform-alpha-derivative} \\
G(\beta, \alpha) - \frac{\partial \chip}{\partial
\beta}(\beta, \alpha) 
& = & 
- \alpha \hp(\beta, \alpha) \;.
\label{eq:gauge-transform-beta-derivative}
\end{eqnarray}
Assuming $\alpha \neq 0$, dividing \refeq{eq:gauge-transform-beta-derivative} by $\alpha$ and inserting in
\refeq{eq:gauge-transform-alpha-derivative} provides the new system of differential equations to solve:
\begin{eqnarray}
\hp(\beta, \alpha) 
& = & 
- \frac{1}{\alpha} G(\beta, \alpha) +
\frac{1}{\alpha} \frac{\partial \chip}{\partial
\beta}(\beta, \alpha) \;,
\label{eq:gauge-transform-solution-for-flambda} \\
(1 - \beta) \frac{\partial \chip}{\partial
\beta}(\beta, \alpha) - \alpha \frac{\partial \chip}{\partial
\alpha}(\beta, \alpha) 
& = &  
\alpha F(\beta, \alpha) + (1-\beta) G(\beta, \alpha) \;.
\label{eq:gauge-transform-chi-pde}
\end{eqnarray}
We will parameterize the domain $\Omega$ by trajectories $\bar{\alpha}(\tau)$,
$\bar{\beta}(\tau)$ originating at $\tau = 0$ from a position $(\beta_0,
\alpha_0) \in \Omega$ (with $\alpha_0 \neq 0$) and such that:
\begin{eqnarray}
\frac{\mathrm{d}\bar{\beta}}{\mathrm{d}\tau}(\tau) 
& = & 1 - \bar{\beta}(\tau) \;, \label{eq:def-define-betabar-trajectory}
\\
\frac{\mathrm{d}\bar{\alpha}}{\mathrm{d}\tau}(\tau) 
& = & - \bar{\alpha}(\tau) \;. \label{eq:def-define-betabar-trajectory}
\end{eqnarray}
On such trajectories, \refeq{eq:gauge-transform-chi-pde} writes:
\begin{equation}
\label{eq:gauge-transform-chi-ode}
\frac{\mathrm{d}\chip}{\mathrm{d}\tau}(\tau) =  
\bar{\alpha}(\tau) F\big(\bar{\beta}(\tau), \bar{\alpha}(\tau)\big) + \big(1-\bar{\beta}(\tau)\big) G\big(\bar{\beta}(\tau), \bar{\alpha}(\tau)\big) \;.
\end{equation}
This last equation can directly be solved by quadrature:
\begin{equation}
\label{eq:gauge-transform-quadrature}
\chip\big(\bar{\beta}(\tau), \bar{\alpha}(\tau)\big) = \chip(\beta_0, \alpha_0) +
\int_0^\tau \mathrm{d}\sigma \, \bar{\alpha}(\sigma) F\big(\bar{\beta}(\sigma), \bar{\alpha}(\sigma)\big) + \big(1-\bar{\beta}(\sigma)\big) G\big(\bar{\beta}(\sigma), \bar{\alpha}(\sigma)\big) \;.
\end{equation}
But \refeq{eq:def-define-betabar-trajectory} and
\refeq{eq:def-define-betabar-trajectory} are respectively equivalent to:
\begin{eqnarray}
\bar{\beta}(\tau)
& = & 
1 + (\beta_0 - 1) e^{-\tau} \;,
\label{eq:gauge-transform-betabar-solution} \\
\bar{\alpha}(\tau)
& = & 
\alpha_0 e^{-\tau} \;.
\label{eq:gauge-transform-alphabar-solution}
\end{eqnarray}
The change of variable $\omega = \bar{\beta}(\sigma)$ in
\refeq{eq:gauge-transform-quadrature} is now straightforward:
\begin{equation}
\label{eq:gauge-transform-quadrature-integrated-over-first-dd-variable}
\chip\big(\bar{\beta}(\tau), \bar{\alpha}(\tau)\big) = \chip(\beta_0, \alpha_0) +
\int_{\beta_0}^{\bar{\beta(\tau)}} \mathrm{d}\omega \,
\left[\frac{\alpha_0}{1 - \beta_0} F\left(\omega,
 (1-\omega) \frac{\alpha_0}{1-\beta_0}\right) + G\left(\omega,
 (1-\omega) \frac{\alpha_0}{1-\beta_0}\right)\right] \;.
\end{equation}
Moreover $\bar{\alpha}(\tau)/(1 - \bar{\beta}(\tau))$ is a constant of
motion along the trajectories parameterized by $\sigma$, which brings:
\begin{eqnarray}
\chip\big(\bar{\beta}(\tau), \bar{\alpha}(\tau)\big) 
& = & \chip(\beta_0, \alpha_0) +
\frac{\bar{\alpha}(\tau)}{1 - \bar{\beta}(\tau)} \int_{\beta_0}^{\bar{\beta(\tau)}} \mathrm{d}\omega \, \left[F\left(\omega,
 (1-\omega) \frac{\bar{\alpha}(\tau)}{1-\bar{\beta}(\tau)}\right) \right. \nonumber \\
& & \quad \left. + \frac{1 - \bar{\beta}(\tau)}{\bar{\alpha}(\tau)} G\left(\omega,
 (1-\omega) \frac{\bar{\alpha}(\tau)}{1-\bar{\beta}(\tau)}\right)\right]
 \;.
\label{eq:gauge-transform-quadrature-constant-of-motion}
\end{eqnarray}
From now on we particularize to the case $\beta_0 = 0$. 
The
considered trajectories $\big(\bar{\beta}(\tau), \bar{\alpha}(\tau)\big)_{\tau}$
become:
\begin{eqnarray}
\bar{\beta}(\tau)
& = & 
1 - e^{-\tau} \;,
\label{eq:gauge-transform-betabar-solution-pobylitsa} \\
\bar{\alpha}(\tau)
& = & 
\alpha_0 e^{-\tau} \;.
\label{eq:gauge-transform-alphabar-solution-pobylitsa}
\end{eqnarray}
All values of $(\beta, \alpha) \in \Omega$ with $\beta > 0$ can be reached with
a suitable choice of $\tau > 0$ and $\alpha_0 \in [-1, +1]$. 
We can also check
that $\big(\bar{\beta}(\tau), \bar{\alpha}(\tau)\big) \in \Omega$ for all $\tau
> 0$ and $\alpha_0 \in [-1, +1]$. 
Therefore
\refeq{eq:gauge-transform-quadrature-constant-of-motion} allows the
determination of $\chi$ over the whole domain of interest.

Since $\bar{\alpha}(\tau)/(1 - \bar{\beta}(\tau)) = \alpha_0$ (constant of
motion), \refeq{eq:gauge-transform-quadrature-constant-of-motion} simplifies to:
\begin{equation}
\label{eq:gauge-transform-quadrature-constant-of-motion-without-initial-condition}
\chip(\beta, \alpha) = \chip\left(0,
\frac{\alpha}{1-\beta}\right) + \frac{\alpha}{1 -
\beta} \int_0^{\beta} \mathrm{d}\omega \, 
\left[F\left(\omega,
 (1-\omega) \frac{\alpha}{1-\beta}\right) + \frac{1-\beta}{\alpha} G\left(\omega,
 (1-\omega) \frac{\alpha}{1-\beta}\right)\right] \;.
\end{equation}
Thus,
for
\refeqs{eq:gauge-transform-alpha-derivative}{eq:gauge-transform-beta-derivative}
to hold, there should exist a function of \emph{one variable} $C$ such that:
\begin{equation}
\label{eq:gauge-transform-quadrature-constant-of-motion-with-extra-function}
\chip(\beta, \alpha) = C\left(\frac{\alpha}{1-\beta}\right)
 + \frac{\alpha}{1 -
\beta} \int_0^{\beta} \mathrm{d}\omega \, 
\left[F\left(\omega,
 (1-\omega) \frac{\alpha}{1-\beta}\right) + \frac{1-\beta}{\alpha} G\left(\omega,
 (1-\omega) \frac{\alpha}{1-\beta}\right)\right] \;.
\end{equation}
Since $\chip(\beta, \alpha)$ is $\alpha$-odd, $C$ has to be odd. 
The boundary
property $\chip(\beta, 1-\beta) = 0$ for all $\beta >0$ is met when $C(1) +
\int_0^\beta \mathrm{d}\omega \, (F+G)(\omega, 1-\omega) = 0$, which implies
$C(1) = 0$ and $F+G = 0$ on the boundary of $\Omega$ with $\beta \geq 0$. 
The converse is also true: if $F+G$ vanishes on the border of $\Omega$, and
if $C$ is odd and satisfies $C(1)=0$, then $\chip$ given by the \rhs of
\refeq{eq:gauge-transform-quadrature-constant-of-motion-with-extra-function} obeys the assumed support property.

In particular, a DD in a general $(F, G)$-scheme cannot be cast in the P scheme by
the transformations \refeqs{eq:gauge-transform-F}{eq:gauge-transform-G} if it
does not vanish on the border of $\Omega$. 

Assuming that we can choose $C = 0$, the DD in the P scheme reads:
\begin{equation}
\label{eq:gauge-transform-generalized-mueller-formula}
\hp(\beta, \alpha) = 
- \frac{1}{\alpha} G(\beta, \alpha) +
\frac{\partial}{\partial \beta} \int_0^\beta
\frac{\mathrm{d}\omega}{1 - \beta} \, 
\left[F\left(\omega,
 (1-\omega) \frac{\alpha}{1-\beta}\right) + \frac{1-\beta}{\alpha} G\left(\omega,
 (1-\omega) \frac{\alpha}{1-\beta}\right)\right] \;.
\end{equation} 
Particularizing to the BMKS scheme as the input DD scheme allows to recover the
formulas first written in \refcite{Mueller:2014hsa}:
\begin{equation}
\label{eq:gauge-transform-mueller-formula}
\chi_{\textrm{P}}(\beta, \alpha) = \frac{\alpha}{1 - \beta} \int_0^\beta
\mathrm{d}\omega \, 
\hbmks\left(\omega, (1-\omega) \frac{\alpha}{1-\beta}\right) \;,
\end{equation}
and:
\begin{equation}
\label{eq:relation-between-bmks-and-pobylitsa}
\hp(\beta, \alpha) = - \hbmks(\beta, \alpha) +
\frac{\partial}{\partial \beta} \int_0^\beta
\frac{\mathrm{d}\omega}{1 - \beta} \, 
\hbmks\left(\omega, (1-\omega)
\frac{\alpha}{1-\beta}\right) \;.
\end{equation}

Conversely, any $\chi_C(\beta, \alpha) = C\big(\alpha/(1-\beta)\big)$ is a
solution of:
\begin{equation}
\label{eq:gauge-transform-chi-homogeneous-pde}
(1 - \beta) \frac{\partial \chi_C}{\partial
\beta}(\beta, \alpha) - \alpha \frac{\partial \chi_C}{\partial
\alpha}(\beta, \alpha) 
= 0 \;.
\end{equation}
By making repeated use of $(1-\beta) \partial/\partial \beta
\big(\alpha/(1-\beta)\big) = \alpha \partial/\partial \alpha
\big(\alpha/(1-\beta)\big)$, it is easy to check that any $\chip$ as in
\refeq{eq:gauge-transform-quadrature-constant-of-motion-with-extra-function}
(with $C(1) = 0$) is a solution of \refeq{eq:gauge-transform-chi-pde}.

The extra function $C$ in
\refeq{eq:gauge-transform-quadrature-constant-of-motion-with-extra-function}
brings a contribution $h_{\textrm{P}, \textrm{C}}$ to the Pobylitsa DD
$\hp$ through \refeq{eq:gauge-transform-solution-for-flambda}:
\begin{equation}
\label{eq:dd-pobylitsa-additive-contribution}
h_{\textrm{P}, \textrm{C}}(\beta, \alpha) = \frac{1}{\alpha}
\frac{\partial}{\partial \beta} C\left(\frac{\alpha}{1 - \beta}\right) =
\frac{1}{(1-\beta)^2} C'\left(\frac{\alpha}{1 - \beta}\right) \;,
\end{equation}
where $C'$ denotes the first derivative of the function of one variable $C$.
$h_{\textrm{P}, \textrm{C}}(\beta, \alpha)$ vanishes for $\beta < 0$ if $C'$ is
zero, and $C$ is itself null (since it is an odd function). The only function
$\hp$ such that \refeq{eq:def-pobylitsa-gauge} holds is thus expressed in
\refeq{eq:gauge-transform-generalized-mueller-formula}.

Following similar steps, it is easy to derive a transformation from a $(F, G)$-scheme to the BMKS scheme.
We look for an $\alpha$-odd function $\chibmks(\beta, \alpha)$,
vanishing on the border of $\Omega$, such that:
\begin{eqnarray}
F(\beta, \alpha) + \frac{\partial \chibmks}{\partial
\alpha}(\beta, \alpha) 
& = & 
\beta \hbmks(\beta, \alpha) \;,
\label{eq:gauge-transform-alpha-derivative-bmks} \\
G(\beta, \alpha) - \frac{\partial \chibmks}{\partial
\beta}(\beta, \alpha) 
& = & 
\alpha \hbmks(\beta, \alpha) \;.
\label{eq:gauge-transform-beta-derivative-bmks}
\end{eqnarray}
This system can be solved for $\chibmks$:
\begin{equation}
\label{eq:gauge-transform-quadrature-bmks}
\chibmks(\beta, \alpha) = \int_{\beta/(\alpha+\beta)}^{\beta} \mathrm{d}\omega \, 
\left[-\frac{\alpha}{\beta} F\left(\omega,
 \omega \frac{\alpha}{\beta}\right) + G\left(\omega,
 \omega \frac{\alpha}{\beta}\right)\right] \;,
\end{equation}
which can also be found in the literature, \eg in \refcite{Belitsky:2005qn}.

\section{A simple example exhibiting the polynomiality property}
\label{app:gpd-toy-model-vanish-rhombus-border}

We highlight here some subtleties
related to the various DD schemes. 
To apply the discussion of \refsec{sec:families-ocdd} leading to
\refeq{eq:gauge-transform-quadrature-constant-of-motion-with-extra-function}, we
need a DD vanishing on the boundary of $\Omega$. 
We simplify further by
considering a quark GPD (hence the related DD vanishes for negative $\beta$) and
a DD expressed as polynomials to simplify the computations of Mellin moments or
GPDs.
We scrutinize the simple GPD model of \refeqs{eq:toy-model-gpd-dglap}{eq:toy-model-gpd-erbl}, based on the DD \eqref{eq:def-toy-model-dd-bmks} to show what can be expected in a simple example.  
 
Table~\ref{tab:mellin-moments-gpd-example-toy-model-dd} displays the first eleven Mellin moments of $H$,
as well as the contributions to the Mellin moments of the DGLAP and ERBL
regions.
The polynomiality \refeq{eq:def-polynomiality} property manifestly hold for
this set of Mellin moments.

\begin{table*}
\begin{center}
\begin{tabular}{c|c|c|c}
\hline
\hline
$n$	& $\int_{+\xi}^{+1} \mathrm{d}x \, x^n \HtoymodelDGLAP(x, \xi)$				&
$\int_{-\xi}^{+\xi} \mathrm{d}x \, x^n \HtoymodelERBL(x, \xi)$	& $\int_{-\xi}^{+1}
\mathrm{d}x \, x^n \Htoymodel(x, \xi)$ \\
\hline
0	& $\frac{(-1+\xi )^2 (1+4 \xi )}{(1+\xi )^2}$							& $-\frac{4 (-2+\xi ) \xi ^2}{(1+\xi )^2}$						& 1 \\
1	& $\frac{(-1+\xi )^2 \left(1+4 \xi +10 \xi ^2\right)}{3 (1+\xi )^2} $					& $ -\frac{4 \xi ^3 (-5+2 \xi )}{3 (1+\xi )^2}$ 						& $ \frac{1}{3} \left(1+2 \xi ^2\right)$ \\
2	& $ \frac{(-1+\xi )^2 \left(1+4 \xi +10 \xi ^2+20 \xi ^3\right)}{7 (1+\xi )^2}$ & 
$ -\frac{4 \xi ^4 (-8+5 \xi )}{7 (1+\xi )^2}$ 					& $ \frac 1 7 (1+2 \xi^2)$ \\
3	& $\frac{(-1+\xi )^2 \left(1+4 \xi +10 \xi ^2+20 \xi ^3+35 \xi ^4\right)}{14
   (1+\xi )^2} $		& $-\frac{4 \xi ^5 (-7+4 \xi )}{7 (1+\xi )^2} $					& $ \frac{1}{14} \left(1+2 \xi ^2+3 \xi ^4\right)$ \\
4	& $ \frac{5 (-1+\xi )^2 \left(1+4 \xi +10 \xi ^2+20 \xi ^3+35 \xi ^4+56 \xi
   ^5\right)}{126 (1+\xi )^2}$		&  $	-\frac{20 \xi ^6 (-10+7 \xi )}{63 (1+\xi )^2}$				& $\frac{5}{126} \left(1+2 \xi ^2+3 \xi ^4\right) $ \\
5	& $ \frac{6+6 \xi ^7 \left(-120+315 \xi -280 \xi ^2+84 \xi ^3\right)}{252
   \left(-1+\xi ^2\right)^2}$					& $ -\frac{20 \xi ^7 (-3+2 \xi )}{21 (1+\xi )^2}$					& $ \frac{1}{42} \left(1+2 \xi ^2+3 \xi ^4+4 \xi ^6\right)$ \\
6	& $ \frac{6+6 \xi ^8 \left(-165+440 \xi -396 \xi ^2+120 \xi ^3\right)}{396
   \left(-1+\xi ^2\right)^2}$					& $ -\frac{20 \xi ^8 (-4+3 \xi )}{33 (1+\xi )^2}$					& $ \frac{1}{66} \left(1+2 \xi ^2+3 \xi ^4+4 \xi ^6\right)$ \\
7	& $\frac{6+6 \xi ^9 \left(-220+594 \xi -540 \xi ^2+165 \xi ^3\right)}{594
   \left(-1+\xi ^2\right)^2}$				& $ -\frac{20 \xi ^9 (-11+8 \xi )}{99 (1+\xi )^2}$					& $ \frac {1}{99} \left(1+2 \xi^2+3 \xi^4+4 \xi^6+5 \xi^8\right)$ \\
8	& $\frac{6+6 \xi ^{10} \left(-286+780 \xi -715 \xi ^2+220 \xi ^3\right)}{858
   \left(-1+\xi ^2\right)^2} $			& $ -\frac{20 \xi ^{10} (-14+11 \xi )}{143 (1+\xi )^2}$				& $ \frac{1}{143} \left(1+2 \xi ^2+3 \xi ^4+4 \xi ^6+5 \xi ^8\right)$ \\
9	& $\frac{5 \left(6+6 \xi ^{11} \left(-364+1001 \xi -924 \xi ^2+286 \xi
   ^3\right)\right)}{6006 \left(-1+\xi ^2\right)^2} $			& $ -\frac{20 \xi ^{11} (-13+10 \xi )}{143 (1+\xi )^2}$				& $ \frac{5}{1001} \left(1+2 \xi ^2+3 \xi ^4+4 \xi ^6+5 \xi ^8+6 \xi
   ^{10}\right)$ \\
10	& $\frac{6+6 \xi ^{12} \left(-455+1260 \xi -1170 \xi ^2+364 \xi
   ^3\right)}{1638 \left(-1+\xi ^2\right)^2}$			& $ -\frac{4 \xi ^{12} (-16+13 \xi )}{39 (1+\xi )^2}$				& $ \frac{1}{273} \left(1+2 \xi ^2+3 \xi ^4+4 \xi ^6+5 \xi ^8+6 \xi
   ^{10}\right)$ \\
\hline
\hline
\end{tabular}
\caption{First 11 Mellin moments of the GPD model defined in 
\refeqs{eq:gpd-toy-model-mellin-moment-erbl}{eq:gpd-toy-model-mellin-moment-dglap} and the
associated contributions of DGLAP and ERBL regions. }
\label{tab:mellin-moments-gpd-example-toy-model-dd}
\end{center}
\end{table*}

Obtaining analytic, all-order expressions for these integrals is
straightforward:
\begin{align}
&\frac{\Gamma(n+6)}{\Gamma(n+2)} (\xi +1)^2 \int_{-\xi}^{+\xi} \mathrm{d}x \, x^n \ \HtoymodelERBL(x, \xi) \nonumber \\
 = &-10 \xi ^{n+2} \Big(-(n+4) \left(2 n^2+16 n-3 (-1)^n+27\right) + \xi(n+3) \left(2n(n+6)+3 (-1)^n+13\right) \Big) \;,
   \label{eq:gpd-toy-model-mellin-moment-erbl} \\
&\frac{1}{20} (n+2) \left(\xi ^2-1\right)^2 \int_{-\xi}^{+1} \mathrm{d}x \, x^n
\HtoymodelDGLAP(x, \xi) \nonumber \\
 = &  
\left((n+2) \xi  \left(\frac{\xi ^2}{n+5}-\frac{3 \xi }{n+4}+\frac{3}{n+3}\right)-1\right) \xi ^{n+2}
+ 6\, \frac{\Gamma(n+3)}{\Gamma(n+6)} \;,
   \label{eq:gpd-toy-model-mellin-moment-dglap}  
\end{align}
where $\Gamma$ is the Euler gamma function.

Following \refeq{eq:gauge-transform-mueller-formula} the function:
\begin{equation}
\label{eq:toy-model-gauge-transform-bmks-to-pobylitsa}
\chi_{\textrm{BMKS}\rightarrow\textrm{P}}(\beta, \alpha) = \frac{5 \beta  ((\beta -3) \beta +3) \left(\alpha ^2-(\beta -1)^2\right)}{(\beta
   -1)^3} \theta(\beta) \;,
\end{equation}
brings the DD $\htoymodelbmks$ to the P scheme:
\begin{equation}
\label{eq:toy-model-dd-pobylitsa}
\htoymodelp(\beta, \alpha) = 5 \left(\alpha ^2 \left(3-\frac{3}{(\beta -1)^4}\right)+(4-3 \beta ) \beta
   +\frac{1}{(\beta -1)^2}-1\right) \theta(\beta) \;.
\end{equation}
Stated differently, this $\htoymodelp$ is such that:
\begin{equation}
\label{eq:toy-model-gpd-in-pobylitsa-gauge}
\Htoymodel(x, \xi) = (1 - x) \int_\Omega \mathrm{d}\beta \mathrm{d}\alpha \,
\htoymodelp(\beta, \alpha) \delta(x - \beta - \alpha \xi) \;.
\end{equation}
However $\htoymodelp$ has a singularity at $\beta = 1$, which makes it
non-summable over $\Omega$ since:
\begin{equation}
\label{eq:toy-model-dd-pobylitsa-gauge-non-summability}
\int_0^1 \mathrm{d}\beta \int_{-1+\beta}^{+1-\beta} \mathrm{d}\alpha \,
\htoymodelp(\beta, \alpha) = \frac{5}{3} \neq -5 = \int_{-1}^{+1}
\mathrm{d}\alpha \int_{0}^{1-|\alpha|} \mathrm{d}\beta \, \htoymodelp(\beta,
\alpha) \;,
\end{equation}
while the DDs $\Fctoymodelp(\beta, \alpha) = (1-\beta) \htoymodelp(\beta,
\alpha)$ and $\Gctoymodelp(\beta, \alpha) = -\alpha \htoymodelp(\beta,
\alpha)$ are summable over $\Omega$. 
Although $\htoymodelp$ being non-summable
does not violate any first principle requirement on GPDs, this feature may
hinder any numerical approximation of $\htoymodelp$.
The singular part of $\htoymodelp$ can be minimally defined as:
\begin{equation}
\label{eq:toy-model-dd-pobylitsa-gauge-singular-part}
\htoymodelpsing(\beta, \alpha) = -15 \left(\frac{\alpha ^2}{(\beta
-1)^4}-\frac{1}{3 (\beta -1)^2}\right) \theta(\beta) \;.
\end{equation}
The \emph{regularized} DD:
\begin{eqnarray}
\htoymodelpreg(\beta, \alpha)
& = &
\htoymodelp(\beta, \alpha) - \htoymodelpsing(\beta, \alpha)
\label{eq:toy-model-dd-regularized-pobylitsa-gauge-def} \\
& = & 
5 \left(3 \alpha ^2-3 \beta ^2+4 \beta -1\right) \theta(\beta) \;,
\label{eq:toy-model-dd-regularized-pobylitsa-gauge} 
\end{eqnarray}
is smooth and summable over $\Omega$.
The related GPD is unchanged in the DGLAP region, but modified
in the ERBL region:
\begin{eqnarray}
\HtoymodelregDGLAP(x, \xi)
& = &
\HtoymodelDGLAP(x, \xi) \;, 
\label{eq:toy-model-regularized-gpd-dglap} \\
\HtoymodelregERBL(x, \xi)
& = &
\HtoymodelERBL(x, \xi) + \frac{5 \left(x^3-\xi ^2 x\right)}{\xi ^3} \;. 
\label{eq:toy-model-regularized-gpd-erbl}
\end{eqnarray}

For the sake of completeness, we briefly sketch\footnote{The transition between
a general $(F, G)$-scheme and the PW scheme has already been described in
many papers and review articles, see \eg \refcite{Belitsky:2005qn}.}
how we can convert our model to the PW scheme.
Using the general definition of the D-term \refeq{eq:def-dterm} to solve 
\refeqs{eq:gauge-transform-F}{eq:gauge-transform-G}, we observe that:
\begin{eqnarray}
\chi_{\textrm{(F, G)}\rightarrow\textrm{PW}}(\beta, \alpha) 
& = &
\int_{-1+|\alpha|}^\beta \mathrm{d}\gamma \, G(\gamma, \alpha) - \theta(\beta)
\int_{-1+|\alpha|}^{+1-|\alpha|} \mathrm{d}\gamma G(\gamma, \alpha)
\label{eq:toy-model-gauge-transform-bmks-to-polyakov-weiss-general-expression}
\\
& = & 
5 \alpha (\left| \alpha \right| +\beta -1) \left(-3 \alpha
   ^2-(\beta -1) \left| \alpha \right| +\left| \alpha \right| ^2+(\beta
   -1)^2\right) \theta (\beta ) \;,
   \label{eq:toy-model-gauge-transform-bmks-to-polyakov-weiss}
\end{eqnarray}
brings the DD $\htoymodelbmks$ to the PW scheme:
\begin{eqnarray}
\Fctoymodelpw(\beta, \alpha) 
& = &
-5 \left(3 \alpha ^2 (4 \beta -3)+9 \alpha ^2 \left| \alpha
   \right| -\left| \alpha \right| ^3-(\beta -1)^2 (4 \beta -1)\right) \theta
   (\beta )  \;.\label{eq:toy-model-dd-F-polyakov-weiss}
   \\
\Gctoymodelpw(\beta, \alpha) 
& = &
5 \alpha  \left(2 \alpha ^2-\left| \alpha \right| -1\right) (\left| \alpha
   \right| -1) \delta(\beta)
   \;.\label{eq:toy-model-dd-G-polyakov-weiss}
\end{eqnarray}
Stated differently, these $\Fctoymodelpw$ and $\Gctoymodelpw$ are such that:
\begin{equation}
\label{eq:toy-model-gpd-in-polyakov-gauge}
\Htoymodel(x, \xi) = \int_\Omega \mathrm{d}\beta \mathrm{d}\alpha \,
\big( \Fctoymodelpw(\beta, \alpha) + \xi \Gctoymodelpw(\beta, \alpha)
\big) \delta(x - \beta - \alpha \xi) \;.
\end{equation} 
The related GPD is unchanged in both DGLAP and ERBL regions.

\bibliography{Bibliography}

\end{document}